\documentclass[10pt,a4paper]{article}
\usepackage[utf8]{inputenc}
\usepackage{amsmath}
\usepackage{amsfonts}
\usepackage{amssymb}

\usepackage{graphicx}
\usepackage{cancel}
\usepackage[a4paper,
            bindingoffset=0.2in,
            left=1in,
            right=1in,
            top=1in,
            bottom=1in,
            footskip=.25in]{geometry}

\begin{document}

\author{
  Stanislav Dubni\v{c}ka $^{1}$ \\
  Anna Z. Dubni\v{c}kov\'{a} $^{2}$\\
  Mikhail A. Ivanov  $^{3}$ \\
  Andrej Liptaj $^{1,\ddagger}$ \\
}

\title{B meson decays in covariant confined quark model}
\maketitle

{ \centering
$^{1}$ Institute of Physics, Slovak Academy of Sciences,\\ Bratislava, Slovakia
\\
$^{2}$ Faculty of Mathematics, Physics and Informatics,\\ Comenius University, Bratislava, Slovakia
\\
$^{3}$ Bogoliubov Laboratory of Theoretical Physics,\\ Joint Institute for Nuclear Research, Dubna, Russia \\
\vspace{0.5cm}
$^{\ddagger}$ Andrej.Liptaj@savba.sk \\
}

\abstract{The aim of this text to present the covariant confined quark model (CCQM) and review its applications to the decays of $B$ mesons. We do so in the context of existing experimental measurements and theoretical results of other authors, which we review also. The physics principles are in detail exposed for the CCQM, the other results (theoretical and experimental) are surveyed in an enumerative way with comments. We proceed by considering successively three categories of decay processes: leptonic, semileptonic and non-leptonic.}

\section{Introduction}
The confinement property of quantum chromodynamics (QCD) implies it is not possible to study the strong force using the scattering of free quarks. The confinement itself being a manifestation of the strong force, one cannot but analyze more complex systems such as hadrons, i.e. bound states of quarks. All hadrons are colorless (white) objects, among them mesons consisting of two quarks only. Even though no stable mesons exist, the meson physics is often seen as the most simple testing ground of QCD.

Various measurement provided us so far with a large amount of experimental data (masses, decay rates) which challenges our ability to provide theoretical predictions. For the above-mentioned reasons, the perturbative calculations performed at partonic level need to be complemented by the so-called hadronic effects, which are non-perturbative in nature and originate in the long-range interaction between quarks and gluons. As of now, we do not have a well-established general method for a reliable computation of hadronic effects for arbitrary processes from first principles.

Our ability to describe mesons and other QCD states without model dependence is limited, yet improves in time. Light meson physics is often treated within the chiral perturbation theory (ChPT) based on an (approximate) flavor chiral symmetry of the QCD which is spontaneously broken. Assuming this symmetry together with constraints from the analyticity and unitarity, phenomenological Lagrangians were proposed in \cite{Weinberg:1978kz}. This allowed to reproduce the results from complicated methods of the current algebra. In \cite{Weinberg:1978kz} the Lagrangians have been given in the leading order only, the extension of this approach which included meson loops was formulated in two original papers \cite{Gasser:1983yg, Gasser:1984gg}. Since, the ChPT proved to be a successful effective field theory approach with remarkable results \cite{Scherer:2005ki, MacHleidt2016}, however the large masses of other quarks besides $u$, $d$ and $s$ exclude the heavy-quark physics from its applicability range.

A different approach is represented by non-perturbative methods, such as the Dyson-Schwinger equations. The latter were formulated decades ago \cite{PhysRev.75.1736, Schwinger:1951ex, Schwinger:1951hq} in terms of an infinite number of coupled differential equations imposed to the Green functions of the theory. With necessary simplifications results were derived first for abelian theories. Then the approach was extended also to the more complicated case of non-abelian theories \cite{Roberts:1994dr}, thus including QCD and hadronic physics. The application to heavy quarks was for the first time presented in \cite{Ivanov:1998ms}.

A distinctive non-perturbative theoretical technique to investigate the strong interaction physics are the QCD sum rules \cite{Shifman:1978bx, Shifman:1978by}. The central object of interest are the  correlation functions of interpolating quark currents treated using the operator product expansion (OPE) and expressed in term of a perturbative continuum contribution and a low energy parameterization. These are then matched assuming the quark-hadron duality. The results are derived in form of sum rules, the uncertainties have to take into account various necessary approximations. Among others, the results for leptonic decay constants and hadron transition form factors have been derived \cite{deRafael:1997ea, Colangelo:2000dp}.

In the domain of heavy meson physics (which we are interested in) a specific tool is available: the approximate realization of the heavy quark symmetry gives rise to the heavy quark effective theory (HQET) \cite{Isgur:1989vq, Isgur:1990yhj,Neubert:1993mb}. The symmetry appears when the mass of the heavy quark goes to infinity and is the combination of a heavy quark flavor symmetry and the independence of hadronic properties on the heavy quark spin state. It allows for important simplifications and leads to results expanded in the inverse of the heavy quark mass.

An important model-independent approach with possibly very broad applicability is represented by numerical QCD calculations on the lattice. Here an important progress was made over last decades \cite{FlavourLatticeAveragingGroupFLAG:2021npn}, nowadays predictions of form factors in weak decays of heavy particles become available \cite{Gambino:2020crt,Desiderio:2020oej,Boyle:2022uba,Gambino:2022dvu}. The potential of the method is immense, since, as is evident from \cite{Workman:2022ynf}, the bulk of the experimental data in high-energy physics is related to hadrons and explaining them at few percent level accuracy would be a triumph. 

However we are not at this point now and the possibility for lattice calculations to become the mainstream of theoretical predictions will depend on the future developments. Thus, despite the important achievements of the lattice QCD, model dependent methods remain the most popular and versatile tools in making QCD predictions with hadronic effects included. This is mainly due to the fact that the lattice QCD remains limited to a narrow set of specific processes while the model framework can be usually easily adopted to various settings, making thus predictions more easy to produce. This is especially true with relation to the B factories, i.e. very high-luminosity accelerator facilities nowadays in operation where a large number of various heavy hadron decays is registered and measured. Many of these approaches can be described as "quark" models, since they describe the hadron by considering its valence quarks using some specific assumptions or ans\"{a}tze (see e.g. \cite{Ebert:1997nk, Ebert:2002pp}).

The Nambu-Jona-Lasinio (NJL) model based on the ideas of
Y.~Nambu and G.~Jona-Lasinio ( Refs.~\cite{Nambu:1961tp,Nambu:1961fr}
of original papers) is widely used in the low-energy phenomenology of
light quarks ($u,d,s$).  The hadron masses are generated due to the
spontaneous breaking of chiral symmetry where the pion plays a role
of the Goldstone boson. This approach found many applications
in light meson physics due to the simplicity of calculations, for
review, see, e.g. Ref.~\cite{Klevansky:1992qe}. Some efforts
have been made to extend the NJL model for applications to heavy mesons
with taking into the account the heavy quark symmetry \cite{Guo:2012tm,Luan:2015goa}.
In our early paper \cite{Anikin:1995cf}, which was a predecessor of the CCQM,
a clear relation of the so-called compositeness condition (addressed later) with
the requirement of the correct normalization of the kinetic term in the NJL Lagrangian after the spontaneous breaking of the chiral symmetry was shown.

As far as quark models are concerned, for weak decays they are usually combined with a perturbative computation at the quark level. Here, it is customary to use an effective four-fermion theory derived using the OPE and governed by the low-energy Hamiltonian
\begin{align}
\mathcal{H}_\text{eff.}^{b \to q} = \frac{G_F}{\sqrt{2}} V_{tb}V_{tq}^* \sum_i C_i(\mu) Q_i(\mu) \label{eq:Heff}
\end{align} 
here written for the $b \to q\in\{s,d \}$ transition. $Q_i(\mu)$ are local operators expressed in terms of quark fields, $C_i(\mu)$ are the Wilson coefficients which can be evaluated perturbatively, $V_{ij}$ are Cabibbo - Kobayashi - Maskawa (CKM) matrix elements and $\mu$ is the QCD renomalization scale. Its value is set to a typical momentum transfer which is for weak decays significantly smaller that the $W$ mass. Thus $W$ is effectively removed from  (\ref{eq:Heff}), it enters in computations of $C_i(\mu)$. An excellent overview of weak  decays is given in \cite{Buchalla:1995vs}.

The heavy decay processes are of a special interest for the particle physics community for several reasons \cite{BaBar:2014omp}. One of them is the determinations of the CKM matrix elements and the study of related questions such as the CP violation, unitarity triangle, baryogenesis and weak physics in general. Further, B factories are used to search for new exotic states including tetraquarks, pentaquarks, glueballs and so on. The collected data also allowed to study fragmentation processes, test the lepton universality, investigate possible lepton flavor violation and address the questions related to a new, beyond Standard Model (SM) physics \cite{doi:10.1142/12696, Altmannshofer:2021qrr}.

Indeed, various new physics (NP) scenarios \cite{Jager:2017gal,Kumbhakar:2018uty,Chala:2019vzu,Coy:2019rfr,Charles:2020dfl,BHUTTA2022115763,Cai:2021mlt} predict deviations from the SM in B meson decay processes. Because of the very high luminosity the nowadays colliders have, there is a hope that even rare (small in number) deviations from the SM physics can be detected.

We present here how the covariant confined quark model (CCQM) \cite{Branz:2009cd} has been used to investigate the b-physics processes. A dedicated effort was made in previous years and decades to cover most of the measured B meson data, and since they are large in number we believe it is appropriate to review them. We provide in this text the overview of the results from the perspective of the CCQM, but we also point to contributions and achievements from other approaches and authors. Up to some exceptions, the majority of the outcome was formulated in terms of the SM predictions which were compared to data. In this way possible tensions or deviations were identified or hypothesis about the nature of an exotic state were expressed. This then points to possible NP phenomena or better understanding of exotic particles, especially when there is an agreement with other theoretical works too.

The large quantity of various B-related results which have been published in the past does not allow us to review each decay in full details. We therefore define three categories and for each we present a demonstrative calculation with one or two example processes. The categories are leptonic, semileptonic and non-leptonic (radiative) decays.

The text is structured as follows: In Sec. \ref{Sec: CCQM} the general features of the CCQM are presented. The following three sections are dedicated to specific process categories, as mentioned above.  Each has three subsections, one with a general overview, the second presenting in more details the computations for a chosen example process and the third where results obtained within the CCQM framework are summarized. The text ends with conclusion and outlook.

\section{Covariant confined quark model \label{Sec: CCQM}}
The key points for the model construction are
\begin{itemize}
\item Lorentz symmetry and invariant Lagrangian,
\item compositeness and double counting,
\item confinement of quarks,
\item gauge symmetry and inclusion of electromagnetic (EM) fields,
\end{itemize}
which we address is this order. In an additional subsection we also briefly describe the computational techniques.

\subsection{Lagrangian}
To construct a theory with Lorentz symmetry one naturally recurs to a Lagrangian formulation. So is done for the CCQM which is an effective field approach where both, quark  and hadronic fields occur. The quark-meson interaction term is written as 
\begin{align}
\mathcal{L}_{int}  = g_{M}M\left(x\right)J_{M}\left(x\right)+\text{ H.c. }, \quad
J_{M}\left(x\right)  =\int dx_{1}\int dx_{2}F_{M}\left(x;x_{1},x_{2}\right)\overline{q}_{2}\left(x_{2}\right)\varGamma_{M}q_{1}\left(x_{1}\right), \label{Lagr}
\end{align}
where $M$ represents the mesonic field, $q$ the quark one, $g_{M}$ is their coupling and $\text{H.c.}$ stands for the Hermitian conjugate. The interpolating quark current $J_M$ is non-local and the integral over the positions $x_1$, $x_2$ of constituent quarks is weighted by a vertex function $F_M$. The symbol $\varGamma_{M}$ represents a combination of gamma matrices depending on the spin of $M$. For a scalar $M$ one has $\varGamma_{M} = 1$, for pseudoscalar $\varGamma_{M} = \gamma^5$ and for a vector particle the expression is $\varGamma_{M} = \gamma^\mu$. In the latter case the mesonic field has a Lorentz index too ($M_\mu$) and the indices are contracted.

It's interesting to see what happens in the case of local interation when
$F_M(x;x_1,x_2)=\delta(x-x_1-x_2)$. Then one clearly observes that the interaction
Lagrangian given by Eq. (\ref{Lagr}) together with free meson and quark Lagrangians
corresponds to the NJL model after bosonization.

The explicit form of $F_{M}$ is driven by two requirements. First the positions of quarks are constrained so as to make the hadron be situated in their barycenter. For this a delta function is introduced where the weights in its argument depend on the constituent quark masses $w_i = m_i/(m_1+m_2)$. Second, to manifestly respect the Lorentz symmetry, the remaining dependence is written as a function of the spacetime interval
\begin{align}
F_{M}\left(x;x_{1},x_{2}\right)=\delta\left(x-w_{1}x_{1}-w_{2}x_{2}\right)\varPhi_{M}\left[\left(x_{1}-x_{2}\right)^{2}\right].
\end{align}
Further steps in the construction of $F_{M}$ are done with respect to the computational convenience. $\varPhi_{M}$ is assumed to have a Gaussian form in the momentum representation
\begin{align}
\varPhi_{M}\left[\left(x_{1}-x_{2}\right)^{2}\right]=\int\frac{d^{4}k}{\left(2\pi\right)^{4}}e^{-ik\left(x_{1}-x_{2}\right)}\widetilde{\varPhi}_{M}\left(-k^{2}\right),\quad
\widetilde{\varPhi}_{M}\left(-k^{2}\right)=e^{k^{2}/\varLambda_{M}^{2}}, \label{eq:vertexFn}
\end{align}
where $\varLambda_{M}$ is a free parameter of the model related to the meson $M$. The square of the momentum in the argument of the exponential becomes negative in the Euclidean region $k^2 = -k_E^2$ which implies an appropriate fall-off behavior and removes ultraviolet divergences in Feynman diagrams. The question of other possible function forms of $\varPhi_{M}$ was addressed in \cite{Anikin:1995cf}, where four different ansatzes were tested, each having a meaningful physical interpretation. The dependence of the results on the function form was found to be small.

The S-matrix is constructed from the interaction Lagrangian as
$S = T \exp \{i \int d^4x {\mathcal L}_{\rm int}(x)\}$. The calculation of the matrix
elements of $S$ proceeds in a standard manner, first, by
making convolution of the quark fields with a help of T-product, and, second,
by using the Fourier-transforms of quark propagators and vertex functions
to go to the momentum space. Note that we use the ordinary local forms
of the quark propagators $S(k)=1/(m_q-\not\! k)$ in our approach.

Besides the hadron-related $\varLambda_{M}$, the CCQM comprises four "global" parameters: three constituent quark masses and one universal cutoff which plays a role in the quark confinement (as explained later). The values expressed in GeV are
\begin{align}
m_q = m_{u,d} = 0.241,
\quad m_s = 0.428,
\quad m_c = 1.67,
\quad m_b = 5.05,
\quad \lambda = 0.181, \label{par_values}
\end{align}
where one does not distinguish between the two light quarks and use the same mass for both. The values slightly changed in the past, they were few times \cite{Ganbold:2014pua, Gutsche:2015mxa} updated if significant new data become available. They were extracted by over-constrained global fits of the model on available experimental points.

The CCQM does not include gluons. The gluonic effects are effectively taken into account by the vertex function which is adjusted to describe data by tuning the free parameter it contains.

At last we have to mention that the CCQM is suitable for description of various multi-quark states including baryons \cite{Gutsche:2014zna,Ivanov:2020iaq}  and tetraquarks \cite{Dubnicka:2020yxy}. In this text we focus on mesons, the approach is in other cases very similar: the interpolating quark current is constructed for a given number of quarks (more alternatives can be considered) and multiplied by the hadronic field to give the interaction Lagrangian.

\subsection{Compositeness condition}

The interaction of a meson is given by the Lagrangian (\ref{Lagr}): the meson fluctuates into its constituent quarks, these interact and afterwards combine back into a mesonic final state. Yet, (\ref{Lagr}) implies that both, quark and mesons, are elementary and this rises concerns about the double counting.

These questions were addressed by implementing the so-called compositeness condition \cite{Ganbold:2014pua,Branz:2009cd,Dubnicka:2015iwg} which originates in works \cite{Jouvet:1956ii,Salam:1962ap,Weinberg:1962hj} (see \cite{Hayashi:1967bjx} for a review). The interaction of a meson through the creation of virtual quark states implies the mesonic field is dressed, i.e. its vertex and wave function need to be renormalized. This is reflected in the renormalization constant $Z_M$ which can be interpreted as the overlap between the physical state and the corresponding bare state. By requiring $Z_M = 0$ one makes this overlap vanish, i.e. the physical state does not contain bare state and can be regarded as a bound state. As a consequence the quarks exist only as virtual and quark degrees of freedom do not appear on the level of the physical state.

\begin{figure}[t]
\begin{center}
\includegraphics[width= 5 cm]{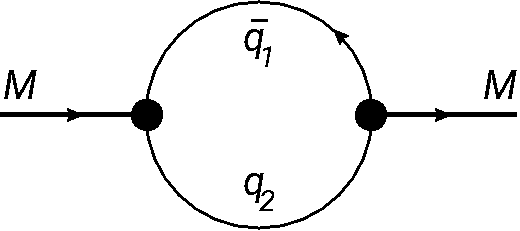}
\caption{Meson mass function diagram.}
\label{Fig_mmf}
\end{center}
\end{figure}

$Z_M$ is expressed in terms of the derivative of the meson mass operator $\Pi_{M}^{'}$  (its scalar part for vector mesons) 
\begin{align}
Z_{M}=1-g_{M}^{2}\Pi_{M}^{'}\left(m_{M}^{2}\right) = 0 \label{eq:Compositeness}
\end{align}
and at one-loop level (Fig. \ref{Fig_mmf}) is given by
\begin{align}
\Pi_{PS}^{'}(p^2) &= \frac{-i}{2p^2} p^\alpha \frac{d}{dp^\alpha}
\int d^4k \, \widetilde{\varPhi}_{PS}^2(-k^2)
\text{tr}\left[ \gamma^5 S_1(k+w_1 p) \gamma^5 S_2(k-w_2p) \right], \\
\Pi_{V}^{'}(p^2) &= \frac{-i}{3} \left( g_{\mu\nu-}
\frac{p_\mu p_\nu}{p^2} \right)
\frac{1}{2p^2} p^\alpha \frac{d}{dp^\alpha}
\int d^4k \, \widetilde{\varPhi}_{V}^2(-k^2)
\text{tr}\left[ \gamma^\mu S_1(k+w_1 p) \gamma^\nu S_2(k-w_2p) \right],
\end{align}
for pseudoscalar and vector mesons respectively. The symbol $S_i$ denotes the quark propagator $S_i = 1 / (m_{q_i}-\gamma^\mu k_\mu)$ and the  differentiation is done using the identity $d\Pi/dp^2= (p^\mu \, d\Pi/dp^\mu) / (2p^2)$.

To reach the equality (\ref{eq:Compositeness}) one profits from the up-to-now undetermined coupling constant $g_M$ and tune its value so that (\ref{eq:Compositeness}) is satisfied. As consequence, the coupling $g_M$  becomes the function of $\Lambda_M$. In this way the number of parameters of the model is reduced and one increases its predictive power and stability. If the values $\Lambda_M$ and $g_M$ are unknown from previous studies, their determination is the first step in the application of the CCQM.

As it will be discussed in the next sections, the adjustable parameters
of the model (quark masses, size parameters and infrared cutoff)
are determined by fitting the experimental data of physical observables.
For instance, in the case of the B-meson, the size parameter is found to be
equal to $\Lambda_B=1.96$~GeV. By using the compositeness condition it
gives the numerical value of the  coupling constant $g_B=4.80$.

\subsection{Infrared confinement}

The CCQM is a successor of the so-called relativistic constituent quark model (see \cite{Faessler:2002ut}) and in \cite{Branz:2009cd} it was proposed to refine the latter by effectively implementing quark confinement into it. This was motivated by data on heavy particles which required an extension to situations where the hadron is heavier than the sum of its constituent quarks. To prevent the decay  into free quarks in such a scenario, a technique inspired by confined propagators is used. Here the propagators are written in the Schwinger representation and a cutoff is introduced in the upper integration limit. The propagator then becomes an entire function
\begin{align}
\frac{S_i(k)}{(m_{q_i}+\gamma^\mu k_\mu)}=\int_0^\infty d\alpha e^{-\alpha(m_{q_i}^2-k^2)}
\to \int_0^{1/\lambda^2} d\alpha e^{-\alpha(m_{q_i}^2-k^2)}
= \frac{1-e^{-(m_{q_i}^2-p^2)/\lambda^2}}{m_{q_i}^2-p^2}, \label{eq:SchwingProp}
\end{align}
where the absence of singularities indicates the absence of a single quark asymptotic state. A modified version of this strategy was adopted and the cutoff was applied to the whole structure $F$ of the Feynman diagram. It can be formally written as
\begin{align}
\Pi = \int_0^\infty d^n\alpha \, F(\alpha_1, \dots , \alpha_2)
= \int_0^{\infty \to 1/\lambda^2} dt \, t^{n-1} \int_0^1 d^n \alpha \, \delta(1-\sum_{i=1}^n \alpha_i) F(t\alpha_1, \dots , t\alpha_2) \label{cutOff}
\end{align}
and can be obtained by inserting the unity $1=\int_0^\infty dt \, \delta(t-\sum_{i=1}^n \alpha_i)$ into the expression on the left hand side. The single cutoff (indicated by the arrow) in the $t$ variable is done in the last step, the remaining integration variables are confined to an $n$ dimensional simplex. After the cutoff is applied the integral becomes convergent for arbitrary values of the kinematic variables meaning that the quark thresholds are removed with quarks never being on the mass shell. The the cutoff value (\ref{par_values}) is the same for all processes.

\subsection{Electromagnetic interactions and gauge symmetry \label{Sec:EM}}
Radiative decays represent another important class of processes measured at heavy meson factories. For their description one has to include the interactions with photons into the CCQM \cite{Branz:2009cd, Branz:2010pq}. Because of the non-local interaction Lagrangian this is not straightforward and requires a dedicated approach. Taking into the account quarks and scalar mesons, the free parts of the Lagrangian are treated in the usual way, i.e. the minimal substitution is used
\begin{align}
\partial^\mu \psi \to (\partial^\mu - ie_\psi A^\mu)\psi, \quad
\partial^\mu \bar{\psi} \to (\partial^\mu + ie_\psi A^\mu)\bar{\psi}, \quad
\end{align}
where $\psi \in \{q,M \} $ and $e_\psi$ is its electric charge in the units of the proton charge. One then gets
\begin{align}
\mathcal{L}^{EM_1} &=
e  A_\mu(x) J_M^\mu(x) + e^2 A^2(x)M^-(x) M^+(x)
+\sum_q e_q A_\mu(x) J_q^\mu(x), \\
J_M^\mu(x) &= i[M^-(x)\partial^\mu M^+(x) - M^+(x)\partial^\mu M^-(x)],
\quad
J_q^\mu(x) = \bar{q}(x)\gamma^\mu q(x).
\end{align}
The compositeness condition formulated above however prevents a direct interaction of the dressed particle, i.e. the meson, with photons: the contributions from the photon-meson tree level diagram and analogous diagrams with self-energy insertions into the external mesonic line determine the renormalization constant $Z$ and $Z=0$ implies they cancel. The interaction thus proceeds only through intermediate virtual states.

The gauging of the non-local interaction (\ref{Lagr}) is done in a manner similar to \cite{Terning:1991yt}. First one multiplies the quark fields in (\ref{Lagr})  by a gauge field exponential
\begin{align}
q_i(x) \to e^{-i e_{q_i}I(x_i,x,P)}q_i(x),
\quad
I(x_i,x,P)=\int_x^{x_i} dz_\mu A^\mu(z), \label{eq:pathGauge}
\end{align}
where $P$ is the path connecting $x_i$ and $x$, the latter being the position of the meson. One can verify that the Lagrangian is invariant under the following gauge transformations
\begin{align}
q_i(x) \to e^{i e_{q_i} f(x)} q_i(x),& \quad
\bar{q}_i(x) \to  \bar{q}_i(x) e^{-i e_{q_i} f(x)},\\
M(x) \to e^{i e_M f(x)}M(x),& \quad
A^\mu(x) \to A^\mu(x) + \partial^\mu f(x),
\end{align}
here $f(x)$ is some scalar function. The apparent path-dependence of the definition (\ref{eq:pathGauge}) is not an actual one: in the perturbative expansion only derivatives of the path integral appear and these are path independent
\begin{align}
\frac{\partial}{\partial x^\mu} I(x,y,P) = A_\mu(x).
\end{align}
The individual terms of the Lagrangian are generated by expanding the gauge field exponential by orders in $A^\mu$. At  first order one has
\begin{align}
\mathcal{L}^{EM_2}(x) = g_M M(x) \iiint dx_1\, dx_2\, dy \,
E^\mu_{M}\left(x;x_{1},x_{2},y\right)
A_\mu(y)\overline{q}_{2}\left(x_{2}\right)\varGamma_{M}q_{1}\left(x_{1}\right),
\end{align}
where $E_M$ is defined through its Fourier transform $\widetilde{E}_M$: $(x_1-x,x_2-x,y-x) \overset{FT}{\leftrightarrow} (p_1,p_2,q)$
\begin{align}
\widetilde{E}^\mu_M (p_1,p_2,q) &=
\sum_{i=1,2} \vartheta_i e_{q_i} w_i(w_i q^\mu + \vartheta_{i+1}2l^\mu)
\int_0^1 dt \widetilde{\Phi}^{'}_M \left[ -t(w_i q + \vartheta_{i+1} l)^2 - (1-t)l^2 \right],\\
l &= w_1 p_1 + w_2 p_2, \quad \vartheta_i = (-1)^{i}.
\end{align}
Symbol $\widetilde{\Phi}^{'}_M$ denotes the derivative with respect to the argument. In corresponding Feynman diagrams the photon is attached to the non-local vertex.

\subsection{Computations}
From the Lagrangian one derives the Feynman diagrams. Gaussian expressions in the vertex function (\ref{eq:vertexFn}) and in the Fock-Schwinger propagator (\ref{eq:SchwingProp}) can be joined into a single exponent which takes a quadratic form in the loop momenta $k$. It can be formally written as $\exp(ak^2 + 2rk+z)$, $a=a(\{\alpha\})$, $r=r(\{\alpha\},\{p\})$, where $\{\alpha\}$ denotes the set of Schwinger parameters and $\{p\}$  external momenta. The exponential is preceded by a polynomial $P$ in loop momenta which originates from the trace of Dirac matrices (numerators of propagators). Since the powers of $k$ can be generated by differentiation with respect to $r$, the loop momenta integration is formally written as
\begin{align}
\int d^4k \, P(k) \exp(ak^2 + 2rk+z)
= \exp(z) P\left( \frac{1}{2}\frac{\partial}{\partial r} \right)
\int d^4k \,  \exp(ak^2+2rk).
\end{align}
Using the substitution $u= k+r/a $, the argument of the exponential is transformed
\begin{align}
\int d^4k \, \exp(ak^2+2rk) = 
\int d^4u \, \exp(au^2-r^2/a) =
\exp(-r^2/a) \int d^4u \, \exp(au^2)
\end{align}
and the integration is performed in the Euclidean region as a simple Gaussian integral. Further, the differential operator and the $r$-dependent exponential can be interchanged 
\begin{align}
P\left( \frac{1}{2}\frac{\partial}{\partial r} \right) \exp \left( -\frac{r^2}{a} \right)
= \exp \left(-\frac{r^2}{a} \right) P\left( -\frac{r}{a} + \frac{1}{2}\frac{\partial}{\partial r} \right) \label{diffOP_ID}
\end{align}
which simplifies the action of the differential operator. One arrives to 
\begin{align}
\int_0^\infty d\alpha_1 \dots \int_0^\infty d\alpha_n F(\alpha_1,\dots, \alpha_n),
\end{align}
where $F$ represents the whole structure of the Feynman diagram including (\ref{diffOP_ID}). A FORM \cite{Ruijl:2017cxj} code is used to treat symbolic expressions: besides computing traces it is also used to repeatedly perform chain rule application in (\ref{diffOP_ID}) and arrive to an explicit formula with no differential operators appearing. The implementation of the infrared confinement as expressed by (\ref{cutOff}) is the last step before the numerical integration.

\section{Leptonic decays of $B$ mesons}

\subsection{Overview}

Large mass difference between heavy mesons and leptons implies, by phase-space arguments,  small branching fractions of pure and radiative leptonic decays. Some of these are further suppressed by CKM elements or helicity. Thus for most leptonic decays only limits have been measured.

At the usual 95\% confidence level a branching fraction measurement is available only for $B_s^0 \to 2 \mu$ \cite{LHCb:2021vsc, ATLAS:2018cur, CMS:2019bbr, Langenegger:2020fns} and $B^\pm \to \tau^\pm \nu_\tau$ \cite{Belle:2015odw, Belle:2012egh, BaBar:2012nus, BaBar:2009wmt}. If the criteria are loosened to (at least) one sigma significance,  additional results can be cited: $B^0 \to 2 \mu$ \cite{LHCb:2021vsc}, $B^+ \to \mu^+ \nu_\mu$ \cite{Belle:2017xhh,Belle:2019iji} and $B^+ \to \ell^+ \nu_\ell \gamma$ \cite{Belle:2018jqd}. The limits are settled \cite{Workman:2022ynf} for $B^+ \to e^+ \nu_e$, $B^+ \to e^+ \nu_e \gamma$, $B^+ \to \mu^+ \nu_\mu \gamma$, $B^+ \to \mu^+ \mu^- \mu^+ \nu_\mu$, $B^0 \to e^+ e^-$, $B^0 \to e^+ e^- \gamma$, $B^0 \to \mu^+ \mu^- \gamma$, $B^0 \to \mu^+ \mu^- \mu^+ \mu^- $ , $B^0 \to \tau^+ \tau^-$, $B^0_s \to e^+ e^-$, $B^0_s \to \tau^+ \tau^-$ and $B^0_s \to \mu^+ \mu^- \mu^+ \mu^- $.

These experimental results motivate various analyses. Pure leptonic decays are considered as theoretically clean with the main source of uncertainty represented by the hadronic effects of the initial state, which are contained in the leptonic decay constant of the hadron. The neutrino production process corresponds, in the leading order, to the annihilation of the constituent quarks into a virtual $W$ meson which subsequently decays. The branching fraction is given by 
\begin{align}
\mathcal{B}(B^+ \to \ell^+ \nu) =
\frac{G_F^2 m_B m_l^2}{8\pi} \left( 1 - \frac{m_l^2}{m_B^2} \right)^2 f_B^2 |V_{ub}|^2 
\tau_{B^+},
\end{align}
where $G_F$ is the Fermi coupling constant, $V_{ij}$ the CKM matrix element and $\tau_{P}$ the lifetime of the particle $P$.

A general information about $B$ leptonic decays is contained in several reviews. Besides \cite{Buchalla:1995vs}, a more specific focus on processes with charged pseudoscalar mesons is given in \cite{Rosner:2015wva} and a summary concerning specifically $B$ decays (leptonic and semileptonic ) is provided in \cite{Dingfelder:2016twb}. The existing theoretical approaches follow two directions. One focuses on the SM contributions at different precision levels, the other is concerned with NP beyond the SM.

\vspace{0.5cm}

Dilepton final states are produced at one loop through box and penguin diagrams. The cross section formula can be found e.g. in  \cite{Buchalla:1993bv}, equation (4.10). The leptonic decays constants of $B$ (and $D$) mesons where determined in a model-independent way using lattice calculations in \cite{Bazavov:2017lyh}. The SM treatment of dilepton decays includes the computation of three-loop QCD corrections \cite{Hermann:2013kca}, the evaluation of the electroweak contributions at the two-loop level \cite{Bobeth:2013tba} and further improvements of theoretical predictions reached by combining additional EM and strong corrections \cite{Bobeth:2013uxa}. The authors of \cite{Beneke:2017vpq} investigated the effect of the virtual photon exchange from scales below the bottom-quark mass and found a dynamical enhancement of the amplitude at the 1\% level. The soft-collinear effective theory approach was used in \cite{Beneke:2019slt} to evaluate the power-enhanced leading-logarithmic QED corrections.
 
  The radiative processes have the advantage of not being helicity suppressed at the price of one additional $\alpha_{\text{EM}}$ factor. A larger number of results can be cited for radiative dilepton production. An evaluation within a constituent quark model was performed in \cite{Eilam:1996vg} to estimate branching fractions, the same observables were predicted by the authors of \cite{Aliev:1996ud,Aliev:1997sd} using the light-cone QCD sum rules and by those of \cite{Geng:2000fs} using the light-front model. Universal form factors related to the light cone wave function of the $B_s$ meson allowed to make estimates in \cite{Dincer:2001hu}. Interesting results were given in \cite{Kruger:2002gf}, where it was shown that the gauge invariance and other considerations allow to significantly constrain the form factor behavior, and also in \cite{Descotes-Genon:2002lal} where the authors have demonstrated that the non-perturbative hadronic effects largely cancel in amplitude ratios of pure leptonic and radiative decays. The impact of the light meson resonances on long-distance QCD effects was studied in \cite{Melikhov:2004mk}. In \cite{Carvunis:2021jga} the authors have identified the effective $B\to \mu \mu \gamma$ lifetime and a related CP-phase sensitive observable as appropriate quantities to study the existing B decay discrepancies.

Also for decays $B \to \gamma l \nu_l$ several studies can be cited. The work \cite{Burdman:1994ip} was concerned with  photon spectrum and the decay rates of the process. The authors of \cite{Korchemsky:1999qb} used the HQET to predict form factors and in \cite{Braun:2012kp} the heavy-quark expansion and soft-collinear effective theory were applied to evaluate the soft-overlap contribution to the photon. The process was also studied in \cite{Beneke:2018wjp}. Here, assuming an energetic photon, the authors aimed to quantify the leading power-suppressed corrections in $1/E_\gamma$ and $1/m_b$ from higher-twist B-meson light-cone distribution amplitudes. The soft-collinear effective theory was the approach adopted in \cite{Lunghi:2002ju,Bosch:2003fc}.

The recent publication \cite{Wang:2021yrr} focused on four-body leptonic B decays: off-shell photon form factors were computed within the QCD factorization formalism and predictions for differential distribution of various observables were presented. Similar processes are addressed also in \cite{Danilina:2019dji, Ivanov:2022uum, Ivanov:2021jsr}.

\vspace{0.5cm}

Although the most tensions with the SM are seen in the semileptonic sector, the pure leptonic decays are of a concern too. The summary papers \cite{Altmannshofer:2021qrr, Geng:2021nhg} mention two tensions. Fist is related to the combined likelihood for $B^0$ and $B^0_s$ decays to $\mu^+ \mu^-$ where the theory-measurement difference reaches $2.3 \sigma$. The other concerns the branching fraction ratio for the  $B_s^0 \to \mu^+ \mu^-$ reaction $R = \mathcal{B}_\text{exp}/\mathcal{B}_\text{SM}$ which deviates from 1 by $2.4 \sigma$. In \cite{Alguero:2021anc} is the difference between the theory and the experiment for the dimuon $B_s$ decay quantified to be $2.2 \sigma$. 

The possible NP contributions are usually assessed by introducing new, beyond SM four-fermion contact operators and the corresponding Wilson coefficients. Once  evaluated in the appropriate NP approach, it is possible to conclude about their effect on the theory-experiment discrepancy, see e.g. \cite{Fleischer:2021yjo}.

An overview of various flavor-violating extensions of the SM also with relation to $B \to \ell \ell$ decay was presented in \cite{DAmbrosio:2002vsn}. In \cite{DeBruyn:2012wk} the $B_s$ dimuon decay was considered and it was argued that the decay width difference between the light and heavy $B_s$ mass eigenstates is a well-suited observable for the detection of NP. The work \cite{Kumbhakar:2018uty} points to the ambiguity in choice of the NP operators that might play role in explaining the tensions in the $B$ semileptonic decays. They show that this ambiguity can be lifted by analyzing the longitudinal polarization asymmetry of the muons in $B_s^* \to \mu \mu $. Various discrepancies in measured data are addressed in \cite{Greljo:2021npi}, among them also dimuon branching fractions. The attempt to explain them is based on lepton-flavored gauge extensions of the SM, a specific construction with a massive gauge boson $X_\mu$ and "muoquark" $S_3$ is presented. Several texts are interested in decays with tau lepton in the final state. In \cite{Hou:1992sy,Akeroyd:2003zr,Crivellin:2012ye} these decays are studied in relation with various alternative scenarios of the Higgs boson model and in \cite{He:2012zp} they are analyzed in the context of non-universal left-right models.

\subsection{Radiative leptonic decay $B_s \to \ell^+ \ell^- \gamma$ in CCQM \label{Sec:leptoCCQM}}
\begin{figure}[t]
\includegraphics[width=0.3\textwidth]{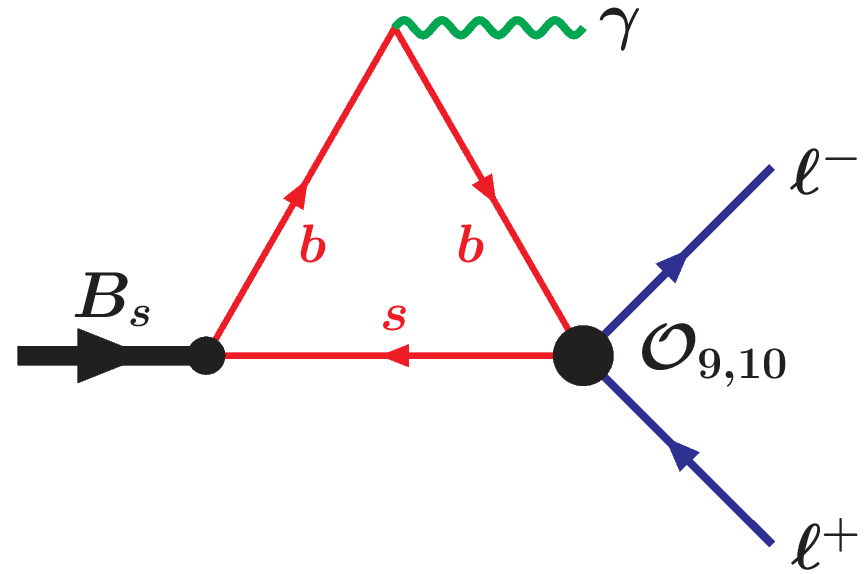}
\hspace{0.5cm}
\includegraphics[width=0.3\textwidth]{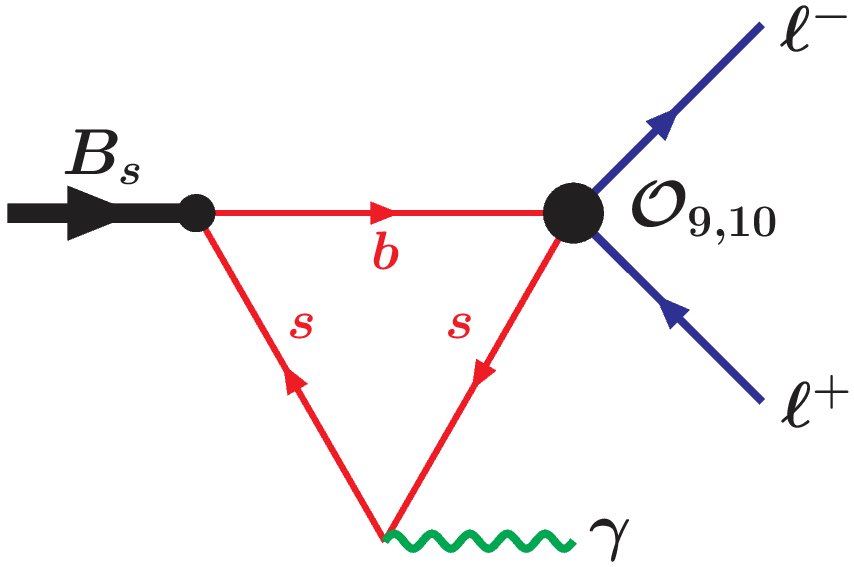}
\hspace{0.5cm}
\includegraphics[width=0.3\textwidth]{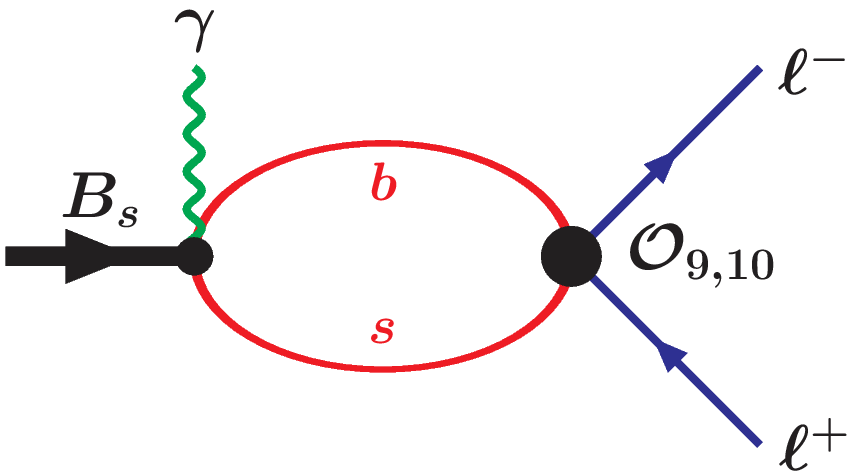}
\\
\includegraphics[width=0.3\textwidth]{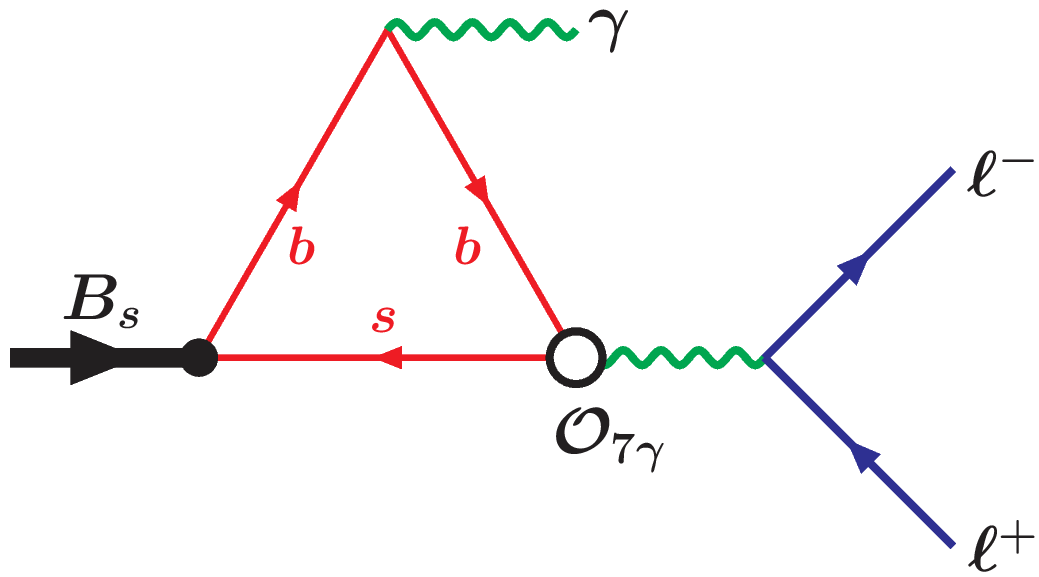}
\hspace{0.5cm}
\includegraphics[width=0.3\textwidth]{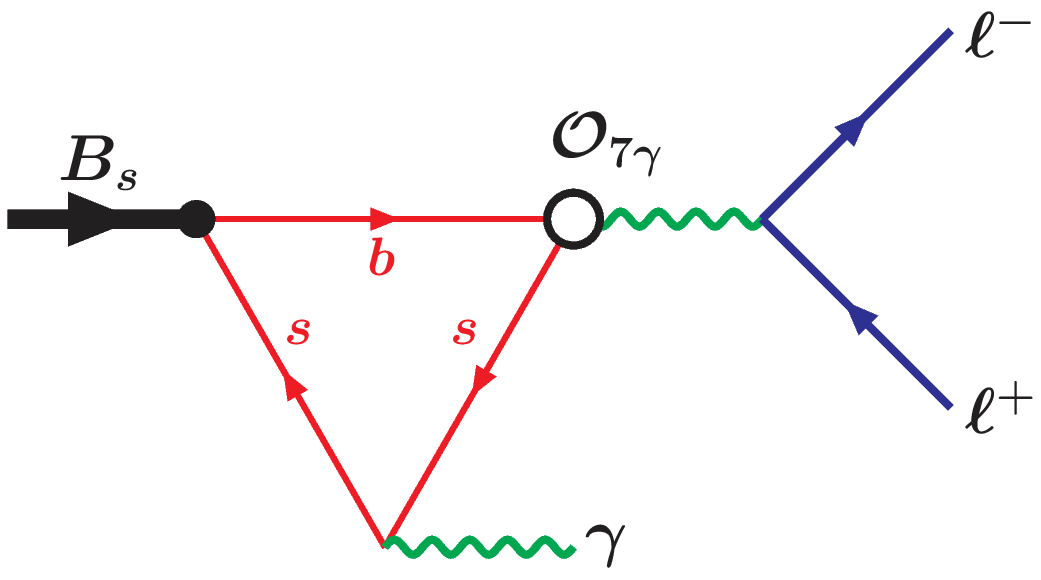}
\hspace{0.5cm}
\includegraphics[width=0.3\textwidth]{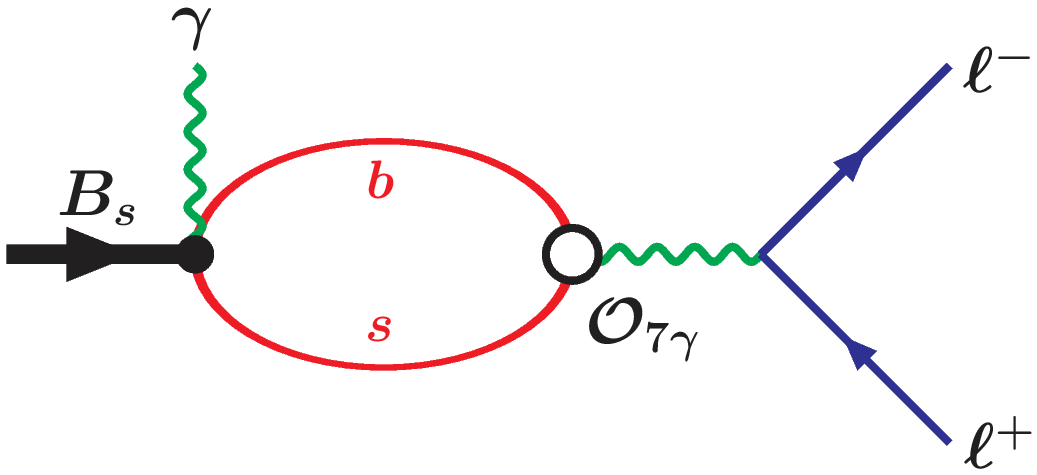}
\caption{Diagrams with the dilepton produced from the $b$-$s$ transition. Figures were originally published in \cite{Dubnicka:2018gqg}. }
\label{Fig:dLfromBS}
\end{figure}

\begin{figure}[t]
\center{
\includegraphics[width=0.3\textwidth]{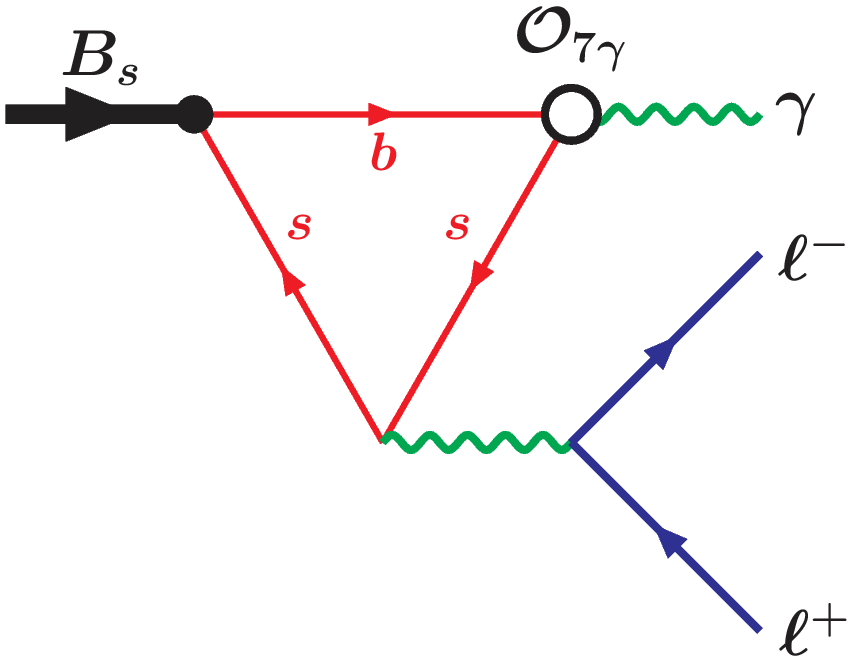}
\hspace{1cm}
\includegraphics[width=0.3\textwidth]{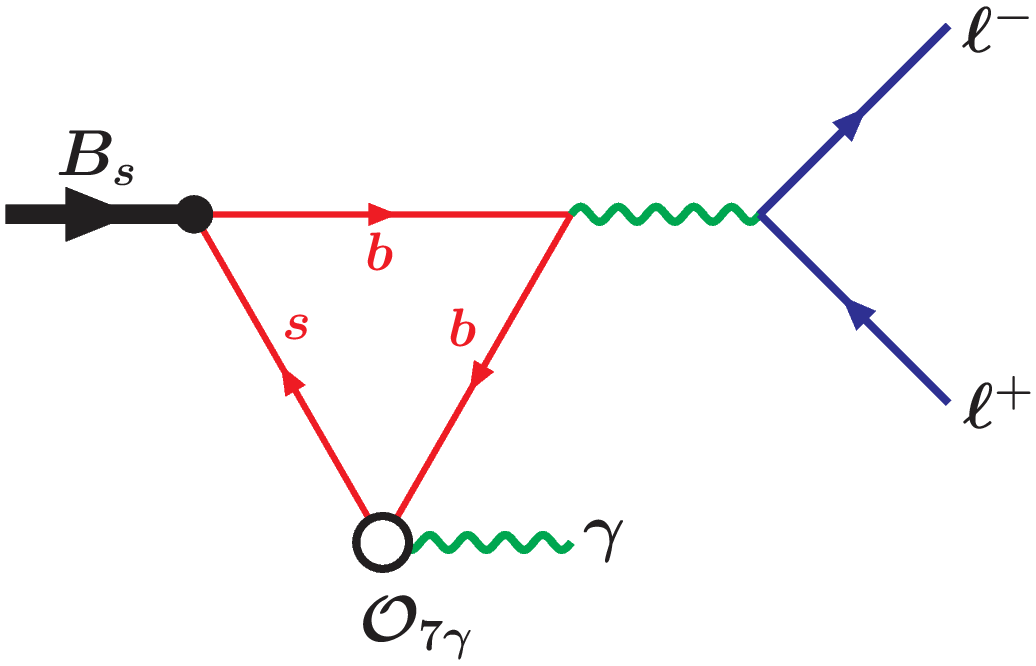}
}
\caption{Diagrams with a real photon produced from the $b$-$s$ transition. Figures were originally published in \cite{Dubnicka:2018gqg}. }
\label{Fig:GfromBS}
\end{figure}

\begin{figure}[t]
\center{
\includegraphics[width=0.3\textwidth]{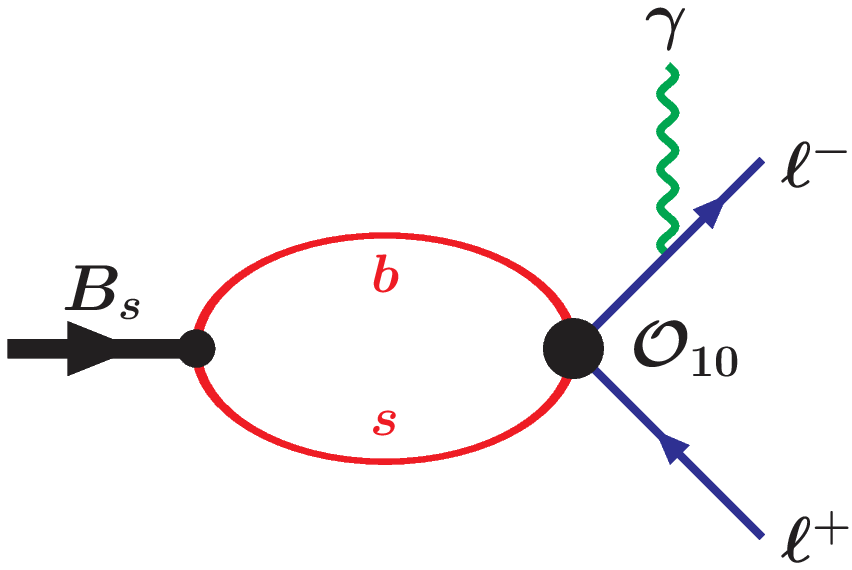}
\hspace{1cm}
\includegraphics[width=0.3\textwidth]{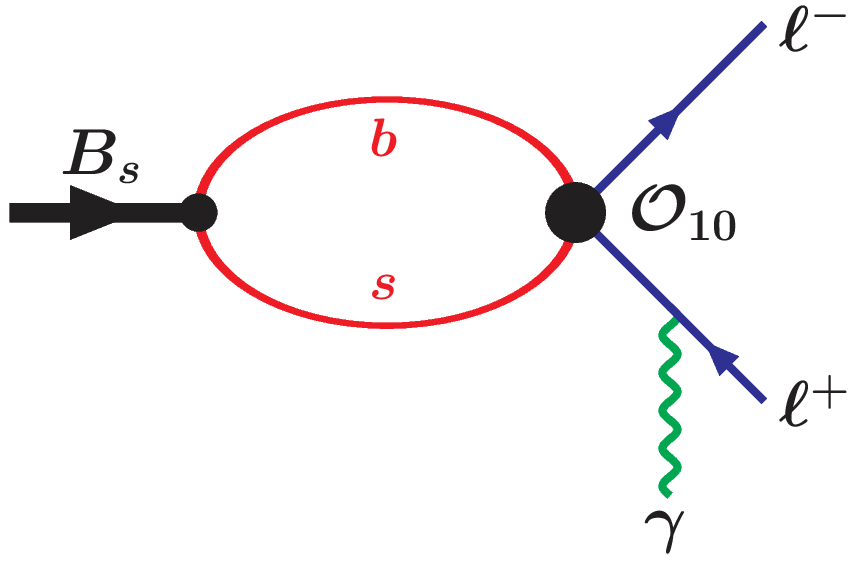}
}
\caption{Final-state radiation diagrams. Figures were originally published in \cite{Dubnicka:2018gqg}.}
\label{Fig:FSR}
\end{figure}

Before reviewing other CCQM results on leptonic $B$ decays we present in more details the evaluation of the branching fraction for $B_s \to \ell^+ \ell^- \gamma$ \cite{Dubnicka:2018gqg}. The computations are in many ways similar to those of other cases and provide an insight of how leptonic and radiative decays are treated within the CCQM. Since $B_s$ is the only hadron, one needs to extend the set of parameters (\ref{par_values}) only by one number, i.e. $\Lambda_{B_s} = 2.05 \, \text{GeV}$ which was settled in previous works. The values of remaining parameters are identical to (\ref{par_values}), see Eq. (8) of \cite{Dubnicka:2018gqg}. Two explicit forms of the effective Hamiltonian (\ref{eq:Heff}) are considered
\begin{align}
\mathcal{H}_{\text{eff.}}^{b \to s \ell^+ \ell^-}
=& \frac{G_F \alpha_{\text{EM}}}{2 \sqrt{2} \pi} V_{tb}V^*_{ts} \bigg[
C_9^\text{eff} \{ \bar{s} \gamma^\mu(1-\gamma^5)b \}(\bar{\ell} \gamma_\mu \ell) \nonumber 
 -\frac{ 2 \tilde{m}_b}{q^2} C_7^\text{eff} \{ \bar{s}i\sigma^{\mu \nu} q_\nu (1+\gamma_5)b \}
(\bar{\ell} \gamma_\mu \ell) \label{eq_effH_gamma1} \\
&+ C_{10}\{\bar{s}\gamma^\mu(1-\gamma_5)b \}(\bar{\ell} \gamma_\mu \gamma_5 \ell)
\bigg],
\\
\mathcal{H}_{\text{eff}}^{b \to s \gamma} =& -\frac{G_F}{\sqrt(2)} V_{tb}V^*_{ts} C_7^\text{eff}
\frac{e \tilde{m}_b}{8 \pi^2} \left[ \bar{s}\sigma_{\mu \nu}(1+\gamma_5)b \right]F^{\mu \nu}, \label{eq_effH_gamma2}
\end{align}
where $\sigma_{\mu \nu} = i[\gamma_\mu,\gamma_\nu]$ and $F^{\mu \nu}$ is the EM field tensor. In (\ref{eq_effH_gamma1}) the dilepton is produced from the weak $b$-$s$ transition, Fig. \ref{Fig:dLfromBS}, in (\ref{eq_effH_gamma2}) the weak transition gives birth to a real photon, Fig \ref{Fig:GfromBS}. An additional set of diagrams depicted in Fig \ref{Fig:FSR} is considered too, where the real photon is emitted as the final-state radiation (FSR).

The tilde notation in (\ref{eq_effH_gamma1})(\ref{eq_effH_gamma2}) indicates the QCD quark mass (different from (\ref{par_values})) which is $\tilde{m_b} = 4.68 \pm 0.03 \, \text{GeV}$ \cite{Bauer:2004ve}. The values of scale-dependent Wilson coefficients were determined in \cite{Descotes-Genon:2013vna} at the matching scale $\mu_0 = 2m_W$ and run to the hadronic scale $\mu_b = 4.8 \, \text{GeV}$. The effective operators are defined through the standard SM operators as follows
\begin{align}
C_7^\text{eff} =& C_7-C_5/3-C_6, \nonumber \\
C_9^\text{eff} =& C_9+C_0[ h(\hat{m}_c,s) + \Omega] - \frac{1}{2}h(1,s)(4C_3+4C_4+3C_5+C_6) \label{def:Ceff} \\
& - \frac{1}{2}h(0,s)(C_3+3C_4)+\frac{2}{9}(3C_3+C_4+3C_5+C_6), \nonumber 
\end{align}
where
\begin{align}
&C_0 = 3C_1+C_2+3C_3+C_4+3C_5+C_6,
\quad
\Omega = \frac{3\pi}{\alpha^2}\kappa \sum_{V_i=\Psi(1s),\Psi(2s)} 
\frac{\Gamma(V_i \to \ell^+ \ell^-)m_{V_i}}{m_{V_i}^2-q^2-im_{V_i}\Gamma_{V_i}}, \nonumber \\
&\hat{m}_c = \tilde{m}_c/m_{B_s},
\quad \tilde{m}_c = 1.27 \pm 0.03 \text{GeV}, 
\quad s=q^2/m_{B_s}^2,
\quad \kappa = 1/C_0, \nonumber \\
&h(0,s) = \frac{8}{27} - \frac{8}{9} \ln\frac{\tilde{m}_b}{\mu} - \frac{4}{9} \ln s + \frac{4}{9}i\pi, \label{eq:VMD} \\
&h(\hat{m}_c,s) = -\frac{8}{9} \left[ \ln \frac{\tilde{m}_b}{\mu} \nonumber
+ \ln \hat{m}_c - \frac{1}{3} - \frac{x}{2} \right]  
- \frac{2}{9}(2+x)\sqrt{|1-x|} \, \Theta(x), \nonumber \\
&\Theta(x)|_{x<1} = \ln \left| \frac{\sqrt{1-x}+1}{\sqrt{1-x}-1} \right| -i\pi,
\quad \Theta(x)|_{x>1} = 2 \arctan \frac{1}{\sqrt{x-1}},
\quad x=\frac{4\hat{m}_c^2}{s}. \nonumber
\end{align}
The $\Omega$ function in $C_9^\text{eff}$ parameterizes, in the standard Breit-Wigner form, the resonant contributions from $\Psi(1s)$ and $\Psi(2s)$ charmonia states.

Amplitudes given by the diagrams in Figs. \ref{Fig:dLfromBS} and \ref{Fig:GfromBS}, where the photon originates from the intermediate QCD-generated states, are labeled as structure dependent and can be described by four $B_s \to \gamma$ transition form factors (see e.g. \cite{Melikhov:2004mk}). Defining momenta as $B_s(p_1) \to \gamma(p_2) \; \ell^+(k_+) \; \ell^-(k_-)$, $q=p_1-p_2$ with $p_1^2 = m_{B_s}^2$, $p_2^2 = 0$, $\epsilon_2^\dagger \cdot p_2=0$ and $k_\pm^2 = m_\ell^2$ one has
\begin{align}
\langle \gamma(p_2,\epsilon_2)|\bar{s} \gamma^\mu b | B_s(p_1) \rangle
&= e (\epsilon_2^\dagger)_\alpha \varepsilon^{\mu \alpha \beta \delta} (p_1)_\beta (p_2)_\delta F_V(q^2)/m_{B_s},  \nonumber \\
\langle \gamma(p_2,\epsilon_2)|\bar{s} \gamma^\mu \gamma_5 b | B_s(p_1) \rangle
&= ie (\epsilon_2^\dagger)_\alpha (g^{\mu \alpha} p_1 p_2 - p_1^\alpha p_2^\mu) F_A(q^2) / m_{B_S}, \nonumber \\
\langle \gamma(p_2,\epsilon_2)|\bar{s} \sigma^{\mu \beta} q_\beta b | B_s(p_1) \rangle
&= ie (\epsilon_2^\dagger)_\alpha \varepsilon^{\mu \alpha \beta \delta} (p_1)_\beta (p_2)_\delta F_{TV}(q^2), \\
\langle \gamma(p_2,\epsilon_2)|\bar{s} \sigma^{\mu \beta} q_\beta \gamma_5 b | B_s(p_1) \rangle
&= e (\epsilon_2^\dagger)_\alpha (g^{\mu \alpha} p_1 p_2 - p_1^\alpha p_2^\mu) F_{TA}(q^2), \nonumber
\end{align}
where $\epsilon$ is the polarization vector. Each of the four introduced form factors can be expressed as sum of contributions from particular Feynman graphs in Figs. \ref{Fig:dLfromBS} and \ref{Fig:GfromBS}. One has
\begin{align}
&F_V = m_{B_s} (e_b \tilde{F}^{b \gamma b}_V + e_s \tilde{F}^{s \gamma s}_V), \nonumber \\
&F_A = m_{B_s} (e_b \tilde{F}^{b \gamma b}_A + e_s \tilde{F}^{s \gamma s}_A
	+ e_b \tilde{F}^{\text{bubble}-b}_A + e_s \tilde{F}^{\text{bubble}-s}_A),
\nonumber \label{Eq:FFdef}\\
&F_{TV} = e_b \tilde{F}^{b \gamma b}_{TV} + e_s \tilde{F}^{s \gamma s}_{TV}
	+ e_b \tilde{F}^{b(\bar{\ell}\ell)b}_{TV} + e_s \tilde{F}^{s(\bar{\ell}\ell)s}_{TV}, \\
&F_{TA} = e_b \tilde{F}^{b \gamma b}_{TA} + e_s \tilde{F}^{s \gamma s}_{TA}
	+ e_b \tilde{F}^{\text{bubble}-b}_{TA} + e_s \tilde{F}^{\text{bubble}-s}_{TA}
	+ e_b \tilde{F}^{b(\bar{\ell}\ell)b}_{TA} + e_s \tilde{F}^{s(\bar{\ell}\ell)s}_{TA}, \nonumber
\end{align}
where "$q \gamma q$" superscript refers to a real photon emission from the quark line, "$bubble$" to the real photon emission from the non-local hadron-quark vertex and "$q(\bar{\ell}\ell)q$" corresponds to the virtual photon emission from the quark line.

The branch point at $q^2 = 4m_s^2$ corresponding to the virtual photon emission from the $s$ quark (left in Fig. \ref{Fig:GfromBS}) is situated well inside the accessible physical $q^2$ region. This leads to the appearance of light vector meson resonance which prevents us to compute the corresponding form factors within the CCQM. An approach inspired by \cite{Kozachuk:2017mdk} is adopted and a gauge-invariant vector-meson dominance model is used to express the form factors in question
\begin{align}
\tilde{F}^{s(\bar{\ell}\ell)s}_{TV,TA} &= \tilde{F}^{s(\bar{\ell}\ell)s}_{TA}(0)
-\sum_V 2 f^{EM}_V G^T_1(0) \frac{q^2/M_V}{q^2-M_V^2+iM_V\Gamma_V}, \label{eq:VMD2} \\
G_1^T: \quad &\langle V(p_2,\epsilon_2)|\bar{s} \sigma^{\mu \nu} b|B_s(p_1)) \rangle \nonumber
=  \\
& \quad
= (\epsilon_2^\dagger)_\alpha 
\left[
\varepsilon^{\beta \mu \nu \alpha} P_\beta G_1^T(q^2)
+\varepsilon^{\beta \mu \nu \alpha} q_\beta G_2^T(q^2)
+\varepsilon^{\alpha \beta \mu \nu} P_\alpha q_\beta \frac{G_0^T(q^2)}{ \left( m_{B_s}+M_V \right)^2 }
\right],
\end{align}
where $P = p_1 + p_2$. With all this objects defined, one can write down the amplitude for the structure dependent part
\begin{align}
\mathcal{M}_\text{SD} =& \frac{G_F}{\sqrt{2}} \frac{\alpha_{EM} V_{tb}V^*_{ts}}{2\pi} e (\epsilon_2^*)_\alpha \bigg\{ \bigg[ \varepsilon^{\mu \alpha \nu \beta} (p_1)_\nu (p_2)_\beta 
\frac{F_V(q^2)}{m_{B_s}} 
- iT_1^{\mu \alpha}\frac{F_A(q^2)}{m_{B_s}}\bigg]
\times \big( C_9^\text{eff} \bar{\ell} \gamma_\mu \ell
\nonumber \\
&+ C_{10} \bar{\ell} \gamma_\mu \gamma_5 \ell \big)
+  \big[ \epsilon^{\mu \alpha \nu \beta} (p_1)_\nu (p_2)_\beta F_{TV}(q^2)
 - iT_1^{\mu \alpha} F_{TA}(q^2) \big]
 \frac{3 \tilde{m}_b}{q^2} C_7^\text{eff} \bar{\ell}\gamma_\mu\ell,
\end{align}
where $T_1^{\mu \alpha} = [g^{\mu \alpha} p_1 p_2 - (p_1)^\alpha (p_2)^\mu ]$. The structure-independent \textit{bremsstrahlung} (Fig. \ref{Fig:FSR}) amplitude takes the form
\begin{align}
\mathcal{M}_\text{BR} = -i \frac{G_F}{\sqrt{2}} \frac{\alpha_{EM} V_{tb}V^*_{ts}}{2\pi} e (\epsilon_2^*)_\alpha
(2 m_\ell f_{B_s} C_{10})
\bar{u}(k_-) \bigg[ \frac{\gamma^\alpha \cancel{p}_1}{t-m_\ell^2}
- \frac{\cancel{p}_1 \gamma^\alpha }{u-m_\ell^2} \bigg] \gamma_5 v(k_+). \label{Amp_BR}
\end{align}
Here $t=(p_2+k_-)^2 $, $u = (p_2+k_+)^2$. To avoid infrared divergences in (\ref{Amp_BR}) a lower boundary on the photon energy has to be introduced $E_\gamma > E_{\gamma \, \text{min}}$ set later, in numerical computations (Table \ref{Tab:rad_BF}),  to $20 \; \text{MeV}$.

The differential branching fraction in $t$ and $s \equiv q^2$ has a general expression
\begin{align}
\frac{d\Gamma}{ds \,dt} = \frac{1}{2^8 \pi^3 m_{B_s}^3} \sum_{\text{pol.}}
\left| \mathcal{M}_\text{SD} + \mathcal{M}_\text{BR} \right|^2,
\end{align}
where one sums over the polarization of photons and leptons, $4m_\ell^2 \le s \le m_{B_s}^2$, $t_- \le t \le t_+$ with $t_\pm = m_\ell^2+(m_{B_s}^2-s)[1 \pm \sqrt{1 - 4m_\ell^2/s}]/2$. The explicit formulas for double and single differential distributions we omit here because of their complexity, they are stated in Eqs. (32)-(38) of \cite{Dubnicka:2018gqg}.

\begin{figure}[t]
\center{
\includegraphics[width=1.0\textwidth]{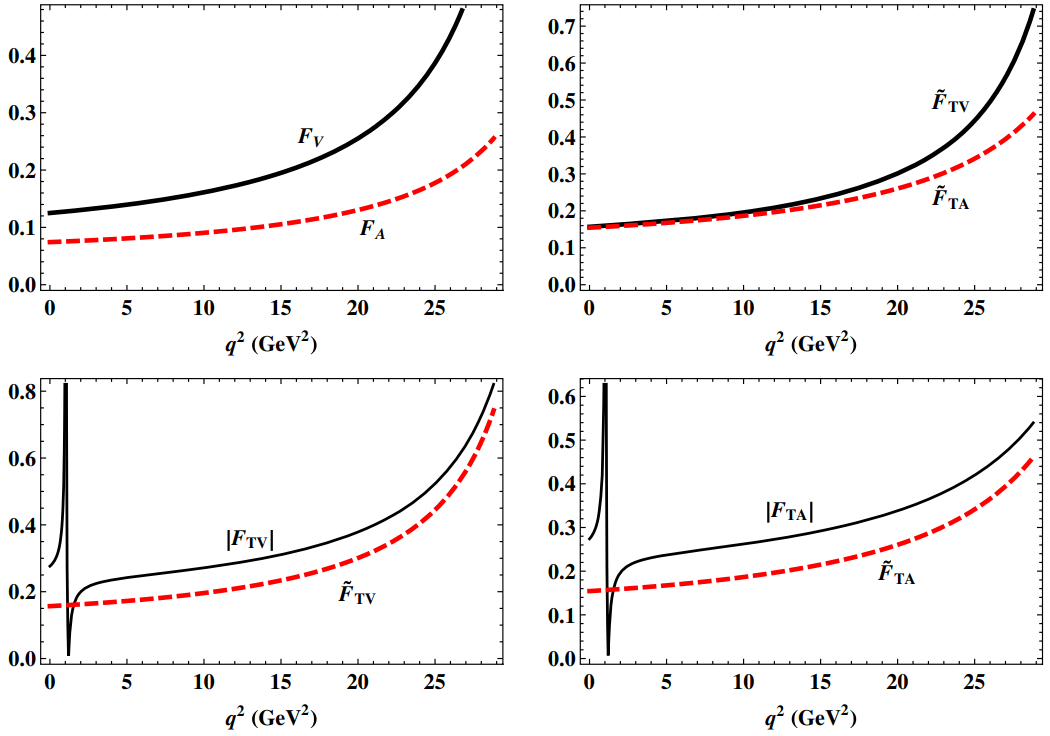}
}
\caption{Transition form factors $B_s \to \gamma$ as defined by (\ref{Eq:FFdef}) and (\ref{eq_noVMD}). Figures were originally published in \cite{Dubnicka:2018gqg}. }
\label{Fig:rad_FFs}
\end{figure}

The form factors predicted by the CCQM model are shown in Fig. \ref{Fig:rad_FFs}. For $F_{TV/TA}$ form factors two scenarios are presented: by including the VMD component (\ref{eq:VMD2}) these form factors become complex and thus their norm is shown. Alternatively, they can be shown without the VMD component as real functions
\begin{align}
\tilde{F}_{TV,TA} \equiv F_{TV} - e_s \tilde{F}^{s(\bar{\ell}\ell)s}_{TV,TA}. \label{eq_noVMD}
\end{align}

\begin{figure}[ht]
\begin{center}
\begin{tabular}{ll} 
\includegraphics[scale=0.5]{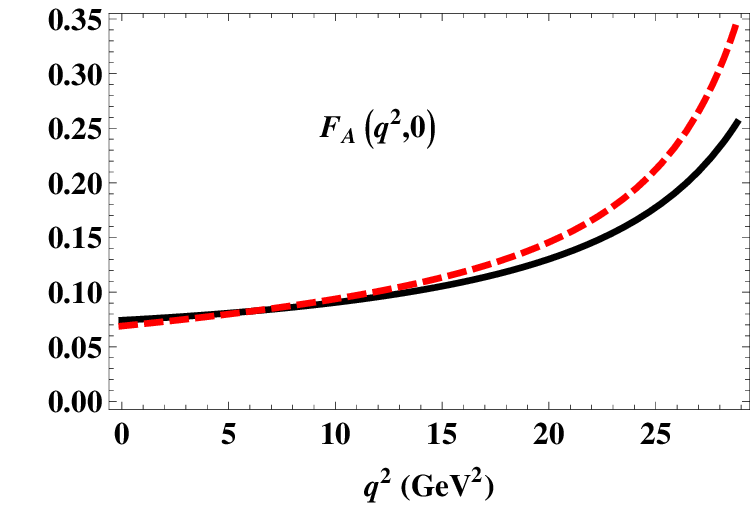} &
\includegraphics[scale=0.5]{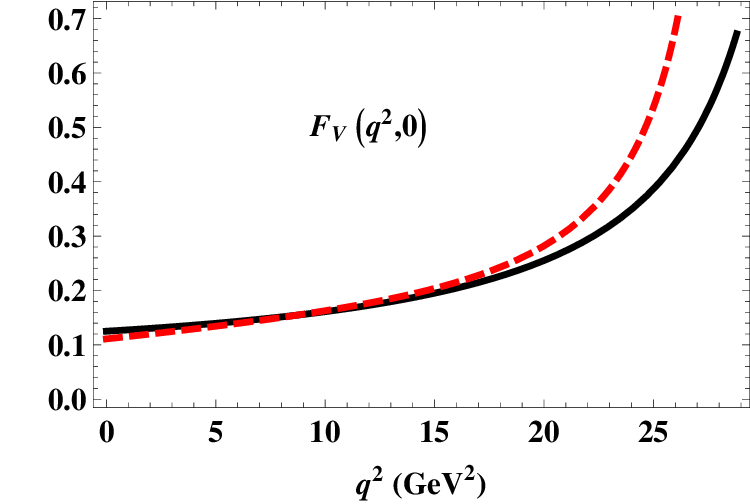} \\ 
\includegraphics[scale=0.5]{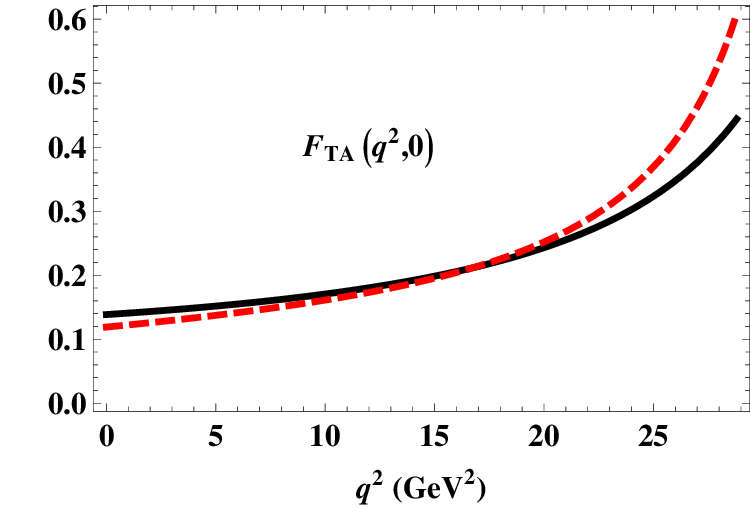} &
\includegraphics[scale=0.5]{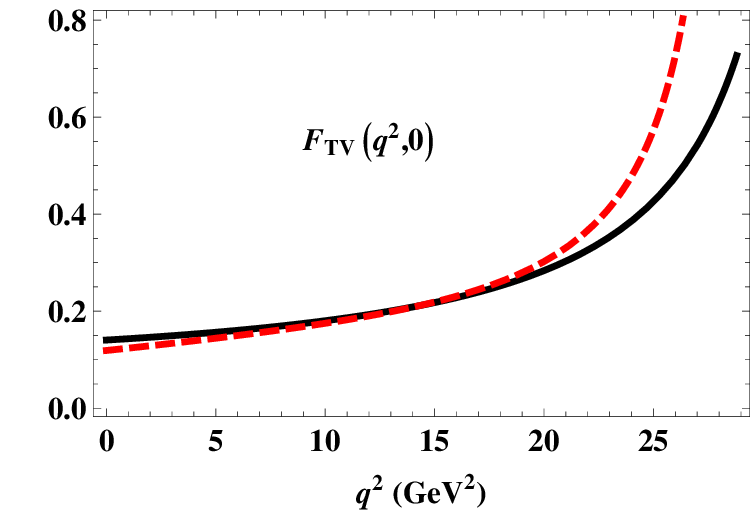}
\end{tabular}
\end{center}
\caption{\label{fig:ffMel}
  Comparison of the form factors $F_i(q^2,0)$ calculated in our model
  (solid lines) with those from Ref.~\cite{Kozachuk:2017mdk} (dashed lines).
Figures are taken from \cite{Dubnicka:2018gqg}.
}
\end{figure}

In Fig.~\ref{fig:ffMel} we compare our form factors with the 
Kozachuk-Melikhov-Nikitin (KMN) form factors calculated in
Ref.~\cite{Kozachuk:2017mdk}. Using the definitions
we can relate our form factors $F_i(q^2)$ to the KMN form factors
$F_i(q^2,0)$ as follows  (see Ref.~\cite{Kozachuk:2017mdk} for more detail):
\begin{align}
F_{V/A}(q^2,0) = F_{V/A}(q^2),
\;
F_{TV/TA}(q^2,0) \equiv \tilde F_{TV/TA}(q^2) =
F_{TV/TA}(q^2)-e_b \tilde{F}_{TV/TA}^{b(\bar{l}l)b}-e_s \tilde{F}_{TV/TA}^{s(\bar{l}l)s}.
\nonumber
\end{align}
One can see that in the low-$q^2$ region ($q^2\lesssim 20$ GeV$^2$) the
corresponding form factors from the two sets are very close.
In the high-$q^2$ region, the KMN form factors steeply increase and
largely exceed our form factors.
It is very interesting to note that our form factors share with the
corresponding KMN ones not only similar shapes (especially in the low-$q^2$
region) but also relative behaviors, i.e., similar relations between the form
factors, in the whole $q^2$ region. Several comments should be made:
(i) our form factors satisfy the constraint $F_{TA}(q^2,0)=F_{TV}(q^2,0)$ at
$q^2=0$, with the common value equal to 0.135; (ii) in the small-$q^2$ region,
$F_{V}(q^2,0)\approx F_{TA}(q^2,0)\approx F_{TV}(q^2,0)$; (iii)
$F_{V}(q^2,0)$ and $F_{TV}(q^2,0)$ are approximately equal in the full
kinematical range and rise steeply in the high-$q^2$ region; and
(iv) $F_{A}(q^2,0)$ and $F_{TA}(q^2,0)$ are rather flat when $q^2\to M_{B_s}^2$
as compared to $F_{V}(q^2,0)$ and $F_{TV}(q^2,0)$. These observations show that
our form factors satisfy very well the constraints on their behavior proposed
by the authors of Ref.~\cite{Kruger:2002gf}.

The analytic ${\cal O}(\alpha_s)$-computation at twist-$1$,$2$
of the $\bar B_{u,d,s} \to \gamma$ form factors have been presented in
\cite{Janowski:2021yvz} within the framework of sum rules on the light-cone.
A fit was provided in terms of a $z$-expansion with correlation matrix and
the form factors extrapolated  to the kinematic endpoint by using
the  $g_{BB^*\gamma}$  couplings as a constraint. 
When comparing with  \cite{Janowski:2021yvz}
the following identification  should be used
\begin{align}
V_{\perp,\parallel} =  F_{V,A} \;, \quad 
T_{\perp,\parallel} =  \tilde F_{TV,TA}  \;.
\end{align}
\begin{figure}[h]
\centering
      \begin{minipage}{0.5\textwidth}
        \centering
        \includegraphics[width=1.\textwidth]{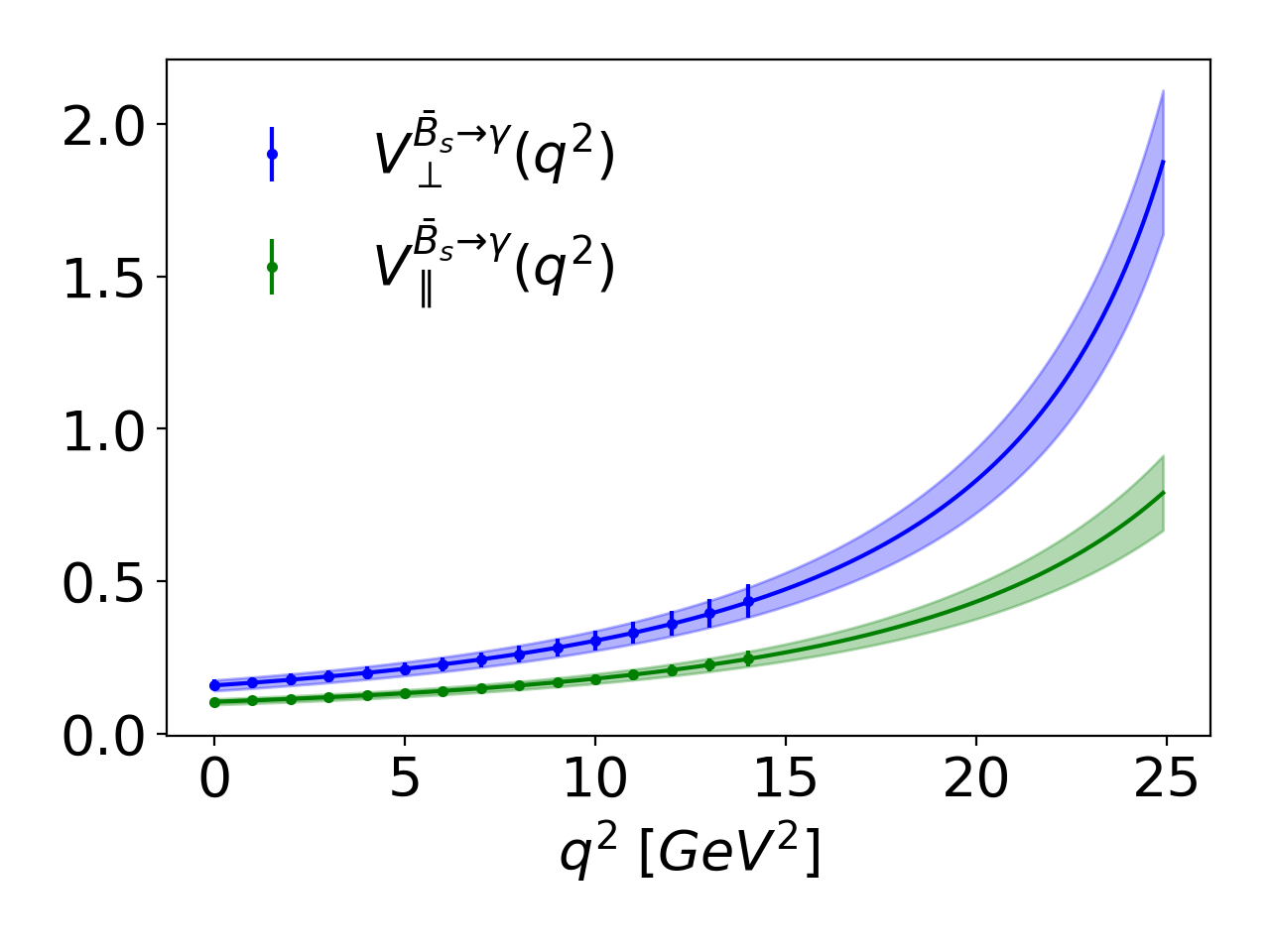}
    \end{minipage}\hfill
    \begin{minipage}{0.5\textwidth}
        \centering
        \includegraphics[width=1.\textwidth]{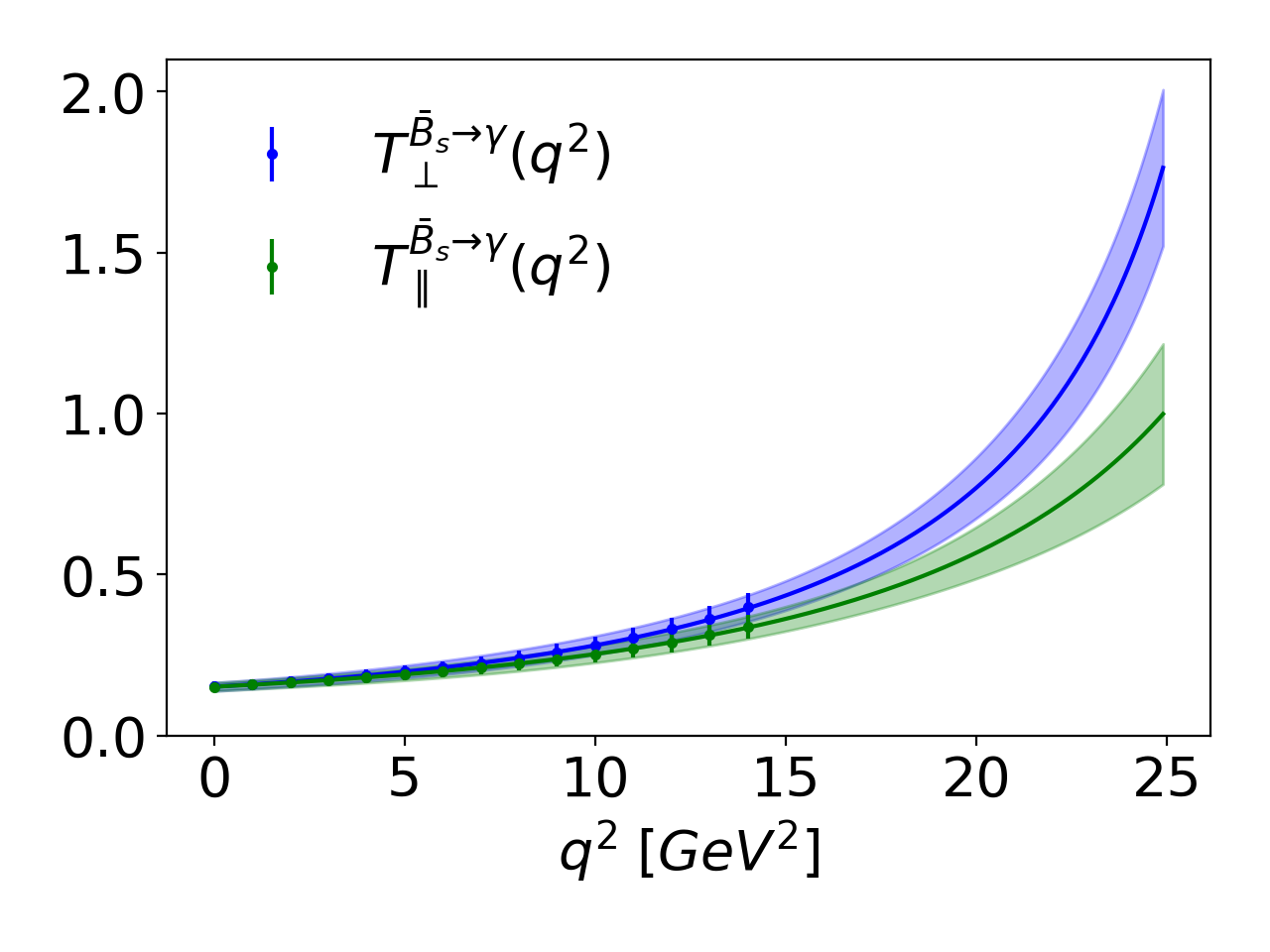}
    \end{minipage}
    \caption{\small Form factors for the $\bar B_s \to \gamma$ transition
      calculated in \cite{Janowski:2021yvz}. 
       Figures are taken from \cite{Janowski:2021yvz}. }      
    \label{fig:FFplots}
\end{figure}
One can see (Fig. \ref{fig:FFplots}) that the  results of  \cite{Janowski:2021yvz} are larger and,
in particular, show an earlier rise.
 The differential branching fractions shown as a function of dimensionless variable $\hat{s} = q^2/m_{B_s}$ are, together with the branching fraction ratio
\begin{align}
r_\gamma(\hat{s}) \equiv 
\frac
{d\mathcal{B}(B_s \to \gamma \mu^+ \mu^-)/d\hat{s}}
{d\mathcal{B}(B_s \to \gamma e^+ e^-)/d\hat{s}} \label{rad_rRatio}
\end{align}
depicted in Fig. \ref{Fig:rad_BF}. 
\begin{figure}[t]
\center{
\includegraphics[width=1.0\textwidth]{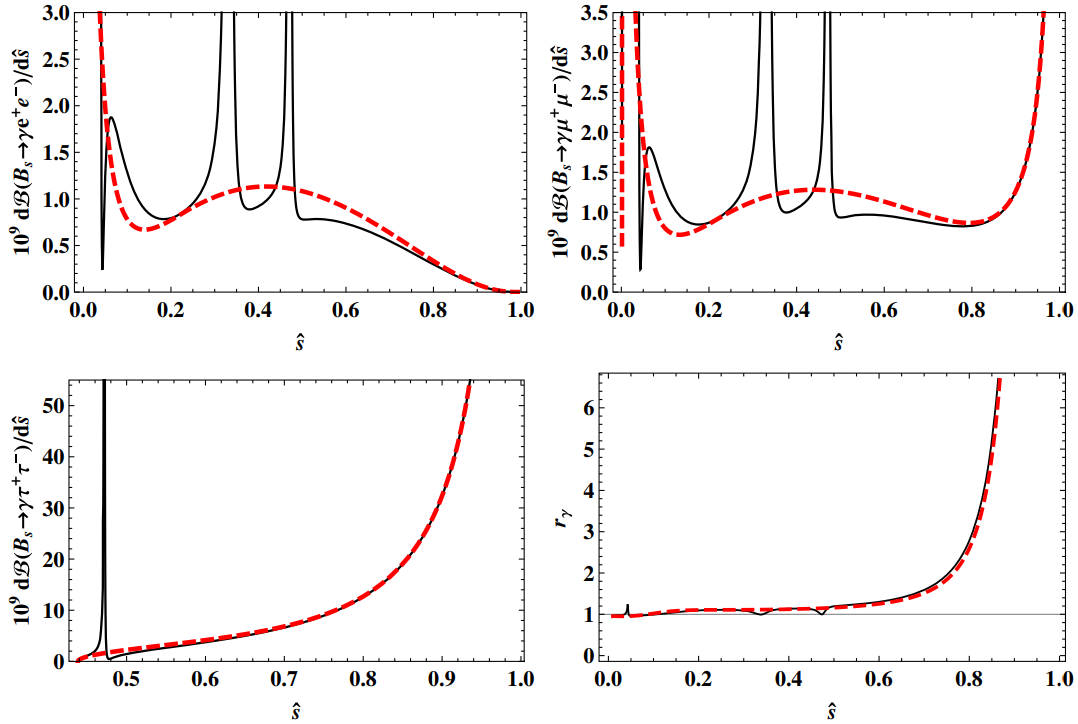}
}
\caption{Differential decay rates for $B_s \to \ell^+ \ell^- \gamma$ and the ratio $\hat{r}$ (\ref{rad_rRatio}) with long-distant contributions included (solid line) and excluded (dashed line). Figures were originally published in \cite{Dubnicka:2018gqg}.}
\label{Fig:rad_BF}
\end{figure}
The total branching fractions for the three lepton flavors are presented in Table \ref{Tab:rad_BF}.
\begin{table}
\centering
\begin{tabular}[t]{c  c  c  c  c}
\hline
\hline
 & Struct. Dep. & Bremst. & Interf. & Sum \\
\hline 
$10^9 \mathcal{B}(B_s \to \gamma e^+ e^-)$ & $3.05 \; (\textbf{15.9})$ & $3.2 \times 10^{-5}$&$-4.8 \; (\textbf{-9.5})\times 10^{-6}$ & $3.05 \; (\textbf{15.9})$\\
$10^9 \mathcal{B}(B_s \to \gamma \mu^+ \mu^-)$ & $1.16 \; (\textbf{10.0})$ & $0.53$&$-7.4 \; (\textbf{-14.4})\times 10^{-3}$ & $1.7 \; (\textbf{10.5})$\\
$10^9 \mathcal{B}(B_s \to \gamma \tau^+ \tau^-)$ & $0.10 \; (\textbf{0.05})$ & $13.4$&$0.30 \; (\textbf{0.18})$ & $13.8 \; (\textbf{13.7})$\\
\hline
\hline
\end{tabular}
\caption{Branching fractions for the three lepton flavors. Values in brackets take into account long-distance contributions. Table was originally published in \cite{Dubnicka:2018gqg}.}
\label{Tab:rad_BF}
\end{table}
The numbers in brackets indicate the results of computations with long-distance contributions included (but one excludes the region of the two low lying charmonia $0.33 \le \hat{s} \le  0.55$), results without the long distance contributions correspond to $\kappa=0$ in (\ref{eq:VMD}). The comparison with theoretical predictions of other authors is shown in Table \ref{Tab:rad_others}.
\begin{table}
\centering
\begin{tabular}[t]{c  c  c  c  c c c c c c}
\hline
\hline
 & CCQM & \cite{Eilam:1996vg} & \cite{Aliev:1996ud} & \cite{Aliev:1997sd} & \cite{Geng:2000fs} & \cite{Dincer:2001hu} & \cite{Melikhov:2004mk} & \cite{Melikhov:2017pwu} & \cite{Wang:2013rfa} \\
\hline 
electron & 15.9 & 6.2 & 2.35 & - & 7.1 & 20.0 & 24.6 & 18.4 & 17.4 \\
muon & 10.5 & 4.6 & 1.9 & - & 8.3 & 12.0 & 18.9 & 11.6 & 17.4 \\
tau & 13.7 & - & - & 15.2 & 15.7 & - & 11.6 & - & - \\
\hline
\hline
\end{tabular}
\caption{Comparison of branching fractions with other theoretical predictions. Table was originally published in \cite{Dubnicka:2018gqg}. }
\label{Tab:rad_others}
\end{table}
The dominant error source of the results was identified to be the uncertainty of the hadronic form factors and the error on the branching fractions was estimated to reach 30\%. One should remark that the resonant peaks induced by light $\phi$ particles lead to significant enhancement of the branching fraction ($\approx $15\%).

In summary, in the presented SM computations within the CCQM the hadronic transition form factors and radiative leptonic branching fractions of the $B_s$ meson were evaluated. The form factors are in a very good agreement with those presented in \cite{Kozachuk:2017mdk} and the branching fraction numbers for light leptons agree with \cite{Melikhov:2017pwu}. For the tau lepton decay mode, where bremsstrahlung dominates, the presented results agree with all other authors. Together, these results from various authors with \cite{Dubnicka:2018gqg} included, reflect our understanding of the SM description of the $B_s \to \ell^+ \ell^- \gamma$ decay process and provide an estimate on the error of theoretical SM predictions, beyond which one can claim NP manifestations.

\subsection{Other CCQM results on $B$ leptonic decay}
The CCQM was applied also to the leptonic decays $B \to \ell^- \bar{\nu}_\ell$ \cite{Ivanov:2015tru} and $B_c^- \to \tau \bar{\nu}$ \cite{Ivanov:2017hun}.

The work \cite{Ivanov:2015tru} provides a SM analysis of pure leptonic and semileptonic decays. Most of the results presented there concern the semileptonic processes, which have richer structure and significant hints for the NP. Yet the results for purely leptonic branching fractions were presented too

\begin{center}
\begin{tabular}{cccc}
$\ell $ & $e$ & $\mu$ & $\tau$ \\
$\mathcal{B}( B^- \to \ell^- \bar{\nu}_\ell )$& $1.16 \times 10^{-11}$ & $0.49 \times 10^{-6}$ & $1.10 \times 10^{-4}$ \\
\end{tabular}.
\end{center}

The numbers are in good agreement with the experimental values for the tau lepton $(1.090 \pm 0.24)\times 10^{-4}$ \cite{Workman:2022ynf} and the muon $(0.53 \pm 0.22)\times 10^{-9}$ \cite{Belle:2019iji}, which became more precisely measured since then, and also with the experimental limit for the electron. The agreement with several theoretical prediction of other authors was shown too. Since the leptonic decay constants are crucial in the description of purely leptonic decays and carry all of the necessary non-perturbavite information, their values have also been listed for $B^{(*)}_{(s,c)}$ and $D^{(*)}_{(s)}$ mesons, see Table I of \cite{Ivanov:2015tru} .

In \cite{Ivanov:2017hun} possible NP contributions were evaluated for chosen leptonic and semileptonic decays. It was assumed that these contributions affect only the third generation of leptons and all neutrinos were considered as left-handed. New, beyond-SM four-fermion operators were introduced in the Hamiltonian (\ref{eq:Heff})
\begin{align}
Q_{V_i} = (\bar{q}\gamma^\mu P_i b)(\bar{\tau} \gamma_\mu P_L \nu_\tau),
\quad Q_{S_i} = (\bar{q} P_i b)(\bar{\tau}  P_L \nu_\tau),
\quad Q_{T_L} = (\bar{q}\sigma^{\mu\nu} P_L b)(\bar{\tau} \sigma_{\mu\nu} P_L \nu_\tau) \label{NPope}
\end{align}
with $\sigma_{\mu\nu} = i[\gamma_\mu,\gamma_\nu]$, $P_{L,R} = (1 \mp \gamma_5)/2$ and $i \in \{L,R \} $ (left, right). The most of the text deals with semileptonic decays where the $R_{D^{(*)}}$ discrepancy is observed (\ref{eq:RD*}). The set of observables was extended to
\begin{align}
R_{\pi(\rho)}=\frac{\mathcal{B}(\bar{B}^0 \to \pi(\rho)\tau\bar{\nu})}
{\mathcal{B}(\bar{B}^0 \to \pi(\rho)\mu\bar{\nu})}, 
\quad R^u_{\tau}=\frac{\tau_{\bar{B}^0}}{\tau_{\bar{B}^{-}}}
\frac{\mathcal{B}(\bar{B}^- \to \tau\bar{\nu})}
{\mathcal{B}(\bar{B}^0 \to \pi\mu\bar{\nu})}, 
\quad R^c_{\tau}=\frac{\tau_{\bar{B}^0}}{\tau_{B^{-}_c}}
\frac{\mathcal{B}(\bar{B}^-_c \to \tau\bar{\nu})}
{\mathcal{B}(\bar{B}^0 \to D\mu\bar{\nu})}, \label{eq:PNobser}
\end{align}
of which the first is meant to analyze the $R$ anomaly also for the $b \to u$ transition and the two remaining concern the leptonic decays. The limits on the Wilson coefficients $C_{V_i, S_i, T_l}$ were extracted assuming that only one of them is dominant at a time (besides the SM ones). Including into the analysis  also the leptonic observable $R^u_{\tau}$ (together with $R_{D^{(*)}}$) it was found that no $C_{S_R,S_L}$ values were allowed (within 2 $\sigma$) and for $C_{V_L,V_R, T_L}$ allowed regions were identified in the complex plane (Fig. 1 of \cite{Ivanov:2017hun}). Further, the leptonic  $\bar{B}^-_c$ branching fractions were evaluated within the SM, $\mathcal{B}(\bar{B}^-_c \to \tau\bar{\nu}) = 2.85\times 10^{-2}$, $\mathcal{B}(\bar{B}^-_c \to \mu\bar{\nu}) = 1.18\times 10^{-4}$ and observables (\ref{eq:PNobser}) were predicted for the SM and NP scenarios. In the latter case the corresponding Wilson coefficient $C_i$ was varied (one at a time) in the allowed region of the complex plain and the impact on the observable was determined. For the leptonic $R^c_{\tau}$ variable the prediction stands
\begin{center}
\begin{tabular}{ccccc}
& $SM$ & $C_{V_L}$ & $C_{V_R}$ & $C_{T_L}$\\
$R^c_{\tau}=$ & $3.03$ & $3.945 \pm 0.735$ & $3.925 \pm 0.815$ & $3.03$.
\end{tabular}
\end{center}
As summary one can say that, within the given scenario, the text translated existing experimental information into the constraints on NP Wilson coefficients. Contributions of some of them ($C_{S_R,S_L}$) were excluded and some ($C_{V_L,V_R, T_L}$) were constrained.

\section{Semileptonic decays of $B$ mesons}

\subsection{Overview}

The experimental information on the semileptonic B decays is much larger than on the pure leptonic decays. The LHCb experiment alone published in the past 10 years more than 35 papers on this topic and the number further increases if other experiments (Belle, BaBar, Belle II) are taken into the account. The same is true for theoretical publications which are large in quantity. With the aim to provide an overview of the CCQM results, we restrain ourselves only to most significant experimental measurements and theoretical predictions of other authors.

The focus of the community is predominantly driven by the so-called flavor anomalies. They are often defined as ratios of branching fractions, the most prominent of them are
\begin{align}
R_{K^{(*)}} = \frac{\mathcal{B}(B \to K^{(*)} \mu^+ \mu^-) }
				   {\mathcal{B}(B \to K^{(*)} e^+ e^- )}, \quad
R_{D^{(*)}} = \frac{\mathcal{B}(B \to D^{(*)} \tau \nu_\tau) }
				   {\mathcal{B}(B \to D^{(*)} \ell \nu_\ell)}, \quad
R_{J/\Psi} = \frac{\mathcal{B}(B \to J/\Psi \tau \nu_\tau) }
				   {\mathcal{B}(B \to J/\Psi \mu \nu_\mu)} \label{eq:RD*}.
\end{align}
The first observable is sensitive to the $b \to s$ quark transition, the two remaining to  $b \to c$. Other quantities measured in semileptonic decays of the $B$ meson are listed for example in Sec. VII of \cite{HeavyFlavorAveragingGroup:2022wzx}.  In these and other observables deviations were seen (see e.g. Tab XVIII  of \cite{Bernlochner:2021vlv} for a nice review) with some of them reaching up to 4$\sigma$, which is naturally interpreted as significant argument in favor of the NP (see e.g. \cite{Crivellin:2018yvo} ). The most recent LHCb measurements nevertheless weaken some of these observations and imply that the discrepancy with the SM may not be so pronounced after all. In \cite{LHCb:2023zxo} the deviation of a correlated observables $R_{D}$ and $R_{D^*}$ from the SM prediction is  $1.9 \sigma$ and the results for $R_K$ and $R_{K^*}$ given in \cite{LHCb:2022qnv} are in agreement with the SM. However, if one includes also older measurements and measurements of different experiments, the situation seem not to be yet solved and discrepancy is still close to $3 \sigma$ \cite{Puthumanaillam}.

\vspace{0.5cm}

The LHCb detector was specifically designed for $b$ physics and the experiment successfully reaches its purpose by being the most important source of the experimental information on  $b$ decays. The measurements of $B \to K^* \ell^+ \ell^-$ were presented in works \cite{LHCb:2013zuf,LHCb:2013ghj,LHCb:2014cxe,LHCb:2015ycz, LHCb:2015svh, LHCb:2017avl, LHCb:2020lmf, LHCb:2020gog, LHCb:2021lvy}. Two of them \cite{LHCb:2017avl,LHCb:2021lvy} study the lepton-flavor universality by measuring $R_{K^*}$, but with no significant deviations from the SM. Most of the remaining works are concerned with angular distributions: the coefficients (noted for a $p$-wave process as $F_L$, $A_{FB}$, $S_{3,\dots,9}$) in front of angular terms which appear in the decay width formula are combined into so-called optimized observables $P_i^{(')}$, and here some significant tensions are seen (e.g. $3\sigma$ in $P_2$ for $q^2$ between 6 and 8 $\text{GeV}^2$ \cite{LHCb:2020gog}).

The semileptonic $B$ decays with the $K$ meson in the final state are addressed in \cite{LHCb:2012juf,LHCb:2014vgu, LHCb:2019hip}. The first publication is concerned with the angular distribution and the differential branching fraction, the two others focus more specifically on the lepton flavor universality question, with an observation of a $2.5\sigma$ deviation from the SM in $R_K$. This was however, as mentioned earlier, undermined by the recent  measurement \cite{LHCb:2022qnv} where no longer the deviation is seen.

The process $B \to D^* \ell^+ \ell^-$ was analyzed in \cite{LHCb:2015gmp,LHCb:2017smo, LHCb:2017rln, LHCb:2023zxo} and no deviation of $R_{D^*}$ from the SM greater than $2\sigma$ was detected. The same is true for the $R_{J/\Psi}$ observable measured in \cite{LHCb:2017vlu}. The decay of the $B_s^0$ particle to $\phi \mu^+ \mu^-$ was studied in \cite{LHCb:2013tgx,LHCb:2015wdu,LHCb:2021zwz}, where, in the last analysis, a disagreement with the SM prediction is observed in the differential branching fraction for $1 \text{GeV}^2 \le q^2 \le 6 \text{GeV}^2$ at the level of $3.6\sigma$.

Various other semileptonic $B$ decays were measured at the LHCb which we do not mention here. An overview of the lepton flavor universality question in $b$ decays at the LHCb was, as of 2022, given in \cite{LHCb:2021trn}.

\vspace{0.5cm}

An additional experimental information on the semileptonic $B$ decays comes from BaBar measurements. Studies of the $B \to D^{(*)} \ell \nu_\ell$ process were presented in \cite{BaBar:2007hvx, BaBar:2012obs, BaBar:2013mob, BaBar:2009zxk, BaBar:2007cke, BaBar:2008zui, BaBar:2019vpl}. In the first three references the question of the lepton flavor universality is addressed ($\ell = \tau$) and the measurement of $R_D$ and $R_{D^*}$ performed. The authors claim a deviation of $2.0 \sigma$ for $R_D$, $2.7 \sigma$ for $R_{D^*}$ and $3.4 \sigma$ for their combination. The four latter references present the measurement of the $|V_{cb}|$ element of the CKM matrix and the analysis of corresponding transition form factors.

The decays with the $K^{(*)} \ell^+ \ell^-$ final state were  addressed in \cite{BaBar:2003szi, BaBar:2006tnv, BaBar:2008jdv, BaBar:2012mrf, BaBar:2008fao, BaBar:2015wkg}. The texts present the measurements of branching fractions, the $R_{K^{(*)}}$ observable, the isospin and CP asymmetries, the forward-backward angular asymmetry of the lepton pair and the $K^*$ longitudinal polarization (and others). Overall, the results are in an agreement with the SM expectations, the anomaly observed for isospin asymmetries in both $K$ and $K^*$ channels in \cite{BaBar:2008jdv} was not later confirmed in \cite{BaBar:2012mrf}.

The BaBar collaboration also published results on semileptonic $B$ decays into light mesons $\pi$ and $\rho$ \cite{BaBar:2006fyo,BaBar:2010efp}. Here the branching fractions and the $|V_{ub}|$  element were determined and also transition form factors were discussed.

Further, BaBar published results on semileptonic decays where hadronic state $X_s$ containing kaons  was produced and measured corresponding branching fractions \cite{BaBar:2004mjt, BaBar:2013qry}. One can also mention the measurement of charmless semileptonic decays \cite{BaBar:2003org, BaBar:2012thb} and the measurement with the electron in the final state \cite{BaBar:2016rxh}, all of which were used to establish the $|V_{ub}|$ value. In \cite{BaBar:2015zkb} the semileptonic decay with five particles in the final state $D^{(*)} \pi^+ \pi^- \ell \nu_\ell$,  was confirmed.

\vspace{0.5cm}

Important contribution to measurements of semileptonic $B$ decays comes form the Belle and Belle II collaborations.

Analyses \cite{Belle:2010tvu, Belle:2015qfa, Belle:2019gij, Belle:2019rba, BelleII:2023lst} investigate both $D$ and $D^*$ decay channels (with $\tau$ and $\nu_\tau$). They measure branching fractions and ratios $R_{D^{(*)}}$, where they do not see significant deviations from the SM expectations. The last work focuses also on the extraction of parameters for the Caprini-Lellouch-Neubert form factor parameterization.

Specifically $D^*$-containing final states are addressed in \cite{Belle:2007qnm, Belle:2010qug, Belle:2016ure, Belle:2016dyj, Belle:2017ilt, Belle:2018ezy}. Also here the objects of interest are the branching fractions and the $R_{D^*}$ observable and, again, no significant deviations from the SM are seen. Works \cite{Belle:2010qug,Belle:2018ezy} present, in addition, the measurement of the $|V_{cb}|$ matrix element and form factor analysis, in works \cite{Belle:2016dyj, Belle:2017ilt} the $\tau$ lepton polarization is measured.

The references \cite{Belle:2015pkj, Belle-II:2022ffa} focus on the $D \ell \nu_\ell$  final state. The first work is concerned with the branching fraction and form factors, in both works  $|V_{cb}|$ is measured. Authors of \cite{Belle:2022yzd} report on the first observation of $B \to \bar{D}_1\ell\nu_\ell$ decay and measure the branching fractions of $B \to \bar{D}^{(*)}\pi\ell^+\nu_\ell$ and $B \to \bar{D}^{(*)}\pi^+\pi^-\ell^+\nu_\ell$ processes.

Production of strange mesons in semileptonic $B$ decays is studied in \cite{Belle:2001oey,BELLE:2019xld} for the $K$ meson, in \cite{Belle:2006gil,Belle:2016fev,Belle:2019oag} for the $K^*$ meson and in \cite{Belle:2009zue} for both, $K$ and $K^*$. Besides branching fractions and $R_{K^{(*)}}$ ratios, some of the works present also measurements of angular and polarization variables and the isospin asymmetry. In general all measured values agree well with the SM predictions, some tensions for the subset of the optimized angular observables $P_i$ were reported in \cite{Belle:2016fev}.

Semileptonic decays to light mesons ($\pi$, $\rho$ and $\eta$) were described in \cite{Belle:2010hep, Belle-II:2022imn, Belle-II:2022fsw, Belle:2021hah}, the works are mostly concerned with the branching fractions and the determination of the $|V_{ub}|$ element of the CKM matrix.

The Belle(II) collaboration also published articles on semileptonic $B$ decays to a general hadronic state $X$ containing the $s$ quark, $X_s$ \cite{Belle:2002qxc, Belle:2005fli}, the $u$ quark, $X_u$ \cite{Belle:2021ymg,Belle:2021eni, Belle:2013hlo} and the $c$ quark, $X_c$ \cite{Belle:2021idw, Belle-II:2023jgq}. The main objects of interest were branching fractions, CKM elements $|V_{ub}|$ and $|V_{cb}|$ and first four moments of the lepton mass squared (for $X_c$). The question of the lepton flavor universality in semileptonic decays to a general hadronic state $X$ was addressed in \cite{Belle-II:2023qyd}.

Other results from different experiments could be cited in the domain of semileptonic $B$ decays, yet the measurements of the above-mentioned B-factories represent the most important data from both, the quantity and quality perspective.

\vspace{0.5cm}

The large number of theoretical works implies strong selection criteria which we base on the impact of the work with some preference for review and pedagogical texts. We have already mentioned nice reviews   \cite{BaBar:2014omp, Buchalla:1995vs, Dingfelder:2016twb, Altmannshofer:2021qrr, Bernlochner:2021vlv} which cover (also) the semileptonic $B$ decays. Further survey papers are \cite{Altmannshofer:2008dz}, where the SM theory and appropriate observables are presented, a pedagogically-written article \cite{Calibbi:2017uvl}, which focuses on the charged lepton flavour violation and also a  generally-oriented texts \cite{Blake:2016olu, Bifani:2018zmi}. One can in addition mention \cite{Cornella:2021sby}, in which $B$ flavor anomalies are discussed and also similarly oriented recent text \cite{London:2021lfn}.

Reliable SM predictions are the starting point for assessing various anomalies. Already decades ago a quark potential model was used to make predictions for semileptonic $B$ and $D$ decays \cite{Isgur:1988gb} with an update several years later \cite{Scora:1995ty}. Decays to $D^{(*)}$ mesons were addressed in \cite{Boyd:1995sq}, the analyticity and dispersion relations were used to produce parametrizations of the QCD form factors with small model dependence. The same authors later published QCD two-loop level computations \cite{Boyd:1997kz} including lepton mass effect, higher resonances and heavy quark symmetry, which further improved the theoretical precision. The heavy quark spin symmetry was used in \cite{Caprini:1997mu} to derive dispersive constraints on $B \to D^{(*)}$ form factors and implications for the determination of $|V_{cb}|$. Semileptonic decays to light mesons $\rho$, $\omega$, $K^{*}$ and $\phi$ were discussed in \cite{Ball:2004rg} in the framework of light-cone sum rules, the authors claim  $10\%$ precision at zero momentum transfer. The angular analysis of the process $\bar{B} \to \bar{K} \ell^+\ell^-$ was presented in \cite{Bobeth:2007dw}. The work is based on the QCD factorization and large recoil symmetry relations and besides angular coefficients it also gives a prediction of $R_K$ and explores the potential of the introduced observables to reach the NP. Taking into the consideration also the excited state $K^*$, the publication \cite{Khodjamirian:2010vf} is dedicated to the charm-loop effect. The results are derived using QCD light-cone sum rules and hadronic dispersion relations and the evaluated charm loop effect, which is claimed to reach up to 20\% , is represented as a contribution to the $C_9$ Wilson coefficient. Lattice QCD was used in \cite{FermilabLattice:2014ysv, Na:2015kha, Harrison:2017fmw} to predict form factors and matrix elements for processes with $D^{(*)}$ mesons. In \cite{Du:2015tda} were the lattice form factors used as input and allowed to determine CKM matrix elements, or, alternatively, constrain the real part of the Wilson coefficients $C_9$ and $C_{10}$. The CKM matrix was also the subject of the work \cite{Alberti:2014yda}, where $|V_{cb}|$ was extracted using the OPE, the expansion in powers of the heavy quark mass and constraints derived from the experimental values on the normalized lepton energy moments. A process with a vector meson particle production $B\to V\ell^+\ell^-$ was considered in \cite{Bharucha:2015bzk} where the authors used light-cone sum rules to predict form factors. The paper \cite{Bigi:2016mdz} has a somewhat review character, it present three common form factor parameterizations, summarizes the  data and the available lattice information (as of 2016) and gives a special emphasis on the unitarity constraints. Then it presents fits to experimental points and to the lattice numbers from which the results on $R_D$ and $|V_{cb}|$ are extracted. Radiative corrections to the $R_{K^{(*)}}$ observables are of a concern to the authors of \cite{Bordone:2016gaq}, their thorough analysis indicates that these observables are indeed well suited to be a probe of NP.
This work \cite{Bordone:2016gaq} was improved in \cite{Isidori:2022bzw} where a full Monte Carlo framework was built to describe QED corrections in $\bar{B} \to \bar{K} \ell^+ \ell^-$. A detailed numerical comparison with those obtained with the general-purpose photon-shower tool PHOTOS has been performed. The charmonium leading logs were fully simulated.
Similar questions related to the same observables are addressed in \cite{Jager:2014rwa}. Still the same observables are, together with the angular observables $P_i$, discussed in a pedagogical way in \cite{Capdevila:2017ert} with special emphasis on the hadronic uncertainties. Coming back to $D$ particles and works published within few years after the first measurements indicating a possible lepton-flavor violation, one can mention \cite{Bigi:2017jbd}, where the coefficients of the Boyd-Grinstein-Lebed form factor parametrization were constrained by analyzing the form factor ratios and their uncertainties in the heavy quark limit. With this knowledge fits to experimental data were performed and $R_{D^*}$ computed. In \cite{Jaiswal:2017rve} two different form factors parameterizations are used to predict $R_{D^*}$ and $|V_{cb}|$. The approach uses, besides data, inputs from the light cone sum rules and lattice and the relations between form factors as given by HQET. To mention more recent theoretical works, one can point to e.g. \cite{Isidori:2020acz, Gubernari:2020eft}, where QED corrections and non-local matrix elements are discussed for $B$ decays to dilepton and a kaon.
It was shown in Ref. \cite{Isidori:2020acz} that there cannot be any hard-collinear logs at the structure dependent level. This is important as in some parts of phase space they are 10-20\% for scalar QED as used by PHOTOS.
The status of the $b \to c \tau \nu$ anomalies as of 2022 is summarized  in \cite{Iguro:2022yzr}, where the models for global fits are based mostly on the HQET and lattice results. The latter are also reviewed the Sec. 8 of \cite{FlavourLatticeAveragingGroupFLAG:2021npn}.

The number of NP papers progressively grew as the evidence for tensions and anomalies became more and more convincing, with the first hints appearing at the beginning of the new millennium. Often, the NP is theoretically addressed by non-SM operators appearing in the effective Hamiltonian. So was done in \cite{Hiller:2003js}, where the approach was applied to the $b \to s$ process. No strong claims were given there, but it was shown that the evaluated NP effects can reach up to 13\% for $R_{K^*}$. The same effective-operator approach was applied in \cite{Fajfer:2012vx} to $b \to c$ transition and the impact of the NP to $B \to D^* \tau \bar \nu_{\tau}$ observables was evaluated. The authors demonstrated that it is significant, i.e. the sensitivity of the process is high enough for the NP to be detected. Effective operators were used also in \cite{Hiller:2014yaa}, where, after the NP operator contributions were discussed, two leptoquark models were proposed to explain two out of three possible scenarios which lead to the observed $R_K$ value. Leptoquarks (vector and scalar, respectively) are also considered in \cite{Fajfer:2015ycq, Crivellin:2017zlb}, both works claim that their theory allows to simultaneously resolve discrepancies appearing in $b \to s$ and $b \to c$ transitions. Still leptoquarks, the authors of \cite{Angelescu:2021lln} investigate single leptoquark extensions of the SM with $ 1\,\text{TeV} \lesssim m_{LQ} \lesssim 2 \, \text{TeV}$ with conclusion that no such scalar leptoquark can be, a vector particle is the only option. The work \cite{Abada:2014kba} uses scenarios with light right-handed neutrinos appearing in low-scale seesaw models as the NP framework for analyzing the lepton flavor violation. Among other results the authors propose observables, i.e. properly chosen branching fraction ratios, which could discriminate between supersymmetric (SUSY) and non-SUSY NP realizations. Further works which analyze the $R_K$ and $R_{K^*}$ anomalies are \cite{Gripaios:2014tna} and \cite{Crivellin:2015mga}, the former assumes a composite Higgs model, the latter uses a two-Higgs-doublet model. At last, let us mention a set of more generally-oriented works \cite{Descotes-Genon:2015uva, Altmannshofer:2017yso, Alguero:2019ptt, Hurth:2021nsi, Geng:2021nhg, Alguero:2021anc} which focus mainly on $b \to s \ell^+ \ell^-$ and which aim to provide model-independent or theoretically clean conclusions. By different approaches they investigate the space for NP parameters and most of them presents arguments in favor of some NP scenario.

\subsection{Semileptonic and radiative decays $B_s \to \phi \ell^+ \ell^-$ and  $B_s\to\phi\gamma$ in CCQM \label{SecCCQMsemi}}

The $B_s \to \phi \ell^+ \ell^-$ and $B_s\to\phi\gamma$ decays were within the CCQM analyzed in \cite{Dubnicka:2016nyy}.  The analysis was done in the light of the LHCb measurements \cite{LHCb:2013tgx, LHCb:2015wdu}, where the second one was recent at that time. The measurement focused on angular observabes and the branching fraction distribution and reported on a deviation from the SM in the latter exceeding $3 \sigma$ for $1 \text{ GeV}^2 \le q^2 \le 6\text{ GeV}^2$. Several years later two new measurements were performed. The work \cite{LHCb:2021xxq} addressed the angular distribution where no significant tensions with the SM were observed, \cite{LHCb:2021zwz} however confirmed the discrepancy from the previous branching fraction measurement. One may put this observation in relation with $R_K$ and $R_{K^*}$ anomalies, which also happen for the $b \to s$ transition, from where the motivation to study this process in more details.

\begin{figure}[t]
\begin{center}
\includegraphics[width= 8 cm]{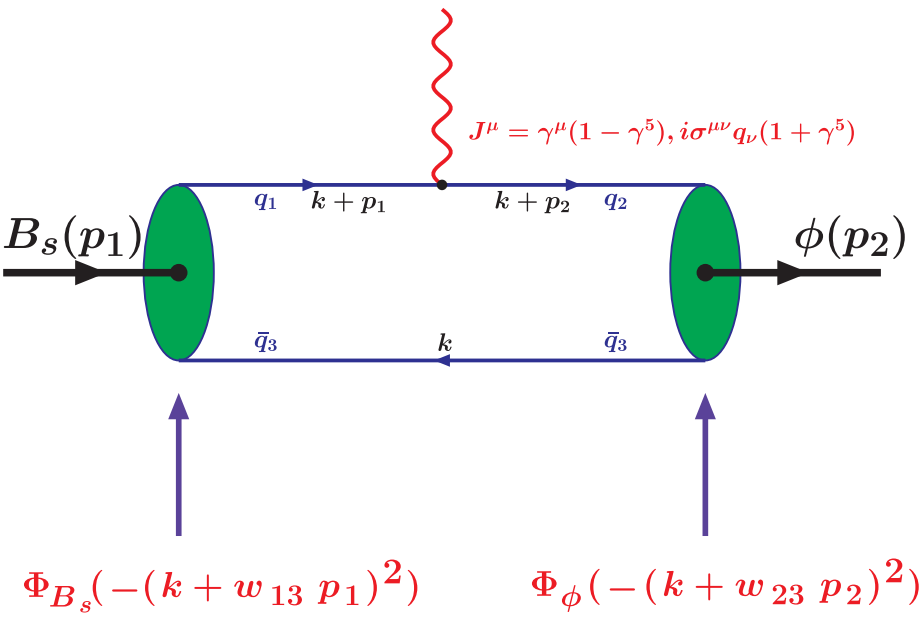}
\caption{$B_s \to \phi$ transition in the CCQM. Figure was originally published in \cite{Dubnicka:2016nyy}.}
\label{Fig_semiLepto}
\end{center}
\end{figure}

\begin{figure}[t]
\begin{center}
\includegraphics[width=0.47\textwidth]{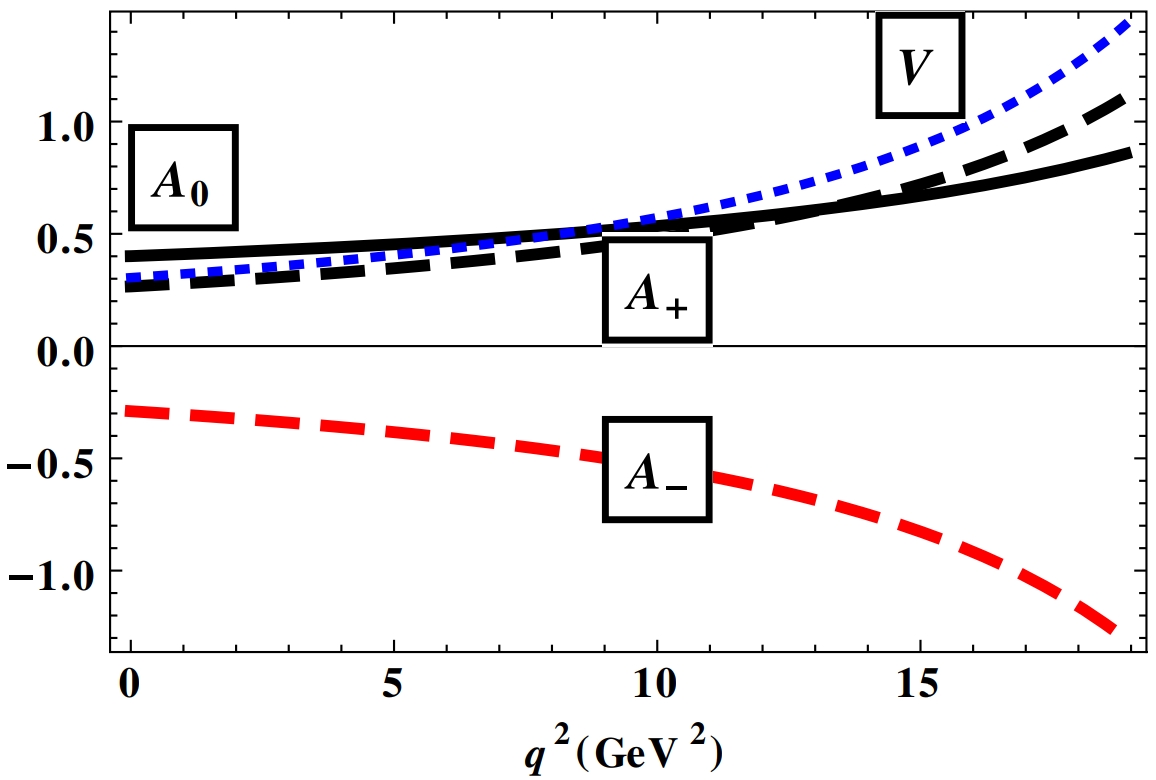}
\hspace{0.5cm}
\includegraphics[width=0.47\textwidth]{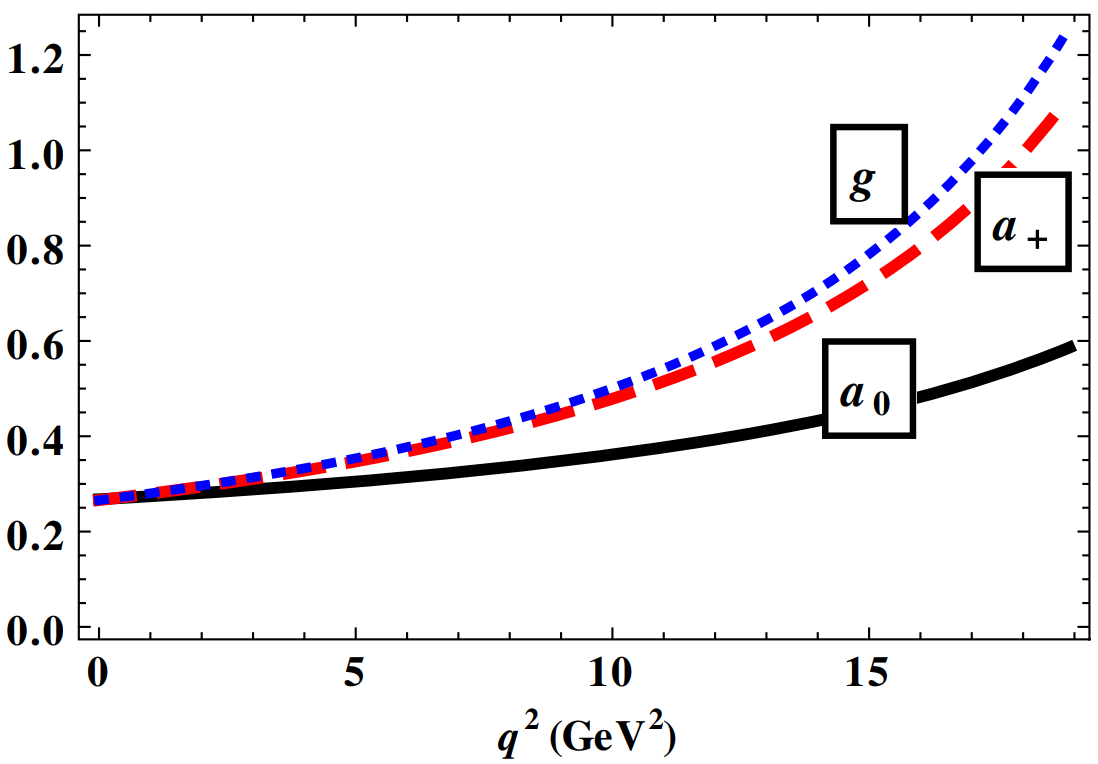}
\caption{Vector and tensor form factors for the $B_s \to \phi$ transition as predicted by the CCQM. Figures were originally published in \cite{Dubnicka:2016nyy}.}
\label{semiFF}
\end{center}
\end{figure}

In \cite{Dubnicka:2016nyy} one analyzes both, the angular coefficients and the differential decay rate distribution. In addition to (\ref{par_values}), the necessary model inputs are
\begin{align}
\Lambda_{B_s} = 2.05  \text{ GeV} \quad  \text{and} \quad \Lambda_\phi = 0.88 \text{ GeV}
\end{align}
determined in prior works. The transition is expressed through two matrix elements
\begin{align}
M_1^\mu = <\phi(p_2, \epsilon)|\bar{s} O^\mu b| B_s(p_1))> ,
\quad
M_2^\mu = <\phi(p_2, \epsilon)|\bar{s} [\sigma^{\mu\nu}q_\nu(1+\gamma^5)] b| B_s(p_1))>,
\end{align}
where $O^\mu = \gamma^\mu(1-\gamma^5)$ and $p_i$ are momenta with $q=p_1-p_2$ and $P=p_1+p_2$. The appearing variables satisfy $p_1^2 = m_{B_s}^2 \equiv m_1^2$, $p_2^2 = m_{\phi}^2 \equiv m_2^2$ and $\epsilon_2^\dagger \cdot p_2 = 0$. In total seven invariant form factors, defined as coefficient functions in front of the Lorentz structures, are necessary to parameterize them
\begin{align}
&M_1^\mu = \frac{\epsilon_\nu^\dagger}{m_1+m_2}
\left[-g^{\mu\nu}P \cdot q A_0(q^2) + P^\mu P^\nu A_+(q^2) 
+ q^\mu P^\nu A_-(q^2) + i\varepsilon^{\mu \nu \alpha \beta} P_\alpha q_\beta V(q^2) \right], \label{M12_vec} \\
&M_2^\mu = \epsilon_\nu^\dagger \left[ -\left(g^{\mu \nu}-\frac{q^\mu q^\nu}{q^2}\right)P \cdot q \, a_0(q^2)
+\left( P^\mu P^\nu - q^\mu P^\nu\frac{P \cdot q}{q^2} \right)a_+(q^2)
+ i \varepsilon^{\mu \nu \alpha \beta } P_\alpha q_\beta \, g(q^2) \right]. \label{M12_ten}
\end{align}
The same amplitudes can be expressed in the CCQM
\begin{align}
M_{1,2}^\mu &=  N_c g_{B_s} g_\phi \int \frac{d^4k}{i(2\pi)^4}
\tilde{\Phi}_{B_s}(-[k+w_{13}p_1]^2) \tilde{\Phi}_\phi(-[k+w_{23}p_2]^2) \times T_{1,2}, \label{M12_CCQM}\\
T_1 &= \text{tr}[O^\mu S_b(k_1+p_1) \gamma^5 S_s(k) \cancel{\epsilon}_2^\dagger S_s(k+p_2)], \label{M12_CCQMa} \\
T_2 &= \text{tr}[\sigma^{\mu \nu} q_\nu (1+\gamma^5)  S_b(k_1+p_1) \gamma^5 S_s(k) \cancel{\epsilon}_2^\dagger S_s(k+p_2)], \label{M12_CCQMb}
\end{align} 
with $S_i$ being quark propagators and $N_c$ the number of colors. The origin of various terms in (\ref{M12_CCQM})-(\ref{M12_CCQMb}) is schematically represented in Fig. \ref{Fig_semiLepto}. Once the model expression (\ref{M12_CCQM}) is evaluated to the level of invariant Lorentz structures, it can be compared to (\ref{M12_vec}) and (\ref{M12_ten}) and form factor expressions read out. Their behavior is shown in Fig. \ref{semiFF}, it determines the necessary model input and completes the model-dependent part of the calculation. 

Let us briefly review also the remaining steps to reach observable quantities. The set of the SM four-fermion operators is written as
\begin{align}
&\mathcal{O}_1 = (\bar{s}_{a_1} \gamma^\mu P_L c_{a_2})(\bar{c}_{a_2} \gamma_\mu P_L b_{a_1}),
&&\mathcal{O}_2 = (\bar{s} \gamma^\mu P_L c)(\bar{c} \gamma_\mu P_L b),
\nonumber \\
&\mathcal{O}_3 = (\bar{s} \gamma^\mu P_L b) \sum_q (\bar{q} \gamma_\mu P_L q),
&&\mathcal{O}_4 = (\bar{s}_{a_1} \gamma^\mu P_L b_{a_2}) \sum_q (\bar{q}_{a_2} \gamma_\mu P_L q_{a_1}),
\nonumber  \\
&\mathcal{O}_5 = (\bar{s} \gamma^\mu P_L b) \sum_q (\bar{q} \gamma_\mu P_R q),
&&\mathcal{O}_6 = (\bar{s}_{a_1} \gamma^\mu P_L b_{a_2}) \sum_q (\bar{q}_{a_2} \gamma_\mu P_R q_{a_1}),
\\
&\mathcal{O}_7 = \frac{e}{8 \pi^2} \tilde{m}_b (\bar{s} \sigma^{\mu \nu} P_R b)F_{\mu \nu},
&&\mathcal{O}_8 = \frac{g_s}{8 \pi^2} \tilde{m}_b (\bar{s}_{a_1} \sigma^{\mu \nu} P_R \textbf{T}_{a_1 a_2} b_{a_2})\textbf{G}_{\mu \nu},
\nonumber
\\
&\mathcal{O}_9 = \frac{e^2}{8 \pi^2} (\bar{s} \gamma^\mu P_L b)(\bar{\ell} \gamma_\mu \ell),
&&\mathcal{O}_{10} = \frac{e^2}{8 \pi^2} (\bar{s} \gamma^\mu P_L b)(\bar{\ell} \gamma_\mu \gamma_5 \ell),
\nonumber
\end{align}
where $P_{L,R} = (1 \mp \gamma^5)$, $a_i$ are color indices (implicit for color singlet currents), $\textbf{T}_{a_1 a_2}$ are generators of the $SU(3)$ color group, $\textbf{G}_{\mu \nu}$ is the gluonic field strength and $g_s$ is the QCD coupling (other symbols have meaning as defined before). Operators $\mathcal{O}_1$ and  $\mathcal{O}_2$ are referred to as current-current operators, $\mathcal{O}_3-\mathcal{O}_6$ are QCD penguin operators,  $\mathcal{O}_{7,8}$ are so-called magnetic penguin operators and $\mathcal{O}_8$ and $\mathcal{O}_9$ operators correspond to semileptonic electroweak penguin diagrams. The transition amplitude takes the form
\begin{align}
\mathcal{M} = \frac{G_F}{2\sqrt{2}} \frac{\alpha |V_{tb}V^*_{ts}|}{\pi}
\bigg[&
C_9^{\text{eff}} \langle \phi | \bar{s} \gamma^\mu P_L b| B_s \rangle (\bar{\ell} \gamma_\mu \ell)
-\frac{2 \tilde{m}_b }{q^2} C_7^{\text{eff}}
\langle \phi | \bar{s} i \sigma^{\mu \nu} q_\nu P_R b | B_s \rangle (\bar{\ell} \gamma_\mu \ell)
\nonumber \\
& + C_{10} \langle \phi | \bar{s} \gamma^\mu P_L b | B_s \rangle (\bar{\ell} \gamma_\mu \gamma_5 \ell)
\bigg].
\end{align}
The Wilson coefficients $C_1-C_6$ are absorbed into the effective coefficients $C_7^\text{eff}$ and $C_9^\text{eff}$, $C_7^\text{eff} = C^7-C_5/3-C_6$ and $C_9^\text{eff}$ is defined by (\ref{def:Ceff})(\ref{eq:VMD}), where, again, the $\bar{c}c$ resonances appear in the Breit-Wigner form and one drops them by setting $\kappa = 0$. The renormalization scale is set to $\mu = \bar{m}_{b,\, \text{pole}}$. Numerical values of Wilson coefficients were taken from \cite{Descotes-Genon:2013vna}, as we described it already in Sec. \ref{Sec:leptoCCQM}. Also the QCD quark masses are the same as in the leptonic-decay section. In addition to the charm loop contribution, one takes into the consideration the two loop effects as computed in \cite{Asatryan:2001zw, Greub:2008cy}. They modify the effective coefficients
\begin{align}
C_7^{\text{eff}} \to C_7^{\text{eff}}-\frac{\alpha_s}{4 \pi}(C_1 F_1^{(7)}+C_2 F_2^{(7)}),
\quad
C_9^{\text{eff}} \to C_9^{\text{eff}}-\frac{\alpha_s}{4 \pi}(C_1 F_1^{(9)}+C_2 F_2^{(9)}),
\end{align}
where the functions $F_{1,2}^{(7,9)}$ were made publicly available by authors of \cite{Greub:2008cy} as \emph{Wolfram Mathematica} code.

\begin{figure}[t]
\begin{center}
\includegraphics[width=0.47\textwidth]{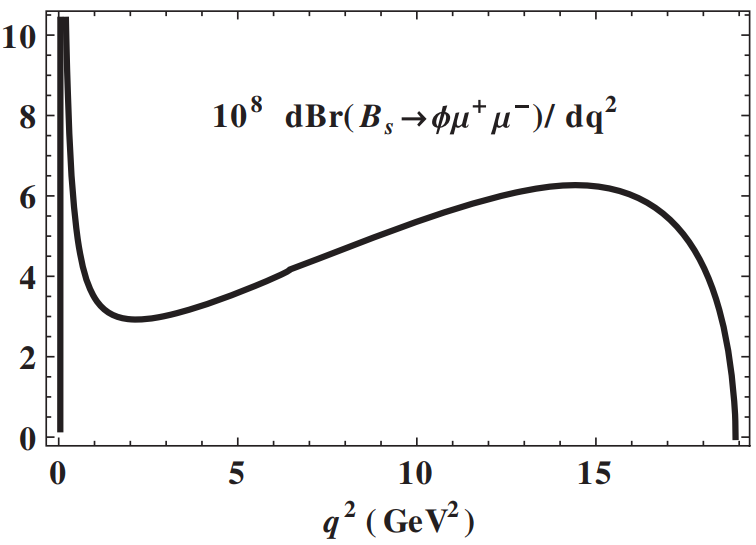}
\hspace{0.5cm}
\includegraphics[width=0.47\textwidth]{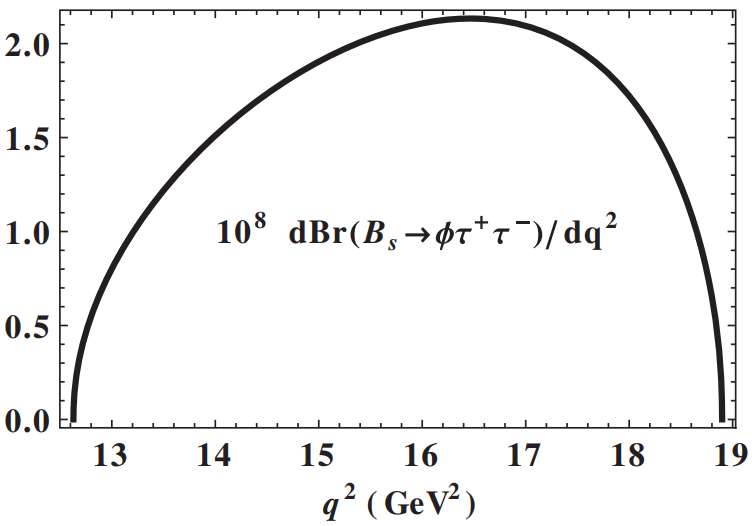}
\\
\includegraphics[width=0.47\textwidth]{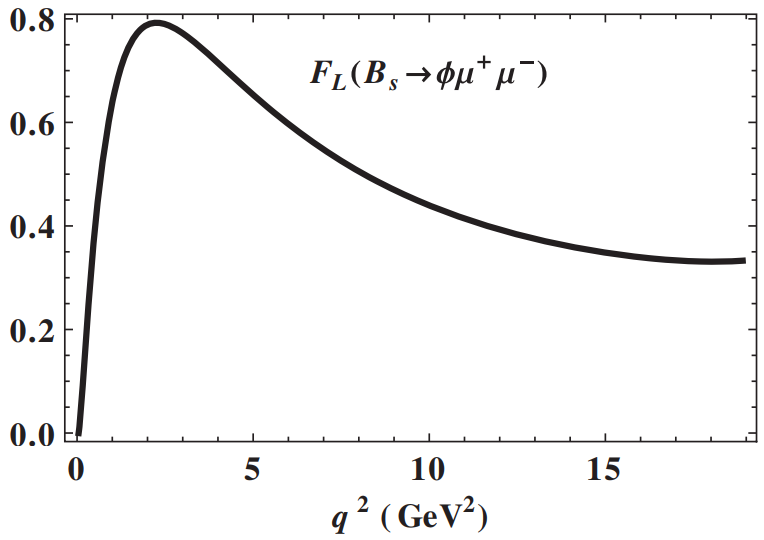}
\hspace{0.5cm}
\includegraphics[width=0.47\textwidth]{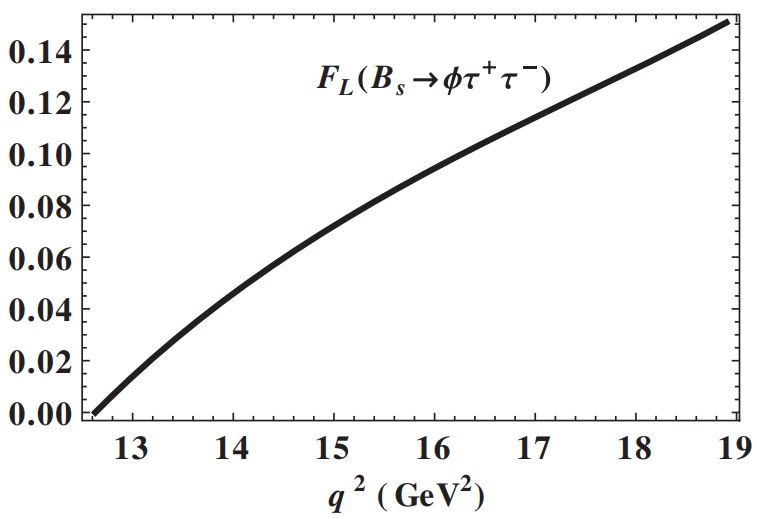}
\\
\includegraphics[width=0.47\textwidth]{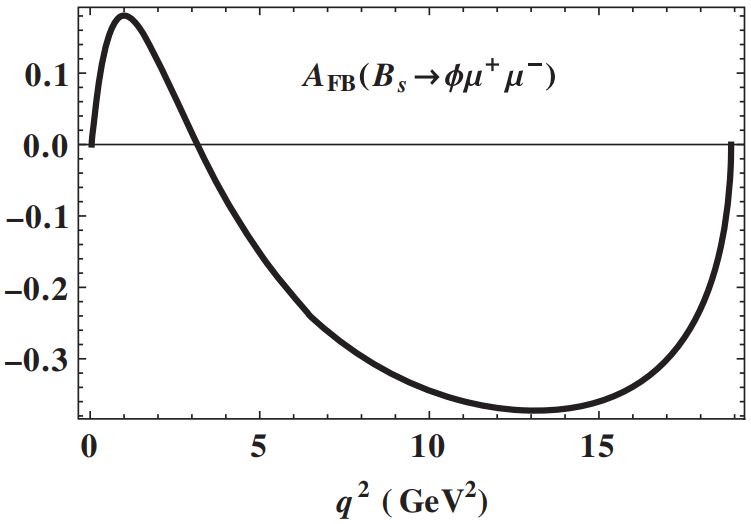}
\hspace{0.5cm}
\includegraphics[width=0.47\textwidth]{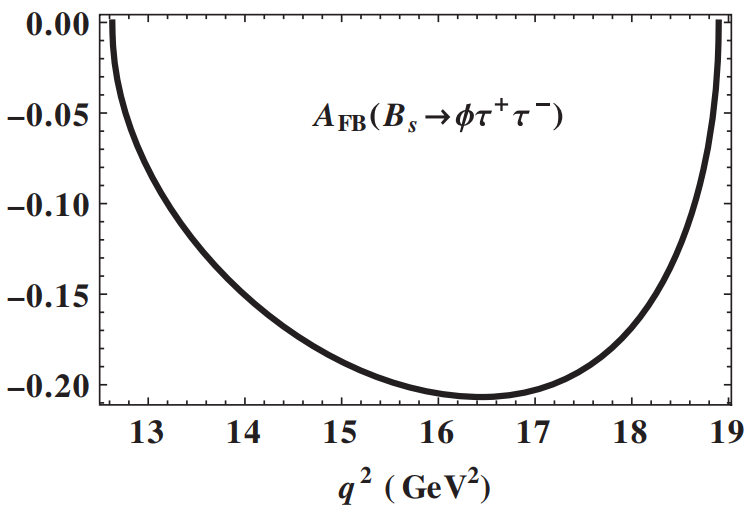}
\caption{Branching fraction, $F_L$ and $A_\text{FB}$  as function of $q^2$ for $\mu$ and $\tau$ in the final state. Figures were originally published in \cite{Dubnicka:2016nyy}.}
\label{semiObs}
\end{center}
\end{figure}

The differential decay rate is then expressed as
\begin{align}
&\frac{d\Gamma(B_s \to \phi \ell \ell)}{dq^2} = \frac{G_F^2}{(2 \pi)^3}
\left( \frac{\alpha |V_{tb}V^*_{ts}|}{2 \pi} \right)^2
\frac{|\mathbf{p_{2}}| q^2 \beta_\ell}{12 m_1^2} \mathcal{H}_\text{tot},
\\
&\mathcal{H}_\text{tot} =
\frac{1}{2} \left( \mathcal{H}_U^{11}+\mathcal{H}_U^{22}+\mathcal{H}_L^{11}+\mathcal{H}_L^{22} \right)
+ \delta_{\ell \ell} \left(
\frac{\mathcal{H}_U^{11}}{2}-\mathcal{H}_U^{22}+\frac{\mathcal{H}_L^{11}}{2}-\mathcal{H}_L^{22}+\frac{3\mathcal{H}_S^{22}}{2}
\right),
\end{align}
where $\delta_{\ell \ell} = 2m_\ell^2/q^2$,
$\beta_\ell = \sqrt{1-2\delta_{\ell \ell}}$
and
$\left|\mathbf{p_{2}}\right|=\sqrt{\lambda^{\text{Källén}}\left(m_{1}^{2},m_{2}^{2},q^{2}\right)}/(2m_{1})$ is the momentum of the $\phi$ meson in the $B_s$ rest frame. The objects $\mathcal{H}_X^{ii}$ represent bilinear combinations of the helicity amplitudes
\begin{align}
\mathcal{H}_U^{ii}=|H^i_{++}|^2+|H^i_{--}|^2, \quad 
\mathcal{H}_L^{ii}=|H^i_{00}|^2, \quad
\mathcal{H}_S^{ii}=|H^i_{t0}|^2,
\end{align}
which are related to the invariant form factors through intermediate functions $A_{+,-,0}^i$ and $V^i$
\begin{align}
&H^i_{t0} = \frac{1}{m_1+m_2} \frac{m_1 |\mathbf{p_{2}}| }{m_2 \sqrt{q^2}}
\lbrace Pq(-A_0^i+A_+^i) + q^2 A_-^i \rbrace, \\
&H^i_{\pm \pm} = \frac{1}{m_1+m_2}( -PqA_0^i \pm 2m_1 |\mathbf{p_{2}}| V^i), \\
&H^i_{00} = \frac{1}{m_1+m_2} \frac{1}{2 m_2 \sqrt{q^2}}
\lbrace  -Pq(m_1^2-m_2^2-q^2)A_0^i + 4 m_1^2 |\mathbf{p_{2}}|^2 A_+^i \rbrace,
\end{align}
with
\begin{align}
&V^1 = C_9^\text{eff} V + C_7^\text{eff} \chi \, g  ,
&&V^2 = C_{10} V,
\\
&A^1_+ = C_9^\text{eff} A_+ + C_7^\text{eff} \chi \, a_+, 
&&A_\pm^2 = C_{10}A_\pm
\\
&A^1_- = C_9^\text{eff} A_- + C_7^\text{eff} \chi Pq \, (a_0 - a_+)/q^2,
&&A^1_0 = C_9^\text{eff} A_0 + C_7^\text{eff} \chi \, a_0,
\\
&A_0^2 = C_{10} A_0, \hspace{5cm} \text{where}
&&\chi = 2 \tilde{m}_b(m_1+m_2)/q^2.
\end{align}
The full description of the $B_s \to \phi \ell \ell$ decay requires, besides the $q^2$, three additional angles, see for example Eq. (2.1) in \cite{ATLAS:2018gqc}, where completely analogous formula is written for fully differential decay rate of $B_d \to K^* \mu^+ \mu^-$. The advantage of the helicity formalism is that the angular observables, i.e. the coefficients in front of various angular terms, have simple expressions. For the longitudinal polarization fraction $F_L$ and the forward-backward asymmetry $A_\text{FB}$ they stand
\begin{align}
&F_L = \frac{1}{2} \beta_\ell^2 \frac{\mathcal{H}_L^{11}+\mathcal{H}_L^{22}}{\mathcal{H}_\text{tot}},
\qquad 
A_\text{FB} = - \frac{3}{4} \beta_\ell \frac{\mathcal{H}_P^{12}}{\mathcal{H}_\text{tot}},
\\
&\text{where } \mathcal{H}_P^{12} =
\text{Re} \left[ H^1_{++} (H^2_{++})^\dagger \right] -
\text{Re} \left[ H^1_{--} (H^2_{--})^\dagger \right].
\end{align}
The CCQM-predicted behavior of the branching fraction and of the two angular observables $F_L$ and $A_\text{FB}$  is, as function of $q^2$, show in Fig. \ref{semiObs}.
The $q^2$-averaged numbers were computed for $F_L$, $A_\text{FB}$, additional angular observables $S_3$, $S_4$ and also for optimized observables $P_1$ and $P_4^{'}$ which are derived from them, $P_1 = 2S_3/(1-F_L)$, $P_4^{'} = S_4/\sqrt{F_L(1-F_L)}$. The results are presented in Tab. \ref{Tab:semiAvgAngular}.
\begin{table}
\centering
\begin{tabular}[t]{c  c  c  c  c}
\hline
\hline
 & $B_s \to \phi \mu^+ \mu^-$ & $B_s \to \phi \tau^+ \tau^-$ & $B_s \to \phi \nu \bar{\nu}$ \\
\hline 
$\mathcal{B}_{tot}$ & $(9.11 \pm 1.82) \times 10^{-7} $ & $(1.03 \pm 0.20) \times 10^{-7} $ & $(0.84 \pm 0.16) \times 10^{-5}$ \\
$\langle A_\text{FB}\rangle $ & $-0.24 \pm 0.05$ & $-0.18 \pm 0.04$ & $\cdot$ \\
$\langle F_L \rangle$& $0.45 \pm 0.09$ & $0.09 \pm 0.02$ & $\cdot$ \\
$\langle P_1 \rangle$& $-0.52 \pm 0.1$ & $-0.76 \pm 0.15$ & $\cdot$ \\
$\langle P_4^{'} \rangle$& $1.05 \pm 0.21$ & $1.33 \pm 0.27$ & $\cdot$ \\
$\langle S_3 \rangle$& $-0.14 \pm 0.03$ & $-0.067 \pm 0.013$ & $\cdot$ \\
$\langle S_4 \rangle$& $0.26 \pm 0.05$ & $0.083 \pm 0.017$ & $\cdot$ \\
\hline
\hline
\end{tabular}
\caption{Total branching fractions and averaged angular observables of selected decay channels for the whole kinematic region. Table contains data originally published in \cite{Dubnicka:2016nyy}.} 
\label{Tab:semiAvgAngular}
\end{table}
The table shows the branching fraction also for  $B_s \to \phi \nu \bar{\nu}$, the corresponding decay formula is indicated in Eqs. (34)-(36) of \cite{Dubnicka:2016nyy}. The text \cite{Dubnicka:2016nyy} also contains predictions for the radiative decay to $\phi \gamma$ and non-leptonic decay to $\phi J/\Psi$ (formulas (38) and (37) there)
\begin{align}
\mathcal{B}(B_s \to \phi \gamma) = (2.39 \pm 0.48) \times 10^{-5},
\quad
\mathcal{B}(B_s \to \phi J/\Psi) = (1.6 \pm 0.3) \times 10^{-3}.
\end{align} The results can be compared to the  actual experimental numbers \cite{Workman:2022ynf}.
\begin{align}
&\mathcal{B}(B_s \to \phi \mu^+ \mu^-) = (8.4 \pm 0.4) \times 10^{-7},
\quad
\mathcal{B}(B_s \to \phi \nu \bar{\nu}) < 540 \times 10^{-5},
\\
&\mathcal{B}(B_s \to \phi \gamma) = (3.4 \pm 0.4) \times 10^{-5},
\quad
\mathcal{B}(B_s \to \phi J/\Psi) = (1.04 \pm 0.04) \times 10^{-3}.
\end{align}
The branching fraction to $\phi \mu^+ \mu^-$ is in good agreement with the SM, in fact the experimental numbers  measured after the publication moved closer to the  published CCQM value. The same is also true for the two non-leptonic decay channels, yet, here a  discrepancy of the order of $2 \, \sigma$ remains.

Coming back to the semileptonic decays, a detailed interval values were presented in Tab. VI of \cite{Dubnicka:2016nyy} for $B_s \to \phi \mu^+ \mu^-$. They mimic the way the experimental measurements are done and they are of the interest because the largest discrepancy observed by \cite{LHCb:2015wdu, LHCb:2021zwz} is the branching fraction on the $q^2$ interval\footnote{In \cite{LHCb:2021zwz,LHCb:2021xxq} the lower interval limit is $1.1 \, \text{GeV}^2$. This effect is considered as negligible because the measured quantities are intensive (not additive), e.g. the branching fraction measurement is $q^2$-averaged (the number of entries in the interval is divided by the integral length).} $1-6 \, \text{GeV}^2$. Also, the table presents the effect of the two-loop contributions by giving the numbers with and without them. We do not reproduce here all of them but focus only on the interval $1 \, \text{GeV}^2 \le q^2 \le 6 \, \text{GeV}^2$ and observables measured on this interval, see Tab. \ref{Tab:semiOnInterval}.
\begin{table}
\centering
\begin{tabular}[t]{c  c  c  c }
\hline
\hline
 & $CCQM, \text{ 2 loop}$ & $CCQM, \text{ 1 loop}$ & Experiment \cite{LHCb:2021zwz,LHCb:2021xxq} (\cite{LHCb:2015wdu})\\
\hline 
$10^7 \mathcal{B}_\text{tot.}$ & $1.56 \pm 0.31$ & $1.64 \pm 0.33$ & $1.41  \pm 0.11 \quad(1.29)$ \\
$F_L$ & $0.69 \pm 0.14$ & $0.71 \pm 0.14$ & $0.715  \pm 0.036 \quad (0.63) $\\
$S_3$ & $-0.034 \pm 0.007$ & $-0.039 \pm 0.008$ & $-0.083  \pm 0.047 \quad (-0.02)$ \\
$S_4$ & $0.17 \pm 0.03$ & $0.19 \pm 0.04$ & $0.155  \pm 0.058 \quad (0.19)$ \\
$S_7$ & $0.0065 \pm 0.0013$ & $0$ & $0.020  \pm 0.059 \quad (-0.03)$ \\
\hline
\hline
\end{tabular}
\caption{Branching fraction and selected angular observables on the interval $1 \, \text{GeV}^2 \le q^2 \le 6 \,\text{GeV}^2$ for $B_s \to \phi \mu^+ \mu^-$. Indicated are the CCQM predictions with and without 2-loop contributions and the experimental value. Table contains a subset of data originally published in \cite{Dubnicka:2016nyy}.} 
\label{Tab:semiOnInterval}
\end{table}
In the table also older measurements are indicated in brackets and one sees that for all indicated observables except $S_3$ the new measurement bring the experimental value closer to the theoretical one. The large error of the $S_3$ measurement implies that both CCQM predictions (1-loop and 2-loop) do not much exceed $1 \, \sigma$ deviation. Considering the 2-loop results one observes that no significant deviations from the experiment are observed, especially in the branching fraction case they bring the value closer to the measurement (w.r.t. one-loop calculations).

As summary we can conclude that the interesting decay channel $B_s \to \phi \ell^+ \ell^-$ was addressed in the framework of the CCQM. Already at the time of the publication the comparison with the LHCb numbers did not allow us to claim NP presence, the major discrepancy in the branching fraction on the $1-6\, \text{GeV}^2$ interval was reduced significantly by the CCQM prediction. This was true also for other discrepancies ($F_L$, $S_4$) seen on other intervals. The new data further decreased the branching fraction discrepancy and with results of the CCQM one cannot talk about a discrepancy any longer.


\subsection{Other CCQM results on semileptonic $B$ decays.}
Quite a few papers were dedicated to the study of semileptonic $B$ decays in the framework of the CCQM. We will not include into the overview older texts, where an earlier version of the model was used \cite{Ivanov:1990tw,Ivanov:1992wx, Ivanov:1999ic, Ivanov:1999tv, Ivanov:2000aj, Faessler:2002ut, Ivanov:2002at, Ivanov:2005fd, Ivanov:2006ni}.

The first text we mention \cite{Branz:2009cd} was already cited several times here. It is a generally oriented text focusing mostly on the model itself and presenting its various aspects, including, for the first time, also the infrared confinement of quarks. A global fit on basic experimental quantities, such as weak leptonic decay constants, was performed in order to determine universal and hadron-specific model parameters. These parameters were used in the same text to predict weak leptonic decay constants (including for $B$ mesons) and Dalitz dacays of several light mesons. The results were encouraging, most of predictions were in a quite good agreement with measured data.

The paper \cite{Ivanov:2011aa} is dedicated to various $B_{(s)}$ decays with, however, emphasis on the  nonleptonic processes. In the first part of the text the global fits are refined and the model parameters are updated. Then, the semileptonic decays are addressed, but only in the context of the universal transition form factors to several final-state mesons (pseudoscalar and vector). The results on form factors are given in form of plots and the comparison with seven other authors based on the value at $q^2 = 0$ is shown in Tab. III.  

Somewhat similar treatment of the semileptonic decays is given in \cite{Dineykhan:2012cp}. Here again the emphasis is on exotic and nonleptonic decays. The semileptonic decays are addressed in the context of transition form factors, similarly to the previous text.

The publication \cite{Issadykov:2015iba} focuses on the semileptonic decays of $B_{(s)}$ to scalar mesons with light masses (below 1 GeV) in the context of the $B \to K^* (\to K \pi) \mu^+ \mu^-$ decay. The CCQM form factors $F_\pm$ and $F_T$ are predicted for the range $0.8 \, \text{GeV} \le \Lambda_S \le 1.5 \, \text{GeV} $ of scalar vector model parameters for the $b \to u$, $b \to d$ and $b \to s$ transitions. The predictions are approximated for $\Lambda_S = 0.8 \, \text{GeV}$ and $\Lambda_S = 1.5 \, \text{GeV}$ by a simplified parameterization which depends on three numbers. They are given in Tab. II of the text, so as to make the results available to other authors. Branching fractions ($\Lambda_S = 1.5 \, \text{GeV}$) for various semileptonic decays $B_{(s)} \to S \ell \ell$, $B_{(s)} \to S \ell \nu_\ell$ are shown in Tab. IV of the work. The text then briefly discusses the role of the scalar $K^*_0(800)$ particle in the cascade decay of the $B$ meson pointing out the fact that the narrow-width approximation is not appropriate and estimating the $S$-wave pollution in the $B \to K^* \ell^+ \ell^-$ decay to 6\%.

The leptonic and semileptonic processes $B \to \ell \bar{\nu}$ and $B \to D^{(*)} \ell^- \bar{\nu}$ are investigated in \cite{Ivanov:2015tru} to address the question of the lepton flavor universality. We have already commented before on the leptonic results, they are entirely linked to the weak decay constant which is for various $B$ and $D$ mesons computed in Tab. I. Semileptonic decay are more demanding and the usual steps are taken: the SM CCQM form factors are determined (also the simplified parameterization is provided) and are used in a helicity formulation to predict the full four-dimensional differential distribution for the decay rate and various $q^2$-dependent distributions for angular and polarization observables. By integration one gets total branching fractions, shown in Tabs. III and IV of the publication, and their ratios $R_D$ and $R_{D^*}$ (Tab. V). The results are favorable to the NP presence, the deviation in $R_{D^{(*)}}$ is not smaller than seen by other authors at that time.

An analogous process with the $K^*$ meson in the final state is the subject of the analysis in \cite{Dubnicka:2015iwg}. The text follows the same logic as the previous one: the model is used to predict form factors and then the helicity formalism is employed to derive various differential distributions. Besides the branching fraction, the empasis is on the angular coefficients $A_{\text{FB}}$, $F_L$ and $P_i^{(')}$, $i = 1-5,8$ depicted in Figs. 7-11 of the publication. The numbers are given for integrated or averaged variables over the whole kinematical range (Tabs. 5 and 6) but also for various intervals (i.e. bins, Tabs. 7,8). The predicted branching fraction exceed the measured values, for what concerns the angular observables  reliable conclusions require more precise experimental data.

The article \cite{Ivanov:2016qtw} analyses possible NP scenarios for $\bar{B}^0 \to D^{(*)}\tau^- \bar{\nu}_\tau$ and in this way differs from the previous ones. The analysis relies on the usual effective Hamiltonian approach where beyond-SM four-fermion operators are introduced with the definition analogous to (\ref{NPope}) where $q \to c$. It is assumed that the NP affects only the leptons of the third generation and the effect of each NP operator is studied separately, with no other NP operator interfering. The form factors are computed in the CCQM framework from where observables quantities are obtained. By the fit to the $R_{D^{(*)}}$ ratios, allowed regions of the complex plane for the Wilson coefficients $V_{L,R}$, $S_L$ and $T_L$ are identified (Fig. 2 of the text). No room was found for the $S_R$ coefficient to explain the observed ratio and thus the corresponding operator was removed from further considerations. Next, full four-fold differential distribution was derived and various $q^2$-differential distributions analyzed: the NP Wilson coefficient was perturbed on the $2 \sigma$ level from the central value and the effect on a given distribution depicted as a gray band around the central line (Figs. 4-9). Depending on what distributions will future measurements provide, the presented results can serve us to identify which NP Wilson coefficients play a role.

The same process is also considered in \cite{Ivanov:2017mrj}, once again in the NP scenario based on the SM-extended effective Hamiltonian. Here the main topic are the longitudinal, transverse, and normal polarization components of the tau lepton and it is argued about their high sensitivity to NP effects. Using a model independent approach and the experimental data, constraints for various NP scenarios are derived and their effect on the polarization observables is investigated. To get numerical results the CCQM form factors are used. The acquired knowledge about the dependence of polarization observables on the NP Wilson coefficients may be useful in future data analysis as a guiding rule to differentiate between various NP scenarios.

Very similar analysis is performed in \cite{Ivanov:2017hun} but for different decays. The text focuses on the processes with light mesons in the final state $\bar{B}^0 \to \pi \tau \bar{\nu}$, $\bar{B}^0 \to \rho \tau \bar{\nu}$ and on the leptonic decay $B_c \to \tau \bar{\nu}$ assuming an SM-extended set of four-fermion operators. It uses the observables (\ref{eq:PNobser}) defined already in the leptonic section and the CCQM-predicted form factors to constrain the introduced NP Wilson coefficients. The effect of their variation on (\ref{eq:PNobser}) and on selected angular observables is analyzed.

Yet another publication which follows the same logic is \cite{Tran:2018kuv}, focusing this time on the decays $B_c \to J/\psi \tau \nu$ and $B_c \to \eta_c \tau \nu$. The observables used to constrain the NP Wilson coefficients are $R_D$, $R_{D^*}$, $R_{J/\psi}$ and $\mathcal{B}(B_c \to \tau \nu)$. With form factors derived in the CCQM assuming the NP, the impact of variation of these coefficients on other branching fraction ratios and angular observables is evaluated. The work provides a detailed comparison of the CCQM form factors with form factors from different approaches.

The work \cite{Issadykov:2018myx} is interested in $B_c \to J/\psi \bar{\ell} \nu_\ell$ and in the hadronic decay $B_c \to J/\psi \pi(K)$. This time a SM calculation is presented, the agreement with the SM is assessed through comparison of measured and predicted values for $R_{J/\psi}$ and two additional observables
\begin{align}
&R_{\pi^+/\mu^+ \nu}=\mathcal{B}(B_c^+ \to J/\psi \pi^+) / \mathcal{B}(B_c^+ \to J/\psi \mu^+ \nu_\mu),
\\
&R_{K^+/\pi^+}=\mathcal{B}(B_c^+ \to J/\psi K^+) / \mathcal{B}(B_c^+ \to J/\psi \pi^+).
\end{align} 
The form factors are evaluated in the CCQM framework and results for a set of semileptonic decays with $J/\psi$ or $\eta$ in the final state are presented (Tab. 2 there). The conclusion regarding the ratios is that an agreement with the SM is reached for $R_{\pi^+/\mu^+ \nu}$ and $R_{K^+/\pi^+}$, but the theoretical prediction for $R_{J/\psi}$ is too low with respect to data.

The semileptonic decays $B \to K^* \mu \mu$, $B_s^0 \to \phi \mu \mu$ and the leptonic decay $B_s \to \mu^+ \mu^-$ are addressed in \cite{Issadykov:2018sgw}. This brief text summarizes selected results and refers to previous papers.

The next paper dedicated to semileptonic decays is \cite{Issadykov:2022iwp}. It analyzes the $B \to K^{(*)} \nu \bar{\nu}$ process, where the current experimental limits on the branching fraction are expected not to be very far from the central value predicted by theory (i.e. the central value may be measured in the future). The CCQM is used to predict hadronic form factors which are then used in the helicity framework to predict branching fractions. The results agree with the experimental limits and also wit most of other authors. Approximately, the value of limits are  only four times higher than the central values predicted by the theory.

\section{Nonleptonic decays of $B$ mesons}

\subsection{Overview}

The number of experimental measurements concerning nonleptonic (or hadronic) $B$ decays is even larger than for semileptonic ones. Again, we briefly review the LHCb results and the results of the two $B$ factories, BaBar and Belle(II), as the most representative. Nevertheless, we do not provide an exhaustive list but mention only works with larger impact.

The question of NP is for hadronic decays less pronounced than for the semileptonic ones, since these are theoretically less clean. Yet, the NP is often mentioned and treated together with some of the usual topics such as (exotic) multiquark states, observations of new decay channels, CP-related measurements, fragmentation fractions or branching fractions determination. In what follows we will try to observe this classification.

\vspace{0.5cm}

The LHCb published several papers reporting the observation of a specific decay channel, some being observed for the first time. This comprises the first observations of $B^0_s \to J/\psi f_0(980)$ \cite{LHCb:2011blx}, $B^+_c \to J/\psi D_s^+$ and $B^+_c \to J/\psi D_s^{*+}$ \cite{LHCb:2013kwl}, $B^+_c \to B^0_s\pi^+$ \cite{LHCb:2013xlg},  $B^+ \to D_s^+  D_s^- K^+$ \cite{LHCb:2022dvn}, $B^0_s\!\to D^{*+}D^{*-}$ \cite{LHCb:2022uvz}, $B^+ \to J/\psi \eta^{\prime} K^+$ \cite{LHCb:2023qca} or $B^0_s\to \chi_{c1}(3872)\pi^+\pi^-$ \cite{LHCb:2023reb}. For the most of these observations some quantitative numbers are given, usually branching fraction ratios to a different decay mode (normalization channel).

A special interest is given to the observation of "resonant structures", i.e. observation of possible exotic multiquark states which are sometimes seen in invariant mass distributions of particles originating from the $B$ disintegration. An important contribution to the exotic physics was done in 2013 when the LHCb  measured, in the $B$ decay channel, the quantum numbers of the $X(3872)$ resonance \cite{LHCb:2013kgk}, previously discovered by Belle. Contemporary texts \cite{LHCb:2012ae}, \cite{LHCb:2014ooi} and \cite{LHCb:2014vbo} analyze the $\bar{B}_s^0\to J/\psi\pi^+\pi^-$ and $\overline{B}^0\to J/\psi \pi^+\pi^-$ spectra, and identify various resonant structures; here only the usual SM resonances are seen. The possible tetraquark character of the $f_0(980)$ invoked in the last text is rejected as inconsistent with data. The situation becomes different in \cite{LHCb:2016axx}, where four resonant structures, possibly tetraquarks, are observed and their quantum numbers are determined. The work \cite{LHCb:2021uow} reports on two exotic particles having $c \bar{c} u \bar{s}$ quark content determined with high significance and also confirms four previously reported states. The authors of \cite{LHCb:2022jad} perform an amplitude analysis of the $B^-\to J/\psi\Lambda\bar{p}$ process, where the $J/\psi\Lambda$ mass spectrum contains a narrow resonance, possibly a strange pentaquark; its quantum numbers are measured. A resonant structure, referred to as $X(3960)$, is also observed in the $B^+\to D_s^+ D_s^- K^+$ decay mode close to the  $D_s^+ D_s^-$ production threshold \cite{LHCb:2022vsv}. It is established to be consistent with a four-quark state $c\bar{c} s \bar{s}$ having quantum numbers $J^{PC} = 0^{++}$. The text \cite{LHCb:2020bls} analyses the spectrum of $B^+\to D^+D^-K^+$ and advances a hypothesis of new charm-strange resonances. Another recent text, \cite{LHCb:2021chn}, also sees a new resonance of mass $4337 \, \text{MeV} $ in the $J/\psi p$ ($J/\psi \bar{p}$) spectrum of the  $B_s^0 \to J/\psi p \bar{p}$ decay. A very recent analysis \cite{LHCb:2023dhg} is concerned with decays of the $B$ mesons to $J/\psi \phi K_S^0$ and presents evidence for $T_{\psi s 1}^\theta$ state in the $J/\psi K_S^0$ invariant spectrum, presumably a tetraquark.

Besides direct investigations of the invariant mass spectrum, many LHCb publications rely, to identify resonant components, on the Dalitz plot and amplitude analysis where further resonances are identified, see \cite{LHCb:2014ioa,LHCb:2015klp, LHCb:2016lxy, LHCb:2016nsl, LHCb:2020pxc, LHCb:2022pjv, LHCb:2022bkt}. The hadronic $B$ decays are also often studied in the context of the CP analysis and weak parameter determination \cite{LHCb:2011aa,LHCb:2012eks,LHCb:2011vae,LHCb:2013ptu,LHCb:2013lcl,LHCb:2013lrq, LHCb:2012xkq, LHCb:2013syl,LHCb:2013odx,LHCb:2014tol, LHCb:2015ups, LHCb:2014iah, LHCb:2022nng, LHCb:2023yjo}. Various topics are addressed in these works: observation of the CP violation in a specific decay, measurement of the CP-violating phase, $B_{(s)}^0$-$\bar{B}_{(s)}^0$ oscillations and determination of the CKM angles. The $B$ decay measurements are also used to determine basic particle quantities, such as production cross sections, branching ratios or fragmentation fractions \cite{LHCb:2012ihf, LHCb:2012ihl, LHCb:2012quo, LHCb:2013vfg, LHCb:2019fns, LHCb:2021qbv, LHCb:2022ioi, LHCb:2022htj, LHCb:2022vns, LHCb:2023evz}. 

\vspace{0.5cm}

The publications of the BaBar experiment fall into similar categories. We choose to mention in more detail the CP-related results which had, in the domain of nonleptonic $B$ decays, the most significant impact. Namely, the violation of the CP symmetry was before the BaBar measurement \cite{BaBar:2001pki} only observed for kaons. The measurement was done for several decay modes of the $B^0$ particle, for each decay the CP asymmetry $A_{\text{CP}}$ was measured. The latter was defined in terms of a decay-time distribution $f_\pm(\Delta t)$ for $B$ and $\bar{B}$ decaying into the common final state. The results were derived for the $\text{sin}(2\beta)$ quantity, where $\beta$ is an angle of the unitarity triangle constructed from the CKM matrix elements and its deviation from zero measures the CP violation. The significance of the measurement reached 4 $\sigma$ level. The CP-violation topic was then discussed in further publications for the neutral \cite{BaBar:2001ags, BaBar:2002zub, BaBar:2002epc,BaBar:2004gyj,BaBar:2004god, BaBar:2004xhu, BaBar:2007wip, BaBar:2009byl, BaBar:2018cka} and also charged $B$ meson \cite{BaBar:2008lpx, BaBar:2010uep,BaBar:2012iuj, BaBar:2012idb}. Both, indirect (i.e. involving particle-antiparticle oscillations) and direct CP violation was seen with relevant significance. Several texts present measurements were the branching fraction and the CP asymmetries were addressed at the same time \cite{BaBar:2001kqz, BaBar:2002sjz,BaBar:2003hgn, BaBar:2004zpa, BaBar:2006xjk, BaBar:2012fqh}. Besides the direct CP violation measurements, the closely related measurements of the CKM angles $\alpha$ and $\gamma$ were presented in \cite{BaBar:2004usm,BaBar:2005rek, BaBar:2007cku, BaBar:2008inr}.

The BaBar collaboration also investigated, in a variety of publications \cite{BaBar:2001iyi,BaBar:2001kpq, BaBar:2003zor,BaBar:2004uwv,BaBar:2005xzz,BaBar:2005sdl,BaBar:2005pcw,BaBar:2006hyf,BaBar:2007uoe,BaBar:2007rbr,BaBar:2007yhb,BaBar:2012eja}, the usual quantities which characterize decays, i.e. branching fractions, angular observables and branching fractions. The  related topic of resonances and exotic states were subject to numerous analysis. The resonances were investigated by invariant mass spectra or the Dalitz-plot method, as presented in \cite{BaBar:2005qms,BaBar:2007cmo, BaBar:2009vfr}. Concerning exotic states, most of the BaBar results are related to the $X(3872)$ particle \cite{BaBar:2004iez, BaBar:2004oro, BaBar:2004cah, BaBar:2006fjg, BaBar:2005xmz,BaBar:2008qzi, BaBar:2008flx, BaBar:2010wfc, BaBar:2019hzd} and  present related searches, observations and measurements in various decay modes. The state $Y(3940)$, first discovered at Belle, was observed also (as a product of a $B$ decay) and its mass and width were determined.

\vspace{0.5cm}

The Belle experiment was very successful in search for various exotic states, tetraquarks and pentaquarks. Not all were related to hadronic decays of the $B$ meson, but the most cited result \cite{Belle:2003nnu} was. It presents the discovery of the $X(3872)$ particle seen in the $\pi^+ \pi^- J/\psi$ spectrum of $B^\pm \to K^\pm \pi^+ \pi^- J/\psi$. Another achievements were the detection of tetraquark candidates $Z(4430)$ \cite{Belle:2007hrb} and $Y(3940)$ \cite{Belle:2004lle}, both among the decay products of $B$. In addition to these, further publications on this topic were issued \cite{Belle:2005ogu, Belle:2006olv, Belle:2008qeq, Belle:2008fma,Belle:2009lvn,Belle:2011vlx,Belle:2011wdj, Belle:2013ewt, Belle:2014nuw, Belle:2023zxm}, all related to nonleptonic $B$ decays. The physic program regarding the CP violation and the weak physics in general is also very present at Belle. The collaboration published the $B^0$ CP-violation paper \cite{Belle:2001zzw} only a short time after BaBar did. Yet, it drew a lot of attention as an independent measurement of the $\text{sin}(2 \beta)$ parameter. The measurement was updated later in \cite{Belle:2002ghj}, direct CP violation was reported in \cite{Belle:2004mad,Belle:2004nch}. Many additional papers were published by Belle where various CP parameters (CKM angles) and weak-physics related processes were studied \cite{Belle:2001qdd, Belle:2003qkg, Belle:2003vik, Belle:2004bbr, Belle:2005lnu, Belle:2005qtf, Belle:2005rpz, Belle:2006lys,Belle:2005grh, Belle:2006pxp,Belle:2006xxx, Belle:2006dlp, Belle:2006yry, Belle:2008alz, Belle:2010xyn,Belle:2012paq,Belle:2012dmz,Belle:2020cio,Belle-II:2023grc,Belle-II:2023bps,Belle-II:2023nmj}.

Naturally, the research at Belle is devoted also to branching fraction measurements of different B decay modes \cite{Belle:2001pbl,Belle:2001otw, Belle:2002ryx, Belle:2003ibz, Belle:2004opq,  Belle:2003nsh, Belle:2004drb, Belle:2009nth, Belle:2020fbd,Belle-II:2023cbc}, observation and analysis of new decay channels \cite{Belle:2002bro, Belle:2002bnx, Belle:2002fay, Belle:2003fgr, Belle:2003guh, Belle:2003lsm, Belle:2003pwf, Belle:2007hht, Belle:2017jrt}, polarization studies \cite{Belle:2003ike, Belle:2005lvd} and photon energy spectra analysis in radiative events \cite{Belle:2004stc, Belle-II:2022hys}.

\vspace{0.5cm}

The large amount of data on hadronic $B$ decays motivates the theorists to describe observations and prove our understanding of the underlying physics to be correct. The exotic multiquark states have a specific character from the perspective of $b$ physics: as a matter of fact many of them originate from nonleptonic $B$ decays, yet, these decays seen as exotic production processes, are not addressed very frequently. They often have a larger number of hadrons in the final state (three or more) and thus large phase space and technically complicated description. The exotic particles are usually treated in the scenario where they represent the initial state (for the CCQM model see \cite{Dubnicka:2020yxy}) and thus are not in the scope of this text (are not $B$ mesons). The emphasis of the theoretical overview is therefore on the remaining topics: branching fractions and weak-interaction physics.

The theoretical grounds to describe (not only) hadronic $B$ decays were laid decades ago. The CP violation in the SM stems from the flavor mixing through the CKM matrix which has an irreducible complex phase, as formulated in the pioneering works \cite{Cabibbo:1963yz, Kobayashi:1973fv}. This rapidly lead to first theoretical predictions. In \cite{Bander:1979px} the expectation of a small but measurable CP non-invariance in $B$ meson decays was expressed. The authors of \cite{Carter:1980tk} argued, studying the on-shell transitions in heavy meson cascade decays, that the effect may not be so small after all and propose methods to detect the CP violation in the $B$ sector. The latter topic is also discussed in \cite{Bigi:1981qs}, where mainly the non-leptonic decay modes are addressed.

In parallel the issues related to the asymptotic behavior and quark interactions were considered. The nice review \cite{Chernyak:1983ej} addressed the question of the power behavior of amplitudes and its relation to mesonic wave functions and quantum numbers. As results quantitative conclusions are made for hadronic form factors, large angle scattering processes and other related quantities. The highly cited paper \cite{Godfrey:1985xj} presents a relativistic extension of the quark model based on one-gluon exchange and a linear confining quark potential. It is used to describe mesons, their spectroscopy and decays, and succeeds to large extent. The work \cite{Bauer:1986bm} studies (among others also) $B$ decays in the framework  of the valence quark model; the model assumes factorization and good results are obtained especially for nonleptonic processes. Following works further sharpen the QCD SM prediction; the next-to-leading QCD corrections are computed in \cite{Buras:1991jm}, the implications of the heavy quark symmetry are analyzed  in \cite{Isgur:1991wq}, the generalized factorization hypothesis and its impact on the structure of non-factorizable corrections are presented in \cite{Neubert:1997uc} and three-loop anomalous dimensions at the next-to-leading order in $\alpha_s$ for weak radiative $B$ decays are computed in \cite{Chetyrkin:1996vx}. The role of the charm penguin diagrams in the $B$ decay to pions was evaluated by the authors of \cite{Ciuchini:1997hb} and a next-to-leading order evaluation of the branching fraction and photon spectrum of the $B \to X_s + \gamma $ process was presented in \cite{Kagan:1998ym}.

Coming back to the CP symmetry, one can mention the publication \cite{Dunietz:1986vi}, where large time-dependent CP asymmetries in the $B^0 - \bar{B}^0$ system are predicted or \cite{Gronau:1990ka} where it is shown that the theoretical uncertainty associated with penguin diagrams in the $B^0 \to \pi \pi$ decay can be reduced by considering isospin relations.

An important issue addressed by various authors is the factorization validity, often assumed for hadronic matrix elements of the four-fermion operators.  In \cite{Ali:1998eb} a theoretical investigation of $B$ branching fractions is undertaken and branching fraction ratios of selected two-body hadronic $B$ decays are proposed as factorization experimental tests. The article \cite{Beneke:2000ry} is focused on the factorization for heavy-light final states. Such decays are treated in the heavy quark limit and the validity of the factorization ansatz is in this scenario proven at the two-loop order. In the similar context the authors of \cite{Beneke:2001ev} study processes with two light mesons ($K$, $\pi$) in the final state. They argue that in the heavy quark limit the hadronic matrix elements of nonleptonic $B$ meson decays can be computed from first principles which helps to reduce the errors on the weak phases $\alpha$ and $\gamma$.  Very similarly is oriented the paper \cite{Bauer:2001cu}, where the proof of the  factorization is provided for $B^- \to D^0 \pi^-$ and $B^0 \to D^+ \pi^-$. The topic of the factorization is further treated in \cite{Beneke:2003zv}, where decays $B \to PP$ and $B \to PV$ are addressed, and also in \cite{Bauer:2004tj}, where soft-collinear effective theory is used to prove factorization for $B$ decaying to two light particles ($\pi$, $K$, $\rho$, $K^*$).

One should also mention new physics searches. The paper \cite{Buras:2003dj} studies the $B \to \pi \pi$ process from which it extracts relevant hadronic parameters. These are then used, under plausible assumptions, to predict $B \to \pi K$. Those observables (for the latter process) which have small EW penguin contributions seem to agree with the experiment, those with significant contributions do not. This might indicate NP in the W penguin sector. Similar ideas are developed also in \cite{Buras:2004ub}. A related topic, the final state interactions in hadronic B decays, is treated in \cite{Cheng:2004ru}. Indeed, when considering the $B$ decays to light mesons, there are, generally speaking, some difficulties to describe the data. To disentangle possible NP, all SM effects need to be considered, rescattering included. The latter is here treated in a phenomenological way in terms of off-shell meson exchange.

Let us briefly mention other works of interest: papers \cite{Ball:2004rg,Ball:2004ye} apply the light-cone sum rules to tackle $B$ decays to light vector and pseudoscalar mesons respectively, the authors of \cite{Gorbahn:2004my} compute, at next-to-next-to-leading order of QCD, the effective Hamiltonian for non-leptonic $|\Delta F| = 1$ decays, and the text \cite{Beneke:2006hg} focuses on the $B$ decays to two vector particles in the framework of the QCD factorization. At last we can mention the paper \cite{Charles:2015gya} which summarizes the status of our CKM matrix knowledge based on a global fit to various (leptonic, semileptonic, hadronic) data.

\subsection{Nonleptonic $B$ decays in CCQM \label{Sec_CCQM_had}}

\subsubsection*{Decay $B_s \to J/\psi \eta^{(')}$}

\begin{figure}[t]
\begin{center}
\includegraphics[width = 0.3\textwidth ]{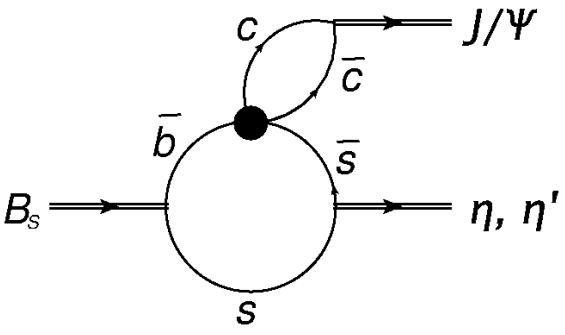}
\hspace{1cm}
\includegraphics[width = 0.3\textwidth ]{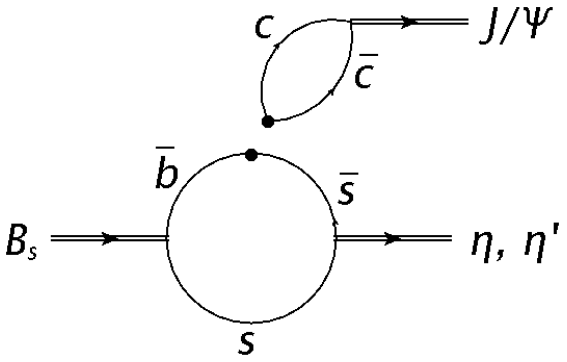}
\\
(a)\hspace{5cm}(b)
\caption{The $B_s \to \eta^{(')} J/\psi$ decay as a $B_s$ transition to the  $\bar{s}s$ component of $\eta^{(')}$ (a) in the factorization picture (b). Figure was originally published in \cite{Dubnicka:2013vm}.}
\label{Fig:had_eta}
\end{center}
\end{figure}

We have chosen to demonstrate the CCQM approach on two hadronic processes to point out various aspects of the model application. The first one is $B_s \to J/\psi \eta^{(')}$ \cite{Dubnicka:2013vm}, were a fit to the data was performed so as to determine the model input parameters. The $\eta^{(')}$ mesons are described as a superposition of light ($q=u,d$)  and strange components, $\eta = -\sin \delta (\bar{q}q) - \cos \delta (\bar{s}s)$ and $\eta^{'} = \cos \delta (\bar{q}q) - \sin \delta (\bar{s}s)$ where $\delta = \varphi_P - \pi/2$, $\varphi_P = 41.4^\circ  $ \cite{KLOE:2006guu}. The considered decay was treated within the na\"ive factorization picture in the leading order, meaning it was described as a $B_s \to \eta^{(')}$ transition where only the $\bar{s}s$ component of the latter is taken into the account, see Fig. \ref{Fig:had_eta}.
The necessary inputs for the decay width formula ($P=\eta, \, \eta^{'}$)
\begin{align}
\Gamma(B_s \to J/\Psi + P ) = \frac{G_F^2}{4\pi}|V_{cb}V_{cs}^\dagger | C_W^2 f^2_{J/\Psi} |\bold{q}_P|^3 \zeta^2_P [F_+^{B_s \eta^{(')}}(m^2_{J/\Psi})]^2,
\; \zeta_\eta = \cos{\delta}, \; \zeta_{\eta^{'}} = \sin{\delta}
\end{align}
are the leptonic decay constants $f_{J/\Psi} \equiv f_V$ and the transition form factor $F_+$
\begin{align}
 m_V f_V \epsilon^\mu_V = N_c g_V \int \frac{d^4k}{(2\pi)^4 i}
\tilde{\Phi}(-k^2) \text{tr}[O^\mu S_1(k+w_1 p) \cancel{\epsilon}_V S_2(k-w_2p)],
\quad
p^2 = m_V^2,
\end{align}
\begin{align}
\langle P_{q_1,q_3} (p_2) | \bar{q}_2 O^\mu q_1 | B_{\bar{q}_3,q_2}(p_1) \rangle
=& F_+(q^2) P^\mu + F_-(q^2)q^\mu,
\label{scalarFF} \\
=& N_c g_B g_P \int \frac{d^4}{(2\pi)^4 i}
\tilde{\Phi}_B(-[k+w_{13}p_1]^2)
\tilde{\Phi}_P(-[k+w_{23}p_2]^2)
\nonumber \\
&\times \text{tr}[O^\mu S_1(k+p_1) \gamma^5 S_3(k) \gamma^5 S_2(k+p_2)], \nonumber
\end{align}
where the Wilson coefficient is given by $C_W = C_1+C_2/N_c+C_3+C_4/N_c+C_5+C_6/N_c$ and the meaning of other symbols is analogous to Sec. \ref{Sec:leptoCCQM} and \ref{SecCCQMsemi}. The results are derived in the large $N_c$ limit $N_c \to \infty$. To get to the form factor and the decay constants one needs to know the model $\Lambda$ parameters $\Lambda_\eta^{\bar{q}q}$, $\Lambda_\eta^{\bar{s}s}$, $\Lambda_{\eta^{'}}^{\bar{q}q}$ and $\Lambda_{\eta^{'}}^{\bar{s}s}$, four in total if one treats $q$ and $s$ components as independent. They can be derived from various processes where they play a role, so, in addition to the two studied decay channels, also $\eta \to \gamma \gamma$, $\eta^{'} \to \gamma \gamma$, $\varphi \to \eta \gamma$, $\varphi \to \eta^{'} \gamma$, $\rho^0 \to \eta \gamma$, $\omega \to \eta \gamma$, $\eta^{'} \to \omega \gamma$, $B_d \to J/\Psi + \eta$ and $B_d \to J/\Psi + \eta^{'}$ have been chosen.
Fitting all together 11 processes, the optimal-fit parameters were determined
\begin{align}
\Lambda_\eta^{\bar{q}q} = 0.881\,\text{GeV}, \quad
\Lambda_\eta^{\bar{s}s}=1.973\,\text{GeV}, \quad
\Lambda_{\eta^{'}}^{\bar{q}q}=0.257\,\text{GeV}, \quad
\Lambda_{\eta^{'}}^{\bar{s}s}=2.797\,\text{GeV},
\end{align}
other model parameters were taken from previous works, namely $\Lambda_{B_s} = 1.95 \; \text{GeV}$, $\Lambda_{B_d} = 1.88 \; \text{GeV}$ and $\Lambda_{J/\Psi} = 1.48 \; \text{GeV}$. Also hadron-independent parameters (\ref{par_values}) were tuned to different values, see Eq. (6) of  \cite{Dubnicka:2013vm}. With these in hand one computes results, see Tab. \ref{Tab:hadronicA}.
\begin{table}
\centering
\begin{tabular}[t]{c  c  c  }
\hline
\hline
Observable & CCQM & Exp.\cite{Workman:2022ynf} \\
\hline 
$\Gamma(\eta \to \gamma \gamma)$ & $0.380$ keV & $0.515 \pm 0.020$ keV \\
$\Gamma(\eta^{'} \to \gamma \gamma)$ & $3.74$ keV & $4.34 \pm 0.14$ keV \\
$\Gamma(\eta^{'} \to \omega \gamma)$ & $9.49$ keV & $4.74 \pm 0.15$ keV \\
$\Gamma(\rho \to \eta \gamma)$ & $53.07$ keV & $ 44.22 \pm 0.24$ keV \\
$\Gamma(\omega \to \eta \gamma)$ & $6.21$ keV & $ 3.91 \pm 0.06$ keV \\
$\Gamma(\varphi \to \eta \gamma)$ & $42.59$ keV & $ 55.28 \pm 0.17$ keV \\
$\Gamma(\varphi \to \eta^{'} \gamma)$ & $0.276$ keV & $ 0.26 \pm 0.001 $ keV \\
$\mathcal{B}(B_d \to J/\Psi + \eta)$ & $16.5 \times 10^{-6}$ & $ (10.8 \pm 2.3)\times 10^{-6} $ \\
$\mathcal{B}(B_d \to J/\Psi + \eta^{'})$ & $12.2 \times 10^{-6}$ & $ (7.6 \pm 2.4)\times 10^{-6} $ \\
$\mathcal{B}(B_s \to J/\Psi + \eta)$ & $4.67 \times 10^{-4}$ & $ ( 4.0 \pm 0.7 )\times 10^{-4} $ \\
$\mathcal{B}(B_s \to J/\Psi + \eta^{'})$ & $4.04 \times 10^{-4}$ & $ (3.3 \pm 0.4)\times 10^{-4} $ \\
\hline
\hline
\end{tabular}
\caption{Decay widths and branching fractions for various processes with $\eta$ and $\eta^{'}$ mesons as predicted by the CCQM. Table contains a subset of data originally published in \cite{Dubnicka:2013vm}.}
\label{Tab:hadronicA}
\end{table}
Generally speaking the discrepancies in terms of standard deviations are rather large, yet the model roughly (within the factor 2) reproduces the data. There might be reasons to the differences one needs to understand, e.g. a gluoniun contribution to the $\eta^{'}$ state \cite{KLOE:2006guu} could weaken the largest disagreement for $\Gamma(\eta^{'} \to \omega \gamma)$. As pointed out in \cite{Dubnicka:2013vm}, other models on the market do not seem to perform better than us.

The Belle and LHCb collaborations also measured the ratio
\begin{align}
R = \frac{\mathcal{B}(B_s \to J/\Psi + \eta^{'})}{\mathcal{B}(B_s \to J/\Psi + \eta)}
= \begin{cases}
      0.73 \pm 0.14, & \text{Belle \cite{Belle:2012aa}} \\
      0.90 \pm 0.1,  & \text{LHCb \cite{LHCb:2012cw}} \\
      0.86, & \text{CCQM} \\
    \end{cases}.
\end{align}
Here the CCQM number reproduces well the measurements and through the predicted form factors adds a non-trivial factor 0.83 to the model-independent part of the calculation
\begin{align}
R^{\text{theor}} = \left( \frac{|\bold{q}_{\eta^{'}}|^3}{|\bold{q}_{\eta}|^3} \tan^2(\delta) \right)
\times \left( \frac{F_+^{B_s\eta'}}{F_+^{B_s\eta}} \right)^2
= 1.04... \times 0.83... \approx 0.86.
\end{align}
The overall precision of results is not fully satisfactory and further efforts may be done to investigate the discrepancies. Yet, besides the results themselves we wanted, in this subsection, also to point to the methodology we adopt in the CCQM for determining the model inputs.

\subsubsection*{Decay $B \to D_{(s)}^{(*)}h$, $(h=\pi,\rho)$}

The second process we want to review is the $B_d$ decay to a $D$ meson and a light particle \cite{Dubnicka:2022gsi}. The interest here comes form the observation confirmed by other authors too, that the predictions systematically overshoot the data, which might indicate the NP.

\begin{figure}[t]
\begin{center}
\includegraphics[width = 0.2\textwidth ]{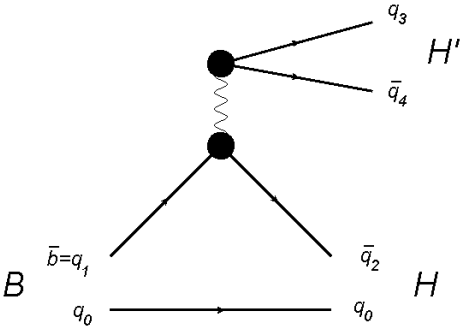}
\hspace{0.5cm}
\includegraphics[width = 0.2\textwidth ]{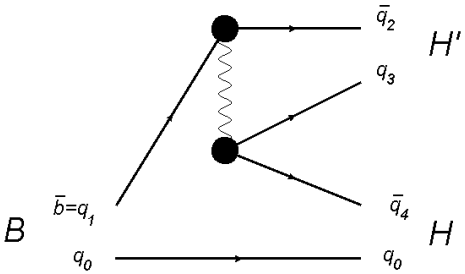}
\hspace{0.5cm}
\includegraphics[width = 0.45\textwidth ]{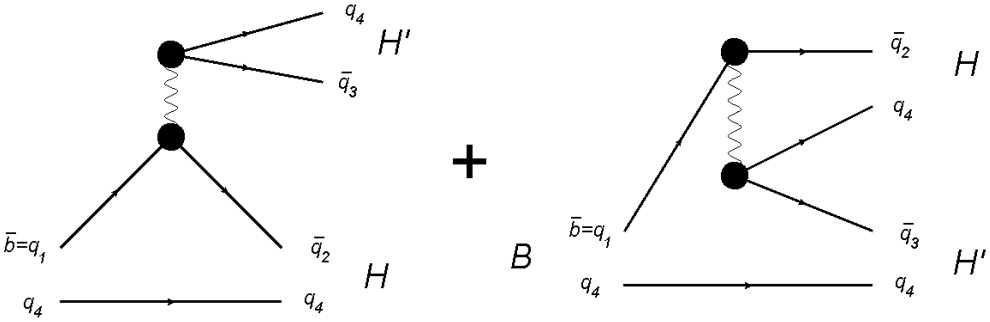}
\\
(a) \hspace{3cm} (b) \hspace{4.3cm} (c)\hspace{1.7cm}
\caption{$B$ decays to two hadrons: color favored $D_1$ (a), color suppressed $D_2$ (b) and their interference $D_3$ (c). Figures were originally published in \cite{Dubnicka:2022gsi}.}
\label{Fig:hadBDh}
\end{center}
\end{figure}

\begin{figure}[t]
\begin{center}
\includegraphics[width = 0.35\textwidth ]{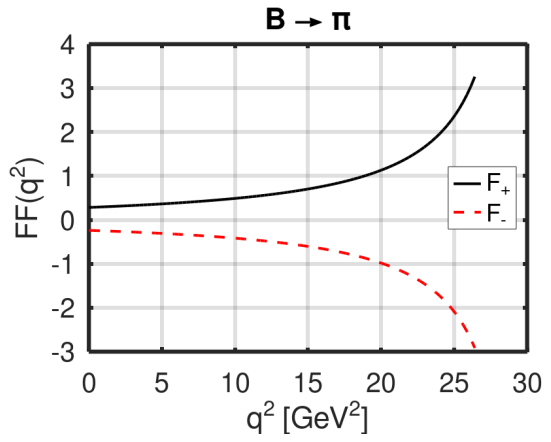}
\hspace{0.5cm}
\includegraphics[width = 0.35\textwidth ]{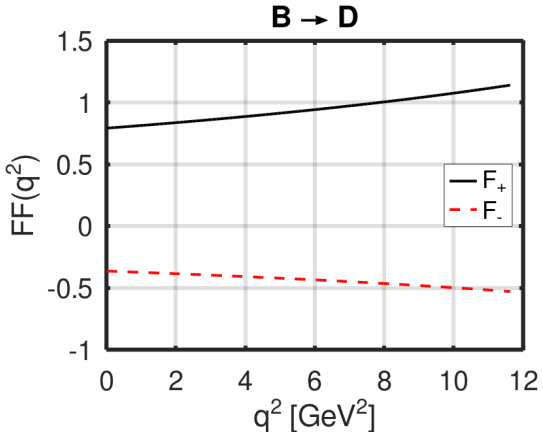}
\\
\includegraphics[width = 0.35\textwidth ]{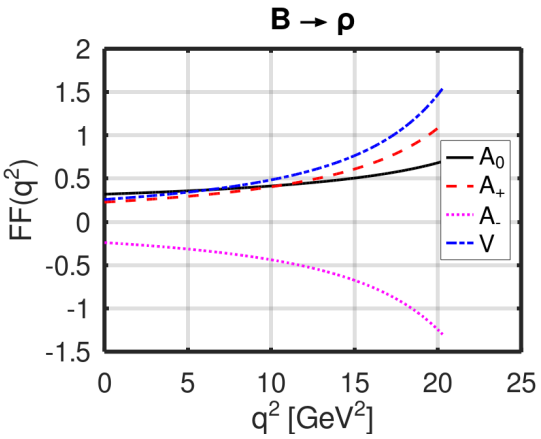}
\hspace{0.5cm}
\includegraphics[width = 0.35\textwidth ]{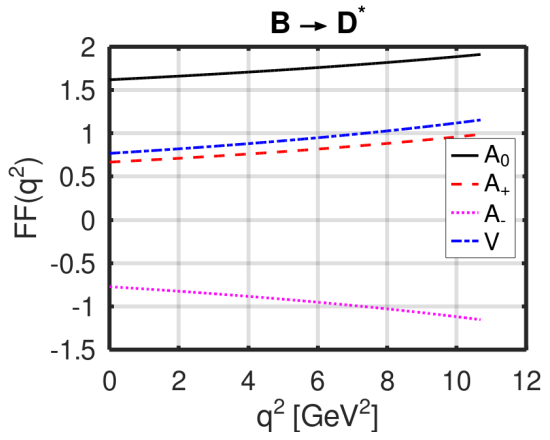}
\caption{Transition form factors as predicted by the CCQM. Figures were originally published in \cite{Dubnicka:2022gsi}.}
\label{Fig:hadBDhFF}
\end{center}
\end{figure}

\begin{figure}[t]
\begin{center}
\includegraphics[width = 0.50\textwidth ]{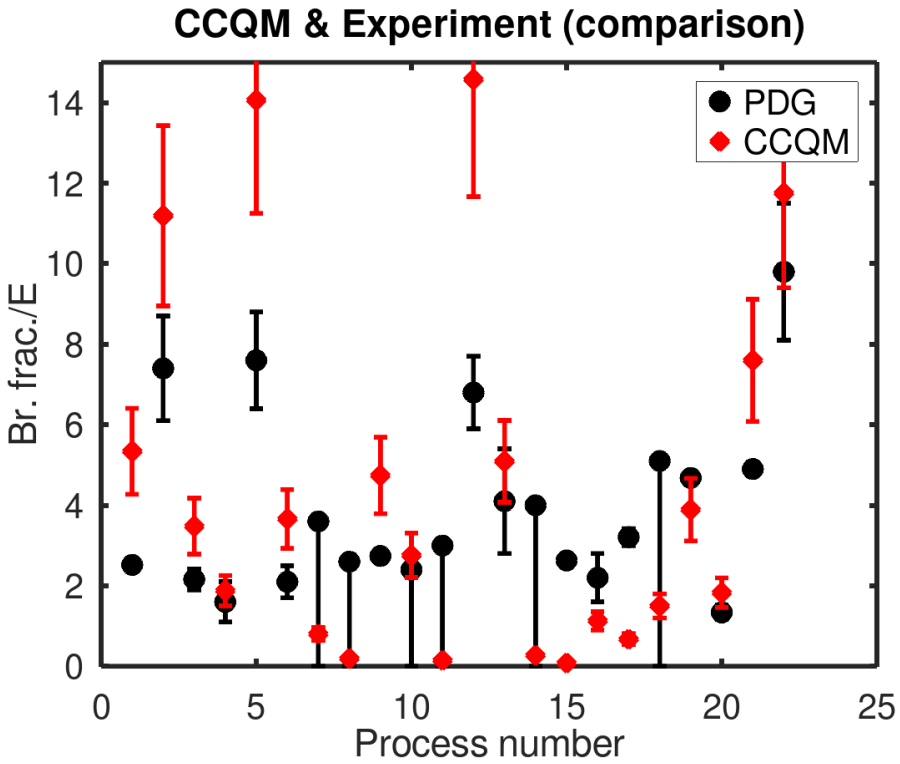}
\caption{The comparison of CCQM predictions and data. Processes are numbered as in Tab. 
\ref{Tab:BDhres}. Figure was originally published in \cite{Dubnicka:2022gsi}.}
\label{Fig:hadResults}
\end{center}
\end{figure}

The processes is described in the leading order and na\"ive factorization framework. These decays correspond to rich set of various spin states and diagram topologies, as is summarized in Fig. \ref{Fig:hadBDh} and Table \ref{Tab:hadTableDro}. One labels by $D_{1,2,3}$ the diagram structure (color favored, color suppressed and their interference ), where within each group, various spin configurations are present (labeled $A,\dots, D$).
\begin{table}
\centering
\begin{tabular}[t]{c  c  c  c}
\hline
\hline
Spin structure & $D_1$ diagram & $D_2$ diagram & $D_3$ diagram \\
\hline \\[-1.0em]
(A) & $\underline{B^0 \to D^-} +\pi^+$ & $\underline{B^0 \to \pi^0} + \bar{D}^0$ & $\underline{B^+ \to \bar{D}^0} + \pi^+$ \\
$\underline{PS \to PS} + PS $ &$\underline{B^0 \to \pi^-} +D^+$ & &\\
& $\underline{B^0 \to \pi^-} + D_s^+$ & & \\
&$\underline{B^+ \to \pi^0} + D_s^+$ && \vspace{0.2em}\\
\cline{2-4} 
\\[-1.0em]
(B) & $\underline{B^0 \to D^-} +\rho^+$ & $\underline{B^0 \to \pi^0} + \bar{D^{*}}^0$ & $\underline{B^+ \to \bar{D}^0} + \rho^+$ \\
$\underline{PS \to PS} + V $ & $\underline{B^0 \to \pi^-} +D_s^{*+}$&& \\
& $\underline{B^+ \to \pi^0} + D^{*+}$ & & \\
& $\underline{B^+ \to \pi^0} + D_s^{*+}$ & & \vspace{0.2em} \\
\cline{2-4}
\\[-1.0em]
(C) & $\underline{B^0 \to D^{*-}} +\pi^+$ & $\underline{B^0 \to \rho^0} + \bar{D}^0$ & $\underline{B^+ \to \bar{D^{*}}^0} + \pi^+$ \\
$\underline{PS \to V} + PS $ & $\underline{B^0 \to \rho^-} +D_s^{+}$&& \\
& $\underline{B^+ \to \rho^0} + D_s^{+}$ & & \vspace{0.2em} \\
\cline{2-4}
\\[-1.0em]
(D) & $\underline{B^0 \to D^{*-}} +\rho^+$ & $\underline{B^0 \to \rho^0} + \bar{D^{*}}^0$ & $\underline{B^+ \to \bar{D^{*}}^0} + \rho^+$ \\
$\underline{PS \to V} + V $ & $\underline{B^0 \to \rho^-} +D_s^{*+}$&& \\
& $\underline{B^+ \to \rho^0} + D_s^{*+}$ & & \vspace{0.2em} \\
\hline
\hline
\end{tabular}
\caption{Studied decays arranged with respect to the spin structure and diagram topology. Underlined parts correspond to the transition of the spectator quark (in case of $D_3$ to the first diagram of Fig. \ref{Fig:hadBDh}(c) ). Table was originally published in \cite{Dubnicka:2022gsi}.}
\label{Tab:hadTableDro}
\end{table}
Using the leading order operators
\begin{align}
Q_1 = [(\bar{q}_1)_{i_1} (q_2)_{i_2}]_{V-A} [(\bar{q}_3)_{i_2} (q_4)_{i_1}]_{V-A}, \quad 
Q_2 = [(\bar{q}_1)_{i_1} (q_2)_{i_1}]_{V-A} [(\bar{q}_3)_{i_2} (q_4)_{i_2}]_{V-A},
\end{align}
where $i_j$ are color indices and $[q_1q_2]_{V-A} = \bar{q}_1 \gamma^\mu(1-\gamma^5)q_2$, one can derive form factors. They are in the case of the scalar-to-scalar transition given by (\ref{scalarFF}), for the scalar-to-vector form factor the expression stands
\begin{align}
&\langle V_{q_3,q_2} (p_2,\epsilon) | \bar{q}_1 O^\mu q_2 | B_{q_3,q_1}(p_1) \rangle =
\\
&\qquad
= \frac{\epsilon_\nu^\dagger}{m_B+m_V}
\left[
-g^{\mu \nu} P \cdot q A_0(q^2)
+P^\mu P^\nu A_+(q^2)
+ q^\mu P^\nu A_-(q^2)
+ \epsilon^{\mu \nu \alpha \beta} P_\alpha q_\beta V(q^2) \nonumber
\right].
\end{align}
The obtained form factors are shown in Fig. \ref{Fig:hadBDhFF}, the hadron-specific and universal CCQM parameters used in their prediction are summarized in Table II of \cite{Dubnicka:2022gsi}.
The corresponding decay-width formulas (see \cite{Dubnicka:2022gsi}, page 3) then allow one to get results summarized in Tab. \ref{Tab:BDhres}.
\begin{table}
\centering
\begin{tabular}[t]{ccccccc}
 &  &  &  &  &  & \tabularnewline
\hline 
\hline 
 & Process & Diagram & $\mathcal{B}_{\mathrm{CCQM}}/\mathrm{E}$ &  & $\mathcal{B}_{\mathrm{PDG}}/\mathrm{E}$ & $\mathrm{E}$\\
\hline 
1 & $B^{0}\to D^{-}+\pi^{+}$ & $D_{1}$ & $5.34 \pm 0.27$ &  & $2.52\pm0.13$ & $10^{-3}$\\
2 & $B^{0}\to\pi^{-}+D^{+}$ & $D_{1}$ & $11.19 \pm 0.56$ &  & $7.4\pm1.3$ & $10^{-7}$\\
3 & $B^{0}\to\pi^{-}+D_{s}^{+}$ & $D_{1}$ & $3.48 \pm 0.17$ &  & $2.16\pm0.26$ & $10^{-5}$\\
4 & $B^{+}\to\pi^{0}+D_{s}^{+}$ & $D_{1}$ & $1.88 \pm 0.09$ &  & $1.6\pm0.5$ & $10^{-5}$\\
5 & $B^{0}\to D^{-}+\rho^{+}$ & $D_{1}$ & $14.06 \pm 0.70$ &  & $7.6\pm1.2$ & $10^{-3}$\\
6 & $B^{0}\to\pi^{-}+D_{s}^{*+}$ & $D_{1}$ & $3.66 \pm 0.18$ &  & $2.1\pm0.4$ & $10^{-5}$\\
7 & $B^{+}\to\pi^{0}+D^{*+}$ & $D_{1}$ & $0.804 \pm 0.04$ &  & $<3.6$ & $10^{-6}$\\
8 & $B^{+}\to\pi^{0}+D_{s}^{*+}$ & $D_{1}$ & $0.197 \pm 0.01$ &  & $<2.6$ & $10^{-4}$\\
9 & $B^{0}\to D^{*-}+\pi^{+}$ & $D_{1}$ & $4.74 \pm 0.24$ &  & $2.74\pm0.13$ & $10^{-3}$\\
10 & $B^{0}\to\rho^{-}+D_{s}^{+}$ & $D_{1}$ & $2.76 \pm 0.14$ &  & $<2.4$ & $10^{-5}$\\
11 & $B^{+}\to\rho^{0}+D_{s}^{+}$ & $D_{1}$ & $0.149 \pm 0.01$ &  & $<3.0$ & $10^{-4}$\\
12 & $B^{0}\to D^{*-}+\rho^{+}$ & $D_{1}$ & $14.58 \pm 0.73$ &  & $6.8\pm0.9$ & $10^{-3}$\\
13 & $B^{0}\to\rho^{-}+D_{s}^{*+}$ & $D_{1}$ & $5.09 \pm 0.25$ &  & $4.1\pm1.3$ & $10^{-5}$\\
14 & $B^{+}\to\rho^{0}+D_{s}^{*+}$ & $D_{1}$ & $0.275 \pm 0.01$ &  & $<4.0$ & $10^{-4}$\\
\cline{2-7}
15 & $B^{0}\to\pi^{0}+\overline{D}^{0}$ & $D_{2}$ & $0.085 \pm 0.00$ &  & $2.63\pm0.14$ & $10^{-4}$\\
16 & $B^{0}\to\pi^{0}+\overline{D}^{*0}$ & $D_{2}$ & $1.13 \pm 0.06$ &  & $2.2\pm0.6$ & $10^{-4}$\\
17 & $B^{0}\to\rho^{0}+\overline{D}^{0}$ & $D_{2}$ & $0.675 \pm 0.03$ &  & $3.21\pm0.21$ & $10^{-4}$\\
18 & $B^{0}\to\rho^{0}+\overline{D}^{*0}$ & $D_{2}$ & $1.50 \pm 0.08$ &  & $<5.1$ & $10^{-4}$\\
\cline{2-7}
19 & $B^{+}\to\overline{D}^{0}+\pi^{+}$ & $D_{3}$ & $3.89 \pm 0.19$ &  & $4.68\pm0.13$ & $10^{-3}$\\
20 & $B^{+}\to\overline{D}^{0}+\rho^{+}$ & $D_{3}$ & $1.83 \pm 0.09$ &  & $1.34\pm0.18$ & $10^{-2}$\\
21 & $B^{+}\to\overline{D}^{*0}+\pi^{+}$ & $D_{3}$ & $7.60 \pm 0.38$ &  & $4.9\pm0.17$ & $10^{-3}$\\
22 & $B^{+}\to\overline{D}^{*0}+\rho^{+}$ & $D_{3}$ & $11.75 \pm 0.59$ &  & $9.8\pm1.7$ & $10^{-3}$\\
\hline 
\hline 
 &  &  &  &  &  & \\
\end{tabular}
\caption{CCQM branching fractions compared to data. Table was originally published in \cite{Dubnicka:2022gsi}.}
\label{Tab:BDhres}
\end{table}
The level of agreement between the model and the data can be visually estimated by looking at Fig. \ref{Fig:hadResults}.
Generally speaking, the description of data is not satisfactory. The agreement within errors is reached for measurements where only limits are given and for few other cases. This might be expected for a subset of the processes since the factorization assumption is not supposed to hold in the scenario where the spectator quark enters the light meson, see \cite{Beneke:2000ry}. Yet one sees an overall overestimation including decays with the spectator quark entering the $D$ meson. This observation joins similar observations made by other authors \cite{Huber:2016xod,Bordone:2020gao,Iguro:2020ndk,Cai:2021mlt}, i.e. it is seen across various approaches which naturally rises the question about the NP. The authors of \cite{Iguro:2020ndk} talk about "novel puzzle" and NP scenarios are advanced to explain it in \cite{Iguro:2020ndk,Cai:2021mlt}.

\subsection{Other CCQM results on nonleptonic $B$ decays.}

The CCQM was also applied to other hadronic decay processes of $B$ mesons. Skipping older publications \cite{Ivanov:2002un,Ivanov:2006ni} with an earlier version of the model, we can mention again the generally oriented text \cite{Ivanov:2011aa} where decay width for $B_s$ going to $D_s^- + D_s^{(*)+}$, $D_s^{*-} + D_s^{(*)+}$ and $J/\Psi + \Phi$ are computed. They are determined within the effective Hamiltonian approach using the helicity formalism from the CCQM-predicted form factors. The numbers are in fair agreement with experimental measurements. The same results are reviewed in paper \cite{Dineykhan:2012cp}, which, in addition, treats the exotic state $X(3872)$ as a tetraquark and evaluates its selected branching fractions.

The work \cite{Dubnicka:2017job} deals with double-heavy $B_c$ particles and their decays to charmonia and various $D$ mesons. Two diagrams contribute in the leading order, in one the $B_c$ spectator quark $\bar{c}$ goes to the charmonium state, in the other it forms the $D$ meson. One thus needs to evaluate form factors of six transitions $B_c \to D,D_s,\eta_c,D^*,D_s^*, J/\Psi$  , their behavior is shown in Fig. 2 of the work and their values at zero are also presented. Next, helicity amplitudes are constructed and branching fractions calculated for in total 8 processes $B_s \to \eta_c + D_{(s)}^{(*)}$ and $B_s \to J/\Psi + D_{(s)}^{(*)}$  (all combinations of brackets). Comparison with the experiment is based on branching fraction ratios $\mathcal{R}(D_s^+/\pi^+)$, $\mathcal{R}(D_s^{*+}/\pi^+)$, $\mathcal{R}(D_s^+/ D_s^+)$ and also $\Gamma_{++}/ \Gamma$ measured by Atlas \cite{ATLAS:2015jep} and LHCb \cite{LHCb:2013kwl}. Here
\begin{align}
\mathcal{R}(A/B) = \frac{\mathcal{B}(B_c^+ \to J/\Psi A)}{\mathcal{B}(B_c^+ \to J/\Psi B)} \label{Rfrac}
\end{align}
and $\Gamma_{++}/ \Gamma$ is the transverse polarization fraction in the $B_c^+ \to J/\Psi + D_s^{*+}$ decay. The results are presented in the Tab. VIII of \cite{Dubnicka:2017job} with no significant deviations from the SM. Yet, as two different sets of Wilson coefficients were investigated, it turned out that the results are quite sensitive to their choice.

Similar processes are addressed in \cite{Issadykov:2018myx}, however with $\pi$ or $K$ in the final state instead of $D$. Consequently only one diagram contributes which is the one corresponding to the transition to charmonium, since all other $\pi/K$ production diagrams from $B_c$ are of a higher order. Also the semileptonic mode to $J/\Psi \mu \nu_\mu$ is investigated so as to define observables $\mathcal{R}(\pi^+/\mu^+\nu)$, $\mathcal{R}(K^+/\pi^+)$, $\mathcal{R}(J/\Psi)$ and $\mathcal{R}(\eta_c)$, see (\ref{eq:RD*}), (\ref{Rfrac}). With the CCQM transition form factors identical to those mentioned previously one gets in total eight decay widths $B_c^+ \to \eta_c + h$, $B_c^+ \to J/\Psi + h$, $h \in \{\pi^+, \rho^+ , K^+, K^{*+}\}$ (Tab. 3 of the publication) and branching fraction ratios which can be compared to the LHCb numbers (Tab. 5 of \cite{Issadykov:2018myx}) and also to other theoretical works. The ratios are in an agreement with measurements except for $\mathcal{R}(J/\Psi)$, which deviates more than $2 \sigma$.

Let us, at last, mention the paper \cite{Ivanov:2022nnq} dedicated to vector particles $B^*$ and $B_s^*$ and their transition to $B_{(s)} \gamma$ and $D_{(s)}^* + V$, $V \in \{ \rho, K^*, D^*, D_s^* \}$. The radiative deexitation processes use the formalism presented in Sec. \ref{Sec:EM} to describe the decay: a photon can be radiated from one of the valence quarks or from the non-local quark-hadron vertex. In the latter case, however, it can be shown that the contribution vanishes due to the anomalous nature of the $V \to P \gamma$ process and so the calculation is simplified. The results on  decay widths of $B^+$, $B^0$ and $B_s^{*0}$, presented in Tab. V of the work, depend on radiative decay constants of the particles given in Tab. IV. For what concerns the decays to two vector particles, the computation proceeds in a usual way, where the CCQM invariant form factors are combined to helicity amplitudes to give branching fractions. Due to small cross sections of the studied processes the experimental numbers are not available and so the CCQM results are compared to other theoretical approaches (Tab. XII of \cite{Ivanov:2022nnq}).

\section{Summary and outlook}
We provided in this text a review of the results of the covariant confined quark model for $B$ decays presented together with a survey of selected experimental and theoretical results. Unlike for other physics models and their achievements mentioned here, we explained in depth the principles of the CCQM  (Sec. \ref{Sec: CCQM}) and presented computational details for chosen processes, namely $B_s \to \ell^+ \ell^- \gamma$ (Sec. \ref{Sec:leptoCCQM}), $B_s \to \phi \ell^+ \ell^- \gamma$ (Sec. \ref{SecCCQMsemi}), $B \to D_{(s)}^{(*)}h$, $(h=\pi,\rho)$ and $B_s \to J/\psi \eta^{(')}$ (Sec. \ref{Sec_CCQM_had}). For the sake of the review the decays were divided into three groups: leptonic, semileptonic and non-leptonic. Although somewhat arbitrary, this division allowed us to demonstrate the application of the CCQM in various situations.  Generally speaking, despite some studies on NP contributions, the CCQM results do not provide strong indications for NP and suggest that further efforts within the SM may be needed.

One should also recall that we presented only a small section of what the CCQM can provide: it was, in many papers, successfully applied to describe baryon, tetraquark and other (than $B$) mesonic states. The quality of the CCQM is also confirmed by the interest of other authors. Narrowing the large number of citations to those related to $B$ decays and referring to the recent version of the model (2010 and later, without conference papers) one sees that the model was noticed by large collaborations (LHCb \cite{LHCb:2021vsc,LHCb:2021awg}, ATLAS \cite{ATLAS:2022aiy}).

The ongoing physics program on existing and future high-luminosity machines implies that the CCQM may also in the future be an appropriate theoretical tool which will contribute to unraveling the questions brought by experiments about the presence of NP or the nature of various (exotic) states. Together with other approaches, it may help to understand model-related uncertainties beyond which new physics observations can be claimed.

\section*{Acknowledgement}
S. D., A. Z. D. and A. L.  acknowledge the support of the Slovak Grant Agency for Sciences VEGA, grant no. 2/0105/21.

\bibliography{BinCCQM_ArXive}{}
\bibliographystyle{unsrt}

\end{document}